\documentclass[fleqn, usenatbib]{mnras}

\usepackage{newtxtext,newtxmath}

\usepackage[T1]{fontenc}

\DeclareRobustCommand{\VAN}[3]{#2}
\let\VANthebibliography\thebibliography
\def\thebibliography{\DeclareRobustCommand{\VAN}[3]{##3}\VANthebibliography}

\usepackage{graphicx}	
\usepackage{amsmath}	

\usepackage{listings}
\usepackage{color}
\definecolor{dkgreen}{rgb}{0,0.6,0}
\definecolor{gray}{rgb}{0.5,0.5,0.5}
\definecolor{mauve}{rgb}{0.58,0,0.82}
\definecolor{golden}{rgb}{0.86,0.65,0.01}
\defcitealias{Rappaport1983}{RVJ}
\defcitealias{Van2019}{V19}
\defcitealias{Sills2000}{S00}
\defcitealias{Garraffo2016}{G16}

\title[Magnetic braking saturates]{Magnetic braking saturates: evidence from the orbital period distribution of low-mass detached eclipsing binaries from ZTF}  

\author[El-Badry et al.]{
Kareem El-Badry$^{1,2,3}$\thanks{E-mail: kareem.el-badry@cfa.harvard.edu}, Charlie Conroy$^{1}$, Jim Fuller$^4$, Rocio Kiman$^{5}$, Jan van Roestel$^{6,4}$, \newauthor Antonio C. Rodriguez$^4$, Kevin B. Burdge$^7$ \\
$^{1}$Center for Astrophysics $|$ Harvard \& Smithsonian, 60 Garden Street, Cambridge, MA 02138, USA\\
$^{2}$Harvard Society of Fellows, 78 Mount Auburn Street, Cambridge, MA 02138\\
$^{3}$Max-Planck Institute for Astronomy, K\"onigstuhl 17, D-69117 Heidelberg, Germany\\
$^4$Department of Astronomy, California Institute of Technology, Mailcode 350-17, Pasadena, CA 91125, USA\\
$^{5}$Kavli Institute for Theoretical Physics, University of California, Santa Barbara, CA 93106, USA \\
$^{6}$Anton Pannekoek Institute for Astronomy, University of Amsterdam, 1090 GE Amsterdam, The Netherlands \\
$^7$MIT-Kavli Institute for Astrophysics and Space Research 77 Massachusetts Ave. Cambridge, MA 02139, USA
}
 
\date{\vspace{-1.0cm}}

\pubyear{2022}

\begin{document}
\label{firstpage}
\pagerange{\pageref{firstpage}--\pageref{lastpage}}
\maketitle

\begin{abstract}
We constrain the orbital period ($P_{\rm orb}$) distribution of low-mass detached main-sequence eclipsing binaries (EBs) with light curves from the Zwicky Transient Facility (ZTF), which provides a well-understood selection function and sensitivity to faint stars.
At short periods ($P_{\rm orb}\lesssim 2$ days), binaries are predicted to evolve significantly due to magnetic braking (MB), which shrinks orbits and ultimately brings detached binaries into contact. The period distribution is thus a sensitive probe of MB.
We find that the intrinsic period distribution of low-mass ($0.1\lesssim M_1/M_{\odot} < 0.9$)  binaries is basically flat (${\rm d}N/{\rm d}P_{\rm orb} \propto P_{\rm orb}^0$),  from $P_{\rm orb}=10$ days down to the contact limit.
This is strongly inconsistent with predictions of classical MB models based on the Skumanich relation, which are widely used in binary evolution calculations and predict ${\rm d}N/{\rm d}P_{\rm orb} \propto P_{\rm orb}^{7/3}$ at short periods. 
The observed distributions are best reproduced by models in which the magnetic field saturates at short periods, with a MB torque that scales roughly as $\dot{J}\propto P_{\rm orb}^{-1}$, as opposed to $\dot{J} \propto P_{\rm orb}^{-3}$ in the standard Skumanich law.
We also find no significant difference between the period distributions of binaries containing fully and partially convective stars. 
Our results confirm that a saturated MB law, which was previously found to describe the spin-down of rapidly rotating isolated M dwarfs, also operates in tidally locked binaries. We advocate using saturated MB models in binary evolution calculations.
Our work supports previous suggestions that MB in cataclysmic variables (CVs) is much weaker than assumed in the standard evolutionary model, unless mass transfer leads to significant additional angular momentum loss in CVs. 
\end{abstract}

\begin{keywords}
binaries: eclipsing -- binaries: close -- stars: rotation -- stars: magnetic field -- novae, cataclysmic variables
\vspace{-0.5cm}
\end{keywords}



\section{Introduction}
\label{sec:intro}

Stellar magnetic fields cause mass lost in a stellar wind to co-rotate with the star, such that the terminal specific angular momentum of the wind is larger than that of the stellar surface. This causes single stars to spin down over time, a process known as magnetic braking \citep[MB; e.g.,][]{Schatzman1962, Weber1967, Mestel1968, Mestel1987}. In close binaries, tides lock rotation periods to the orbital period, preventing spin-down of the component stars. In this case, the angular momentum debt of the stellar wind is instead repaid by loss of orbital angular momentum, ultimately shrinking the binary orbit and increasing the spin rate of the component stars \citep[e.g.][]{Verbunt1981}. 

MB provides a mechanism to bring detached binaries into contact and is thus an important ingredient in the evolution of close binaries. It determines the lifetime, mass transfer rate, and many of the basic observables of cataclysmic variables (CVs; \citealt{Warner_2003, Knigge_2011}). 
Only in short-period CVs with $P_{\rm orb}\lesssim 2$ hours is MB thought to become subdominant to gravitational wave radiation in driving angular momentum loss. Even at short periods, there is evidence of additional angular momentum loss beyond gravitational waves; MB is one possible explanation \citep{Knigge_2011}. MB plays a similarly important role in the formation and evolution of low-mass X-ray binaries (LMXBs) and main-sequence contact binaries \citep{Podsiadlowski2002, Li2004}. 

Despite its importance for binary evolution, MB is poorly understood. The most widely used MB prescriptions are empirical, and were motivated by early observations of the rotation rate evolution of single, solar-type stars. These prescriptions are often used in models of close binaries hosting a compact object and a rapidly-rotating, low-mass star. When extrapolated to the low masses and rapid rotation rates relevant for CVs and LMXBs, prescriptions for the MB torque in the literature differ by several orders of magnitude \citep[e.g.][]{Knigge_2011}. 
The prescription that is by far the most widely used in binary evolution studies \citep{Verbunt1981, Rappaport1983} was calibrated to pioneering measurements of the rotation rates of solar-type stars \citep{Skumanich1972, Smith1979}. More recent studies have found evidence that the rotation period evolution of low-mass main-sequence stars is more complicated than suggested by those early works, particularly at high rotation rates \citep[e.g.][]{Stauffer1997, Barnes2003, McQuillan2013, vanSaders2016, Newton2016}. 

Attempts have also been made to infer a MB law directly from the observed mass transfer rates and period distribution of CVs \citep[e.g.][]{Patterson_1984}. However, such efforts are complicated by several factors: (a) mass transfer rates in CVs are difficult to measure and vary on both short and long timescales, (b) a wide range of mass transfer rates are observed at fixed orbital period in the period range where MB is expected to drive mass transfer \citep[e.g.][]{Townsley2009, Pala2017, Pala2022}, (c) CVs may experience additional angular momentum losses as a consequence of the mass transfer process, which are not easily decoupled from MB,  and (d) observed CV samples are affected by complicated selection effects because their observable properties depend strongly on mass transfer rate, and thus on the strength of MB itself. 

In this paper, we constrain the strength of MB in close binaries by studying the period distribution of short-period main-sequence binaries. This approach has two clear advantages over previous works: (1) the stars involved are rapidly-rotating, low-mass  stars in tidally locked binaries; i.e, the same kind of stars found in CVs and LMXBs, and (2) the selection function of a search for eclipsing main-sequence binaries is straightforward to model. High-quality light curves from ZTF and astrometry from {\it Gaia} allow us to probe lower masses and shorter periods than was possible until recently.

Many previous studies have assembled large samples of EBs and studied their period distribution \citep[e.g.][]{Farinella1979, Giuricin1983, Giuricin1984, Maceroni1991, Rucinski1998, Maceroni1999,  Rucinski2007, Norton2011, Derekas2007}. However, these studies have had poorly understood selection effects and in most cases included binaries with a wide range of masses and evolutionary states, making population modeling difficult. As expected for flux-limited samples, most studies have been dominated by binaries with components more luminous than the Sun, in which MB is expected to be less important.
The primary innovations of our work are well-known distances and stellar masses, a reasonably well-understood selection function, and sensitivity to faint, low-mass sources down to the bottom of the main sequence. 



This paper is organized as follows. Section~\ref{sec:theory} summarizes classical MB theory and the relation between the MB law and the orbital period distribution. Section~\ref{sec:data} describes our search for detached eclipsing binaries (EBs) using {\it Gaia} and ZTF. Section~\ref{sec:results}  presents our constraints on the incompleteness-corrected period distribution and compares to predictions of several magnetic braking models. We discuss the implications of our results for  CVs and LMXBs in  Section~\ref{sec:discussion}, and we summarize the main results in Section~\ref{sec:summary}. Additional details about our modeling and EB sample are provided in the Appendices.

\section{Magnetic braking theory}
\label{sec:theory}

\subsection{Classical models}
The MB model most widely used in binary evolution calculations was developed by \citet{Verbunt1981} and \citet[][hereafter RVJ]{Rappaport1983}. The model was motivated by the observation that among solar-type stars, rotation velocities decline with age, roughly as
\begin{equation}\
    \label{eq:vrot}
    v_{{\rm rot}}\propto t^{-1/2},
\end{equation}
where $t$ is the age of a star \citep{Skumanich1972, Smith1979}. Rotation velocity scales as $v_{\rm rot} \propto R/P_{\rm rot}$, with $P_{\rm rot}$ the rotation period and $R$ the stellar radius. This implies a spin-down $\dot{P}_{\rm rot}  \propto R^2\,P_{\rm rot}^{-1}$, from which one can empirically infer an effective torque due to MB as follows.

The rotational angular momentum of a single star is $J_{{\rm rot}}=2\pi IP_{{\rm rot}}^{-1}\propto MR^{2}P_{{\rm rot}}^{-1}$, where $I$ is its moment of inertia and $M$ its mass. In the absence of changes to $M$ and $R$, this implies $\dot{J}_{{\rm rot}}\propto-MR^{2}P_{{\rm rot}}^{-2}\dot{P}_{{\rm rot}}$. Inserting the expression for $\dot{P}_{\rm rot}$ implied by the Skumanich law (and ascribing the observed spin-down to MB) then implies a MB torque 
\begin{equation}
    \dot{J}\propto-MR^{4}P_{{\rm rot}}^{-3}.
\end{equation}
The proportionality constant can be estimated from the observed normalization of Equation~\ref{eq:vrot}. The most straightforward approach is to scale to solar properties (e.g. \citetalias{Rappaport1983}): 

\begin{equation}
    \label{eq:jdot}
    \dot{J}_{{\rm RVJ}} =-\tau_0 \left(\frac{M}{M_{\odot}}\right)\left(\frac{R}{R_{\odot}}\right)^{4}\left(\frac{P_{{\rm rot}}}{1\,{\rm d}}\right)^{-3},
\end{equation}
where $\tau_0 \approx 6.8\times10^{34}\,{\rm erg}$.\footnote{A simple estimate of $\tau_0$ can be calculated via $\tau_{0}\approx\frac{J_{\odot}}{2t_{\odot}}\left(\frac{P_{{\rm rot},\odot}}{1\,{\rm d}}\right)^{3}\approx 1.05\times 10^{35}\,\rm erg$, where $P_{\rm rot,\odot}\approx 25\,\rm d$ is the Sun's equatorial rotation period, $J_{\odot}= 2\pi \beta_{\odot} M_{\odot} R_{\odot}^2/P_{\rm rot,\odot}$ is its rotational angular momentum, $\beta_{\odot} \approx 0.07$ is a dimensionless parameter related to its internal structure, and $t_{\odot}\approx 4.6\,\rm Gyr$ is its age. However, the Sun's differential rotation, as well as changes in its interior structure during its main-sequence evolution, complicate a precise measurement of this parameter. We adopt $\tau_0 = 6.8\times 10^{34}\,\rm erg$ for consistency with \citetalias{Rappaport1983}.}  Because most of the observed stars in early studies that motivated Equation~\ref{eq:vrot} had similar masses and radii, the dependence on $M$ and $R$ is uncertain. \citetalias{Rappaport1983} parameterized Equation~\ref{eq:jdot} as $\dot{J} =-\tau_0 \left(M/M_{\odot}\right)\left(R/R_{\odot}\right)^{\gamma}\left(P_{{\rm rot}}/1\,{\rm d}\right)^{-3}$ to capture this uncertainty. We will consider the most natural case with $\gamma=4$ in our analysis; this is also a common choice in the literature. A lower value of $\gamma$ would lead to stronger MB in low-mass stars. 

For a single star with constant mass and radius, the  angular momentum reservoir is rotation, with $J_{\rm rot}\propto MR^2/P_{\rm rot}$. 
Equation~\ref{eq:jdot} is then a differential equation that can be solved for $P_{\rm rot}(t)$: for a given initial rotation period, it implies a spin-down with asymptotic $\dot{P}_{{\rm rot}}\propto t^{-1/2}$. If their spin-down evolves according to Equation~\ref{eq:jdot}, the rotation period distribution of a sample of stars with a uniform age distribution will be skewed toward long periods, since individual rotation periods evolve more slowly at long $P_{\rm rot}$.

The \citetalias{Rappaport1983} MB law has been {\it extremely} widely used in models of close binary evolution over the last 4 decades \citep[e.g.][to list only some of the most influential studies]{Kolb1993, Tauris1999, Howell2001, Podsiadlowski2002, Knigge_2011, Istrate2014b, Paxton_2015, Kalomeni2016, Belloni2018, Schreiber2021}. 
It should be emphasized, however, that Equations~\ref{eq:vrot}-\ref{eq:jdot} were inferred from a sample of G-type main sequence stars with rotation periods ranging from about 2 to 30 days. MB is expected to produce significant orbital evolution only at short periods ($P \lesssim \rm few\,\,days$). Moreover, while Equation~\ref{eq:vrot} was motivated by observations of G dwarfs,  most of the binaries in which the role of MB is of greatest interest (e.g. CVs and LMXBs) have K and M type donors.
In short, applying Equation~\ref{eq:jdot} to the donors in CVs and LMXBs requires considerable extrapolation.

\subsection{Application to binaries}
\label{sec:application_binaries}
In close binaries, tides lead to the synchronization of rotation with the orbital period \citep[e.g.][]{Zahn1977}. Observations show that synchronization is efficient in main-sequence binaries with orbital periods below about 10 days, with weak dependence on mass \citep[e.g.][]{Lurie2017}. In a tidally synchronised binary, the MB torque due to winds from either star cannot spin the stars down, since their rotation is locked to the orbital period. The result is that angular momentum is removed from the orbit instead, while the component stars remain tidally locked.  This leads to a decrease of the orbital period and ultimately a spin-up of the component stars, in contrast to the single-star case.

\begin{figure*}
    \centering
    \includegraphics[width=\textwidth]{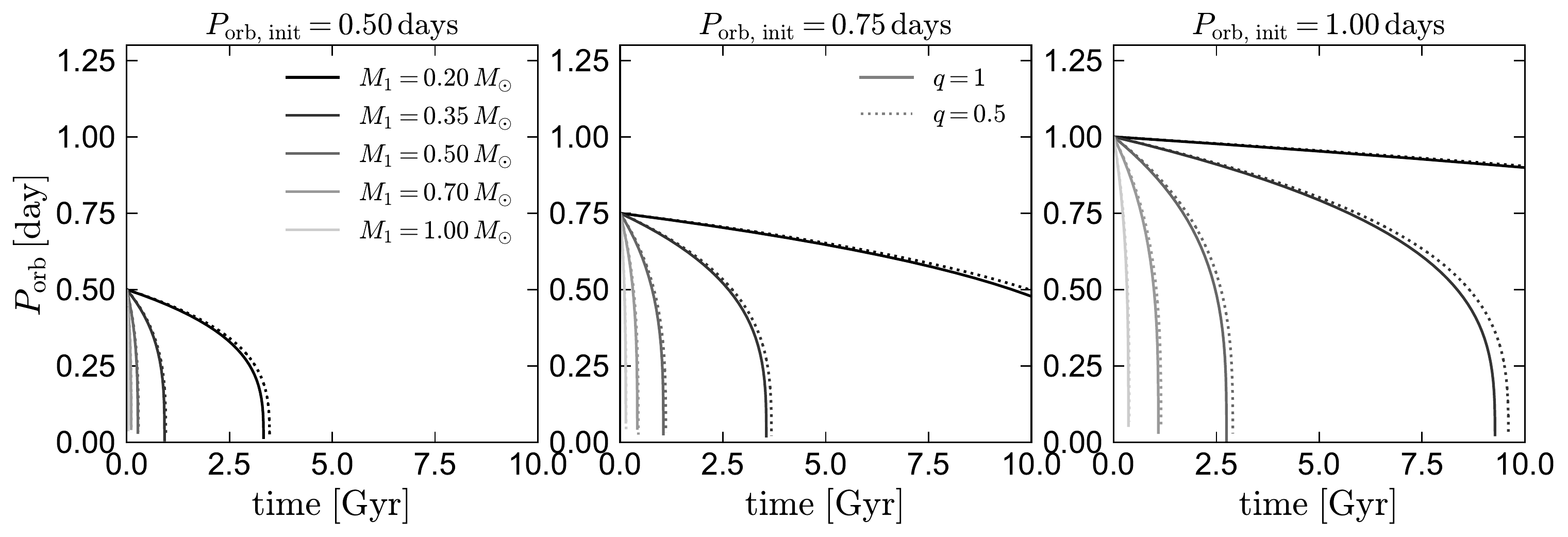}
    \caption{Predicted orbital period evolution of binaries in which both components are tidally locked and subject to a MB torque derived from the Skumanich relation (Equation~\ref{eq:jdot}). The three panels show different initial periods; different colored lines show different primary masses. Solid lines show equal-mass components ($q=1$); dotted lines show $q=0.5$. At fixed initial period, inspiral is faster for higher-mass binaries. Sensitivity to mass ratio is relatively weak. 
    A generic feature of all tracks is that {\it inspiral speeds up at short $P_{\rm orb}$}. That is, the time that a  binary spends in a given period bin decreases with period.} 
    \label{fig:porb_tracks}
\end{figure*}

The orbital angular momentum of a binary is 
\begin{equation}
    \label{eq:jorb}
    J_{{\rm orb}}=\frac{M_{1}M_{2}}{\left(M_{1}+M_{2}\right)}\sqrt{G\left(M_{1}+M_{2}\right)a\left(1-e^{2}\right)},
\end{equation}
where $M_1$ and $M_2$ represent the masses of the component stars, $a$ is the semimajor axis, and $e$ is the eccentricity. We consider a circular orbit ($e=0$), since tidal circularization is also efficient at the short orbital periods of interest. Then in terms of orbital period,

\begin{equation}
    \label{eq:jorb_circ}
    J_{{\rm orb}}=P^{1/3}_{\rm orb}M_{1}^{5/3}\frac{q}{\left(1+q\right)^{1/3}}\left(\frac{G^{2}}{2\pi}\right)^{1/3},
\end{equation}
where $q=M_2/M_1$ is the mass ratio. At fixed $P_{\rm orb}$ and $M_1$, $J_{\rm orb}$ is maximized for equal-mass binaries, with nearly linear dependence on $q$. For all tidally synchronized binaries of interest, the orbital angular momentum dominates over the rotational angular momentum. E.g., in a binary with two equal-mass, equal-radius stars,

\begin{equation}
    \label{eq:jorb_jrot}
    \frac{J_{{\rm orb}}}{J_{{\rm rot}}}=70\times\left(\frac{P_{{\rm orb}}}{1\,{\rm d}}\right)^{4/3}\left(\frac{M}{M_{\odot}}\right)^{2/3}\left(\frac{R}{R_{\odot}}\right)^{-2}\left(\frac{\beta}{0.1}\right)^{-1},
\end{equation}
where $\beta = I/(M\,R^{2})$ and $J_{\rm rot}$ includes the rotational angular momentum of both stars. We therefore consider only orbital angular momentum in subsequent calculations. 

If both components of a binary remove orbital angular momentum according to equation~\ref{eq:jdot} while the binary remains tidally locked, one can predict the evolution of $P_{\rm orb}(t)$ analytically. Details are given in Appendix~\ref{sec:analytic_appendix}. We are primarily interested in low-mass main sequence binaries in which both components have lifetimes longer than a Hubble time, so we assume that the component masses and radii do not evolve.  

Figure~\ref{fig:porb_tracks} shows predicted tracks of $P_{\rm orb}(t)$  for main-sequence binaries with a range of masses and initial periods, assuming the angular momentum loss from each component follows Equation~\ref{eq:jdot}. We assign radii for each mass using solar-metallicity zero age-main sequence MIST models \citep[][]{Choi_2016}, and show two different mass ratios for each primary mass. The predicted inspiral is much more rapid at higher binary masses. For example, a binary containing two $1\,M_{\odot}$ stars in a 1-day orbit is predicted to come into contact within 0.6\,Gyr, while the same inspiral in a binary containing two $0.35\,M_{\odot}$ stars is predicted to take 9 Gyr. This is primarily a consequence of the strong ($\propto R^4$) dependence in $\dot{J}$ predicted by Equation~\ref{eq:jdot}. 

For all initial masses and periods, the inspiral accelerates as the binary shrinks: it takes much longer for a binary to evolve from $P_{\rm orb}=1$ day to 0.5 days, than from 0.5 days to 0 days. For a low-mass main-sequence binary in which both components follow a mass-radius relation $R\propto M$, the expected period evolution follows

\begin{equation}
    \label{eq:dpdt}
    \frac{ {\rm d} P_{\rm orb}}{ {\rm d} t}\propto -P_{\rm orb}^{-7/3}M_{1}^{10/3}\frac{\left(1+q^{5}\right)\left(q+1\right)^{1/3}}{q}.
\end{equation}
Note in particular the strong dependence on $P_{\rm orb}$ and weak dependence on $q$.\footnote{Equation~\ref{eq:dpdt} predicts a factor of 2 variation over $0.3 < q < 1$. It predicts very rapid inspiral in the limit of $q\to 0$ (e.g., a planet-mass companion), reflecting the low available angular momentum in this limit. However, in this case (which does not apply to any binaries in our study) the primary will not be tidally synchronized, and so MB will not actually shrink the orbit.  } At sufficiently short periods, most binary orbits are expected to have evolved due to MB by the time they are observed. In this case, we expect the observed period distribution, ${\rm d}N/{\rm d}P_{\rm orb}$, to be proportional to $\left|{\rm d}P_{{\rm orb}}/{\rm d}t\right|^{-1}$. Thus, a Skumanich-like MB law predicts that at short periods, the period distribution follows

\begin{align}
    \label{eq:dndp}
    \frac{{\rm d}N}{{\rm d}P_{{\rm orb}}}\propto P_{{\rm orb}}^{7/3}.
\end{align}

\subsection{Other magnetic braking prescriptions}
\label{sec:other_MB}
Various alternatives to Equation~\ref{eq:jdot} for the MB torque due to a single star have been proposed. We review several models below and compare them in Figure~\ref{fig:mb_laws}.

\subsubsection{``Saturated'' magnetic braking}
\label{sec:saturated}

Chromospheric activity, coronal X-ray emission, flare activity, and magnetic field strengths in low-mass main-sequence stars are all highly correlated. Observations indicate that all these observables increase with rotation rate up to a mass-dependent critical rotation rate, above which the activity-rotation rate relations reach a plateau \citep[e.g.][]{Stauffer1994, Delfosse1998, Reiners2009, Newton2017, Johnstone2021}. Motivated by these observations, several works have introduced ``saturated'' MB laws, in which the scaling of the MB torque with rotation rate becomes shallower above a given rotation rate. \citet[][hereafter S00]{Sills2000} used the following parameterization:

\begin{equation}
    \label{eq:jdot_sat}
\dot{J}_{{\rm sat}}=\begin{cases}
-\tau_{1}\left(\frac{P_{{\rm rot}}}{1\,{\rm d}}\right)^{-3}\left(\frac{R}{R_{\odot}}\right)^{1/2}\left(\frac{M}{M_{\odot}}\right)^{-1/2}, & P_{{\rm rot}}\geq P_{{\rm crit}}\\
-\tau_{1}\left(\frac{P_{{\rm rot}}}{1\,{\rm d}}\right)^{-1}\left(\frac{P_{{\rm crit}}}{{\rm 1\,d}}\right)^{-2}\left(\frac{R}{R_{\odot}}\right)^{1/2}\left(\frac{M}{M_{\odot}}\right)^{-1/2}, & P_{{\rm rot}}<P_{{\rm crit}}
\end{cases}
\end{equation}
Here $\tau_1=1.04\times 10^{35}\,\rm erg$ is a calibrated constant, and $P_{\rm crit}$ depends on stellar mass. This form of the MB law results in a $\propto P_{\rm rot }^{-3}$ scaling at long periods,  and a $\propto P_{\rm rot }^{-1}$ scaling at short periods. Similar relations were used earlier by \citet[][]{Kawaler1988} and \citet[][]{Chaboyer1995}. 

For the saturation threshold, we adopt $P_{{\rm crit}} = 0.1 P_{\odot}(\tau_{c}/\tau_{c,\odot})$. Here $\tau_{c}$ is the convective turnover timescale, which we calculate from the empirical determination of \citet[][their Equation 11]{Wright2011}, and $P_{\odot} = 28\,\rm days$. The saturation threshold can be more succinctly expressed in terms of the Rossby number, ${\rm Ro}=P_{\rm rot}/\tau_c$: saturation occurs at ${\rm Ro} \lesssim 0.1\,{\rm Ro}_{\odot}$. The convective turnover timescale -- and the period below which magnetic braking is saturated -- increases with decreasing stellar mass for main-sequence stars. For the saturation threshold adopted here, the saturation period increases from $P_{\rm crit} = 2.7\,\rm d$ at $M=1\,M_{\odot}$, to $P_{\rm crit} = 5.4\,\rm d$ at $M=0.6\,M_{\odot}$, to $P_{\rm crit} = 11.9\,\rm d$ at $M=0.3\,M_{\odot}$. This means that essentially {\it all} binaries in the period range where MB sets the period distribution are expected to be in the saturated regime. 

A modified saturated model was proposed by \citet[][]{Matt2015}, who fit the rotation period distributions of cluster and field stars observed by {\it Kepler}. They found 

\begin{equation}
    \label{eq:jdot_m15}
\dot{J}_{{\rm M15}}=\begin{cases}
-\tau_{2}\left(\frac{P_{{\rm rot}}}{1\,{\rm d}}\right)^{-3}\left(\frac{\tau_{c}}{\tau_{c,\odot}}\right)^{2}\left(\frac{R}{R_{\odot}}\right)^{3.1}\left(\frac{M}{M_{\odot}}\right)^{0.5}, & P_{{\rm rot}}\geq P_{{\rm crit}}\\
-\tau_{2}\left(\frac{P_{{\rm rot}}}{1\,{\rm d}}\right)^{-1}\left(\frac{P_{{\rm crit}}}{1\,{\rm d}}\right)^{-2}\left(\frac{\tau_{c}}{\tau_{c,\odot}}\right)^{2}\left(\frac{R}{R_{\odot}}\right)^{3.1}\left(\frac{M}{M_{\odot}}\right)^{0.5}, & P_{{\rm rot}}<P_{{\rm crit}}
\end{cases}
\end{equation}
Here $\tau_2=1.38\times 10^{35}\,\rm erg$ is a calibrated constant, and we use the same $P_{\rm crit}$ and $\tau_c$ described above. This results in a similar MB law to the \citetalias{Sills2000} prescription, but with a  stronger dependence on radius (and thus mass).

In the saturated regime, the same steady-state considerations that led to Equation~\ref{eq:dndp} instead predict
\begin{align}
    \label{eq:dndp_sat}
    \frac{{\rm d}N}{{\rm d}P_{{\rm orb}}}\propto P_{{\rm orb}}^{1/3}.
\end{align}

Another quasi-saturated MB prescription was presented by \citet{Ivanova2003}, who inferred a MB law from the observed scaling of X-ray luminosity with rotation rate in single stars. There is no explicit saturation of the magnetic field in their model, but their analysis implied that at high rotation rates an increased fraction of magnetic field lines are closed, leading to less efficient MB. They found $\dot{J}\propto P_{\rm rot}^{-1.3}$ in the saturated regime, only moderately steeper than the models from \citetalias[][]{Sills2000} and \citet{Matt2015}. A stronger scaling with $R$ in their model makes the predicted MB in low-mass stars significantly weaker than in the \citetalias[][]{Sills2000} and \citet{Matt2015} models.

\subsubsection{Complexity-modulated magnetic braking}
\label{sec:garaffo}
Another class of models, popularized by \citet[][]{Garraffo2015}, proposes that MB becomes inefficient as the complexity (i.e., multipole order) of the magnetic field increases. They find that for quadrupolar and higher-order magnetic fields, the mass loss rate and magnetic open flux decline, and the Alfven surface shrinks, compared to a simple dipolar field. All these factors  diminish angular momentum loss. Observations indeed indicate that the field complexity of low-mass main-sequence stars increases with rotation rate \citep[e.g.][]{Donati2009}, potentially leading to less efficient MB for fast rotators. 

\citet[][hereafter G16]{Garraffo2016} express the effective MB torque in terms of a fitting function calibrated to MHD simulations: 
\begin{equation}
    \label{eq:jdot_g16}
    \dot{J}_{{\rm G16}}=\tau_{3}\left(\frac{P_{{\rm rot}}}{1\,{\rm d}}\right)^{-3}\left(\frac{\tau_{c}}{\tau_{c,\odot}}\right)Q_{J}\left(n\right).
\end{equation}
Here $\tau_3=4.8\times 10^{34}\,\rm erg$, and $Q_J(n)$ is a modulating factor that depends on the magnetic complexity, $n$. \citetalias[][]{Garraffo2016} provide fitting functions: 
\begin{equation}
    \label{eq:QJn}
    Q_{J}\left(n\right)=4.05e^{-1.4n}+\frac{1}{60}\frac{n-1}{n}\left(\frac{B}{1\,{\rm G}}\right)^{-1},
\end{equation}
where $B$ is the surface magnetic field, and 
\begin{equation}
    \label{eq:n_G16}
    n=\frac{a}{{\rm Ro}}+1+b{\rm Ro},
\end{equation}
where $a=0.02$ and $b=2$ are parameters determined by fitting observed rotation rates of single stars in clusters. The value of $b$ only becomes important in the long-period regime and has no effect on our results here. In our simulations and model comparisons later in this paper, we assume $B= 1000\,G$ for $M\leq 0.5\,M_{\odot}$, $B= 100\,\rm G$ for $0.5 < M/M_{\odot}\leq 0.75\,M_{\odot}$, and $B= 10\,\rm G$ for $0.75 < M/M_{\odot}\leq 1\,M_{\odot}$, following typical observed magnetic field strengths \citep[e.g.][]{Reiners2022}. 

The \citetalias[][]{Garraffo2016} MB law results in a torque with a local maximum at a period of order a few days (Figure~\ref{fig:mb_laws}). The model can thus predict a bimodal period distribution for young clusters, as observed \citep[e.g.][]{Garraffo2018}. Stars are born with a range of rotation rates after decoupling their rotation from their protostellar disks. Those with rotation periods below the local maximum are predicted to experience a relatively weak braking torque due to their complex magnetic fields, and thus can remain at high rotation rates for a long time. However, the torque strengthens as the stars spin-down, eventually causing them to move rapidly from the fast-rotator to the slow-rotator sequence when they are near the maximum torque. Such a torque is similar to that required by the phenomenological ``metastable dynamo'' spin-down model introduced by \citet{Brown2014}.

\subsubsection{CARB Magnetic braking}
\label{sec:carb}
Another MB prescription we consider is the ``Convection And Rotation Boosted'' (CARB)  model introduced by \citet[][hereafter V19]{Van2019}. This model assumes a radial magnetic field whose strength is linearly proportional to rotation rate and convective turnover time, and a spherically symmetric wind that co-rotates with the star until it reaches the Alfven radius, which in their model is made smaller by the donor's rotation. This leads to a MB torque 

\begin{multline}
    \label{eq:jdot_carb}
    \dot{J}_{{\rm V19}}=-\frac{2}{3}\dot{M}^{-1/3}R^{14/3}\left(v_{{\rm esc}}^{2}+\frac{8\pi^2R^{2}}{P_{\rm rot}^2  K_{2}^{2}}\right)^{-2/3}\times \\ B_{\odot}^{8/3} \left(\frac{2\pi}{P_{{\rm rot},\odot}}\right)\left(\frac{P_{\rm rot}}{P_{\rm rot,\odot}}\right)^{-11/3}\left(\frac{\tau_{c}}{\tau_{c,\odot}}\right)^{8/3}
\end{multline}
Here $\dot M$ is the mass-loss rate due to winds, $v_{\rm esc}=\sqrt{2GM/R}$ is the escape velocity, $B_{\odot}=1\,\rm G$ is the solar surface magnetic field, and $K_2=0.07$ is a dimensionless constant.  Following \citetalias{Van2019}, we use the prescription from \citet[][]{Reimers1975} with $\eta = 1$ for the wind mass-loss rate, even though this prescription is calibrated to giants. We again calculate $\tau_c$ following \citet[][]{Wright2011}. 

\begin{figure*}
    \centering
    \includegraphics[width=\textwidth]{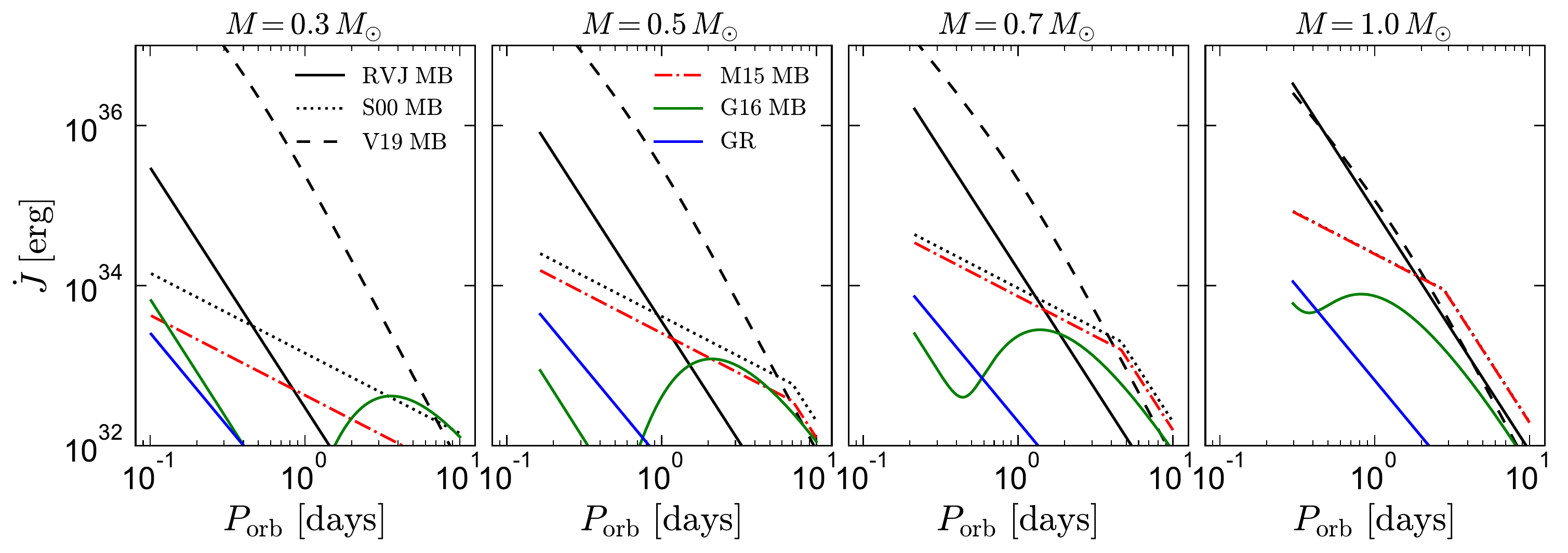}
    \caption{Predicted angular momentum loss for an equal-mass main-sequence binary due to several different MB laws discussed in Section~\ref{sec:other_MB}. Solid blue line shows the torque due to gravitational wave radiation, which will operate on top of any MB. All lines terminate at the period where the component stars overflow their Roche lobes. At $P_{\rm orb} < 1$\,day, the models disagree by several orders of magnitude.}
    \label{fig:mb_laws}
\end{figure*}

For low-mass main-sequence stars, Equation~\ref{eq:jdot_carb} predicts a torque that is significantly stronger than the other models we consider (see Figure~\ref{fig:mb_laws}). The scaling with period is similar to that in the \citetalias{Rappaport1983} model, transitioning from $\dot{J}\propto P_{\rm rot}^{-11/3}$ at slow rotation rates to  $\dot{J}\propto P_{\rm rot}^{-7/3}$ at the fastest rotation rates. \citetalias{Van2019} found that this relation could explain the high mass transfer rates inferred for persistent LMXBs with low-mass donors, for which the predicted rates for other MB  models were too low \citep[][]{Van2019b}. The model was also tested by \citet{Chen2021} and \citet{Soethe2021}, who found that it could better explain the population of millisecond pulsars with helium white dwarf companions than other prescriptions.

\subsection{Comparison of different models}
\label{sec:comparison}
We compare the different MB laws discussed above in Figure~\ref{fig:mb_laws} for equal-mass main-sequence binaries with a range of masses and orbital periods. We use the same values of $\tau_c$ and the solar rotation period in all models for consistency, leading to minor differences between our calculated torques and those in the papers that introduced each model. 

Although all models make similar predictions for solar-mass stars with periods longer than a few days -- likely because they all considered the Sun as a calibration point -- the models diverge at lower masses and higher rotation rates. Both the \citetalias{Rappaport1983} and \citetalias{Van2019} models scale roughly as $\dot{J} \propto P_{\rm orb}^{-3}$ in the period range of interest, but the \citetalias[][]{Van2019} model predicts a stronger torque at low masses, mainly because of the $\dot{J}\propto \tau_c^{8/3}$ scaling. Both the \citetalias{Sills2000} and \citet{Matt2015} models predict a shallower $P_{\rm orb}^{-1}$ scaling at short periods, leading to weaker MB at the shortest periods, where the \citetalias{Garraffo2016} model also predicts strong MB suppression. 
For the models considered here, the torque due to gravitational radiation is almost always negligible compared to MB at all periods where main-sequence binaries are detached. 

It is worth noting that all the MB laws we consider, except the one from \citetalias[][]{Van2019}, are empirically inferred. An alternative approach would be to predict the angular momentum loss theoretically based on models for the stars' winds and magnetic fields. This is in principle straightforward to do, and toy models developed elsewhere for stellar magnetospheres provide some conceptual guidance. But in practice, it is difficult to predict from first principles how the magnetic field strength and geometry, and the mass loss rate due to winds, change as a result of rotation \citep[e.g.][]{Taam1989, Garraffo2015}.

\begin{figure*}
    \centering
    \includegraphics[width=\textwidth]{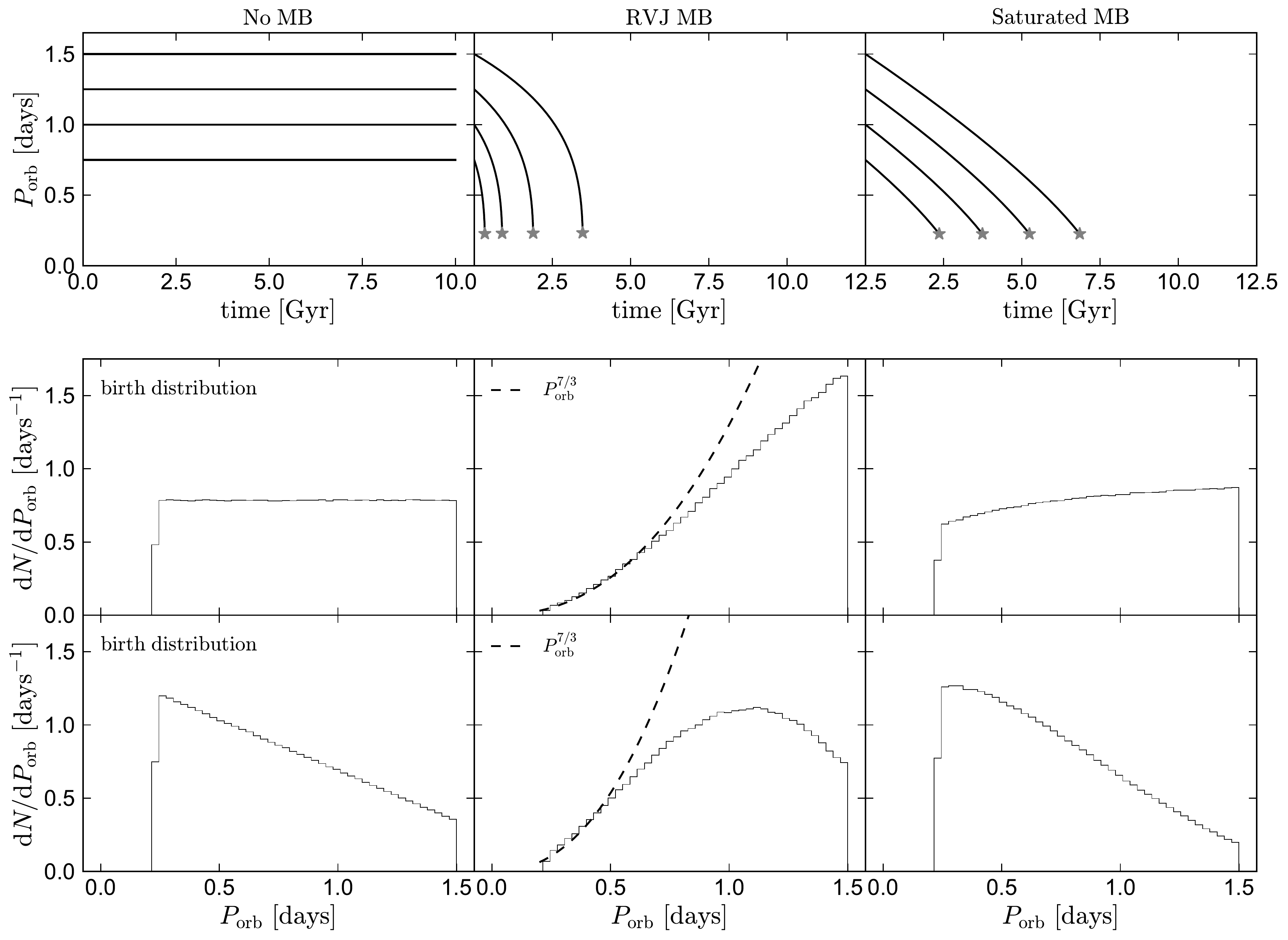}
    \caption{Period evolution and distributions of simulated binaries. Top row shows inspiral tracks for equal-mass binaries ($M_1=M_2=0.75\,M_{\odot}$) with different periods and MB laws. Left panels show no MB, middle panels assume a MB torque given by the \citetalias{Rappaport1983} model (Equation~\ref{eq:jdot}), and right panels assume saturated MB (Equation~\ref{eq:jdot_sat}). Lower panels show the period distribution of a simulated population of binaries, with the two rows showing different birth period distributions. We assume a uniform age distribution between 0 an 10 Gyr and remove binaries from the sample once they come into contact. Irrespective of the birth distribution, the \citetalias{Rappaport1983} MB law predicts a deficit of binaries at short periods, with ${\rm d}N/{\rm d}P_{\rm orb}\propto P_{\rm orb}^{7/3}$ at short periods. For saturated MB, the observable period distribution is more sensitive to the birth distribution, but generally lacks the short-period fall-off predicted for the \citetalias{Rappaport1983} law. }
    \label{fig:simulations}
\end{figure*}

\subsection{Sensitivity of the main-sequence binary period distribution to MB }
\label{sec:motivation}

 The sensitivity of the period distribution to MB is illustrated schematically in Figure~\ref{fig:simulations}, where we compare expected orbital evolution and period distributions of a population of short-period binaries with different MB laws. We consider equal-mass binaries containing $0.75\,M_{\odot}$ stars with a uniform distribution of ages, approximately as expected for a stellar population with constant star formation rate. 

For the \citetalias{Rappaport1983} MB law, the period distribution at short periods is predicted to exhibit a deficit of short-period binaries, with ${\rm d}N/{\rm d}P_{\rm orb}\propto P_{\rm orb}^{7/3}$ (Equation~\ref{eq:dndp}). Figure~\ref{fig:simulations} shows that this conclusion is not very sensitive to the adopted birth period distribution. Even if binaries are preferentially born at the closest periods, their predicted lifetime is sufficiently short that an equilibrium is reached and the period distribution is set by MB. On the other hand, for saturated MB, the orbital period evolves almost linearly with time, and so the observable period distribution is also sensitive to the birth period distribution. In practice, the ``birth'' distribution for short-period binaries refers to the distribution after various dynamical effects (such as three-body interactions in birth environments; \citealt{Fabrycky2007}) have occurred. The main point we wish to convey from Figure~\ref{fig:simulations} is that with a MB torque that increases strongly at short periods (e.g. \citetalias{Rappaport1983} or \citetalias{Van2019}), a strong deficit of short-period detached binaries is expected. We now test this prediction.

\section{Data}
\label{sec:data}

\subsection{Search for eclipsing binaries with ZTF and Gaia}
\label{sec:sample_selection}

Many large samples of EBs have been collected in the literature, with more than half a million binaries cataloged to date \citep[e.g.][]{Paczynski2006, Prsa2011, Soszynski2016, Jayasinghe2018, Chen2020, Rowan2022, GaiaCollaboration2022}.
Our goal here is not to produce a large sample, but a pure one, and one that has a reasonably well-understood selection function.  

We use light curves from ZTF \citep[][]{Bellm2019}. ZTF provides densely sampled, high-quality photometry across the northern sky, with improved sensitivity to faint sources compared to other wide-field surveys. The typical $r$-band  uncertainty of individual photometric epochs is 0.03 mag for 18th mag sources and 0.06 mag for 19th mag sources. 

\subsubsection{Parent samples}
\label{sec:parent}
\begin{figure*}
    \centering
    \includegraphics[width=\textwidth]{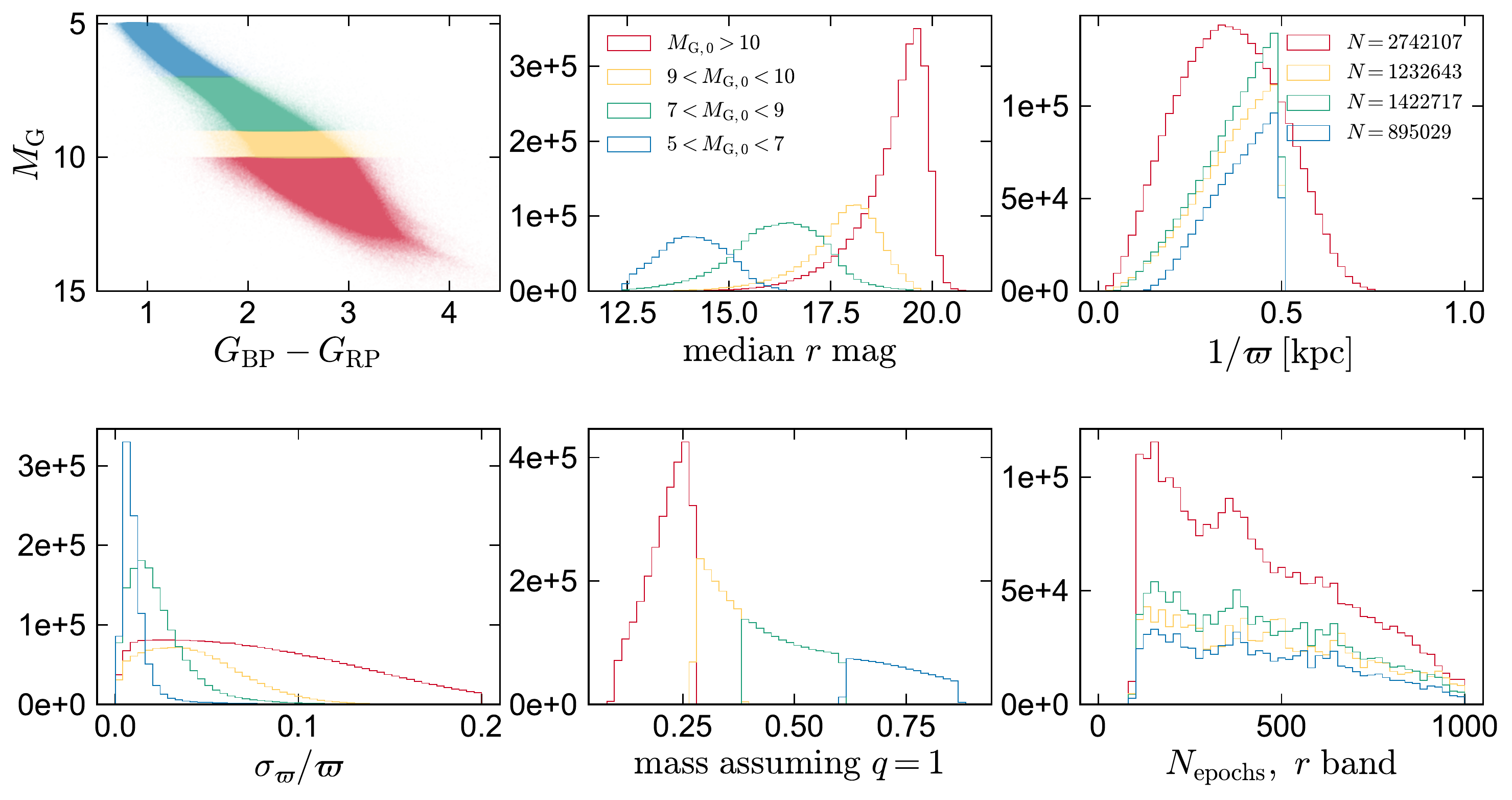}
    \caption{Summary of the samples whose light curves we searched for eclipsing binaries. We divide the lower main sequence into 4 bins by absolute magnitude (upper left), which map to different bins of primary mass (lower center). All sources in the sample have fractional parallax uncertainties below 20\% (lower left) and more than 100 clean epochs of ZTF photometry in the $r$-band (lower right). The three highest-mass bins are limited to distances $d < 0.5\,\rm kpc$ (upper right) and mean magnitudes $G<19$. For the lowest-mass bin, we limit the apparent magnitude to $G<19.5$ and the distance to $d < 1$\,kpc. }
    \label{fig:samples}
\end{figure*}

We assembled 4 different parent samples from {\it Gaia} DR3 within which to search for EBs. The different samples correspond to different extinction-corrected absolute magnitude (and thus, mass) ranges and are summarized in Figure~\ref{fig:samples}. The selection criteria were as follows:
\begin{itemize}
    \item {\it Magnitude limit:} $12.5 \leq G \leq G_{\rm max}$, where $G$ is the \texttt{phot\_g\_mean\_mag} reported in the {\it Gaia} archive. We set $G_{\rm max}=19.5$ for the lowest-mass sample and $G_{\rm max}=19$ for the 3 higher-mass samples, since the lowest-mass stars are faint. The bright limit is to avoid saturated sources, and only removes a small fraction of stars in the highest-mass bin.
    \item {\it Distance limit and uncertainty}: We required {\it Gaia} DR3 \texttt{parallax\_over\_error > 5}, so all sources have well-constrained distances and luminosities. We also require \texttt{parallax} > $\varpi_{\rm min}$, where $\varpi_{\rm min}=2\,\rm mas$ (500 pc) for the 3 highest-mass bins, and 1 mas (1 kpc) for the lowest-mass bin.
    \item {\it Color}: To exclude most white dwarf binaries, cataclysmic variables, and related objects, we remove objects below the main sequence, requiring $M_{\rm G} < 3.25(G_{\rm BP}-G_{\rm RP}) +9.625$. This does not remove  binaries with cool and faint white dwarfs; these are identified later based on light curve shape.
    \item {\it Astrometric fidelity}: We require \texttt{fidelity\_v2} > 0.75, where \texttt{fidelity\_v2} is the {\it Gaia} astrometric reliability diagnostic calculated by \citet{Rybizki2021} to filter out sources with spurious parallaxes.
    \item {\it Uncontaminated photometry}: To minimize spurious variability in the ZTF light curves due to blended sources, we require \texttt{norm\_dG} < -4, where \texttt{norm\_dG} is the blending diagnostic calculated by \citet{Rybizki2021}. This removes sources that have bright neighbors: e.g., a companion at least 0 magnitudes brighter at a distance of 4 arcsec; a companion less than 2 magnitudes fainter at a distance of 2 arcsec, or a companion at least 2 magnitude brighter at a distance of 6 arcsec. 
    \item {\it Light curve sampling}: We required each object to have at least 100 epochs of clean photometry in the $r$-band as of ZTF DR8, where for ``clean'' photometry we required \texttt{catflags} < 32768.  
    \item {\it Sky position}: We excluded sources with declination $\delta < -28$ deg, corresponding roughly to the southern pointing limits for ZTF and PanSTARRS (which the extinction map relies on). 
\end{itemize}

We retrieved the public ZTF light curves in $r-$, $g-$, and $i-$bands of all {\it Gaia} sources passing the cuts above, excluding data with \texttt{catflags} $\geq$ 32768. To increase the number of usable photometric epochs, we combined photometry from the  primary and secondary ZTF grids \citep[see][]{Bellm2019}. In cases where there are multiple ZTF object IDs within $< 0.5$ arcsec of a {\it Gaia} source, we concatenate photometry from both IDs into a single light curve.  

 We estimated extinctions for each source using the 3D dust map from \citet{Green2019}, assuming $A_G= 2.67 E(g-r)$ and $E\left(G_{{\rm BP}}-G_{{\rm RP}}\right)=1.33E\left(g-r\right)$. We neglect the weak color-dependence of these transformations and assume a universal extinction curve.  Because the sources in our sample are nearby, extinction is generally modest. 
 
\begin{table*}
\begin{tabular}{llllllll}
{\it Gaia} DR3 Source ID  & RA [deg] & Dec [deg] & $G\,\rm [mag]$ & $P_{\rm orb}\,\rm [day]$ & $M_{G,0}\,\rm [mag]$ & $\varpi\,[\rm mas]$ & \\
\hline
930924437606993792 & 125.968560 & 48.254674 & 18.64 & 0.0755974 & 11.38  & $3.78 \pm 0.17 $ \\
2085085710293765120 & 305.014398 & 50.560460 & 18.70 & 0.0792846 & 12.92  & $7.01 \pm 0.15 $ \\
3985846989195370752 & 160.993536 & 17.913792 & 18.66 & 0.0853155 & 11.09  & $3.06 \pm 0.22 $ \\
3086450738282401024 & 112.892067 & -0.079707 & 18.25 & 0.0899373 & 10.45  & $2.90 \pm 0.27 $ \\
2480332961921525376 & 26.908116 & -3.919711 & 18.06 & 0.0948654 & 10.91  & $3.81 \pm 0.14 $ \\
119685070499349888 & 52.859499 & 28.654637 & 18.14 & 0.0988626 & 10.30  & $3.63 \pm 0.17 $ \\
5726592458262106240 & 124.272623 & -13.010703 & 18.03 & 0.1020786 & 10.99  & $4.17 \pm 0.14 $ \\
3351801656143671424 & 102.540760 & 13.146829 & 18.92 & 0.1024674 & 11.17  & $2.84 \pm 0.26 $ \\
587277820712312064 & 140.782769 & 8.104268 & 18.72 & 0.1030375 & 11.56  & $3.91 \pm 0.23 $ \\
1331586474764850560 & 249.982391 & 38.975436 & 16.71 & 0.1048172 & 11.98  & $11.37 \pm 0.04 $ \\
$\cdots$ &  &  &  &  &   &  \\
\hline
\end{tabular}
\caption{\label{tab:summary} Eclipsing binary sample from ZTF, ordered by orbital period. The full table contains 3,879 EBs with $P_{\rm orb}<10$ days and is available in machine-readable format.}
\end{table*}

We divided the lower main sequence into four bins of extinction-corrected absolute magnitude, $M_{G,0}$.
Basic properties of the resulting parent samples are summarized in Figure~\ref{fig:samples}. The four bins correspond roughly to primary spectral types M3.5-M8 (red, $M_{G,0} > 10$, $M_1 \lesssim 0.3\,M_{\odot}$); M3.5-M2.5 (yellow,  $9 < M_{G,0} < 10$, $0.3 \lesssim M_1/M_{\odot} \lesssim 0.4\,M_{\odot}$); M2.5-K7 (green,  $7 < M_{G,0} < 9$, $0.4 \lesssim M_1/M_{\odot} \lesssim 0.65\,M_{\odot}$), and K7-K0  (blue,  $5 < M_{G,0} < 7$, $0.65 \lesssim M_1/M_{\odot} \lesssim 0.9\,M_{\odot}$). We estimated these ranges from solar-type MIST isochrones \citep{Choi_2016}, assuming the secondary brightens the unresolved source by 0.5 mag.

The divisions between bins are somewhat arbitrary, but were chosen such that the lowest-mass bin is expected to contain only fully-convective stars\footnote{The fully convective boundary appears to depend somewhat on metallicity, ranging from $M_{G,0} = 9.8$ for metal-poor stars to  $M_{G,0} = 10.2$ \citep[e.g.][]{Jao2018}. Since EBs have extra light from the second star, we expect a cut of $M_{G,0} > 10$ to efficiently select binaries with both components below the full-convective boundary. We verified that using a cut of $M_{G,0} > 10.2$ does not change our conclusions.}, and comparable numbers of EBs are found in each subsample. The mapping from absolute magnitude to mass is not unique, because a range of mass ratios (and ages and metallicities) are found in the sample. In Figure~\ref{fig:samples}, we show primary mass estimates calculated from $M_{G,0}$ assuming $q=1$, solar metallicity, and an age of 1 Gyr. In the mass range of interest, a lower mass ratio of $q\lesssim 0.5$ would imply a primary mass typically 10\% larger.
By design, all stars in all samples are expected to have main-sequence lifetimes exceeding the age of the Universe and are expected to be near the zero-age main sequence.

\begin{figure*}
    \centering
    \includegraphics[width=\textwidth]{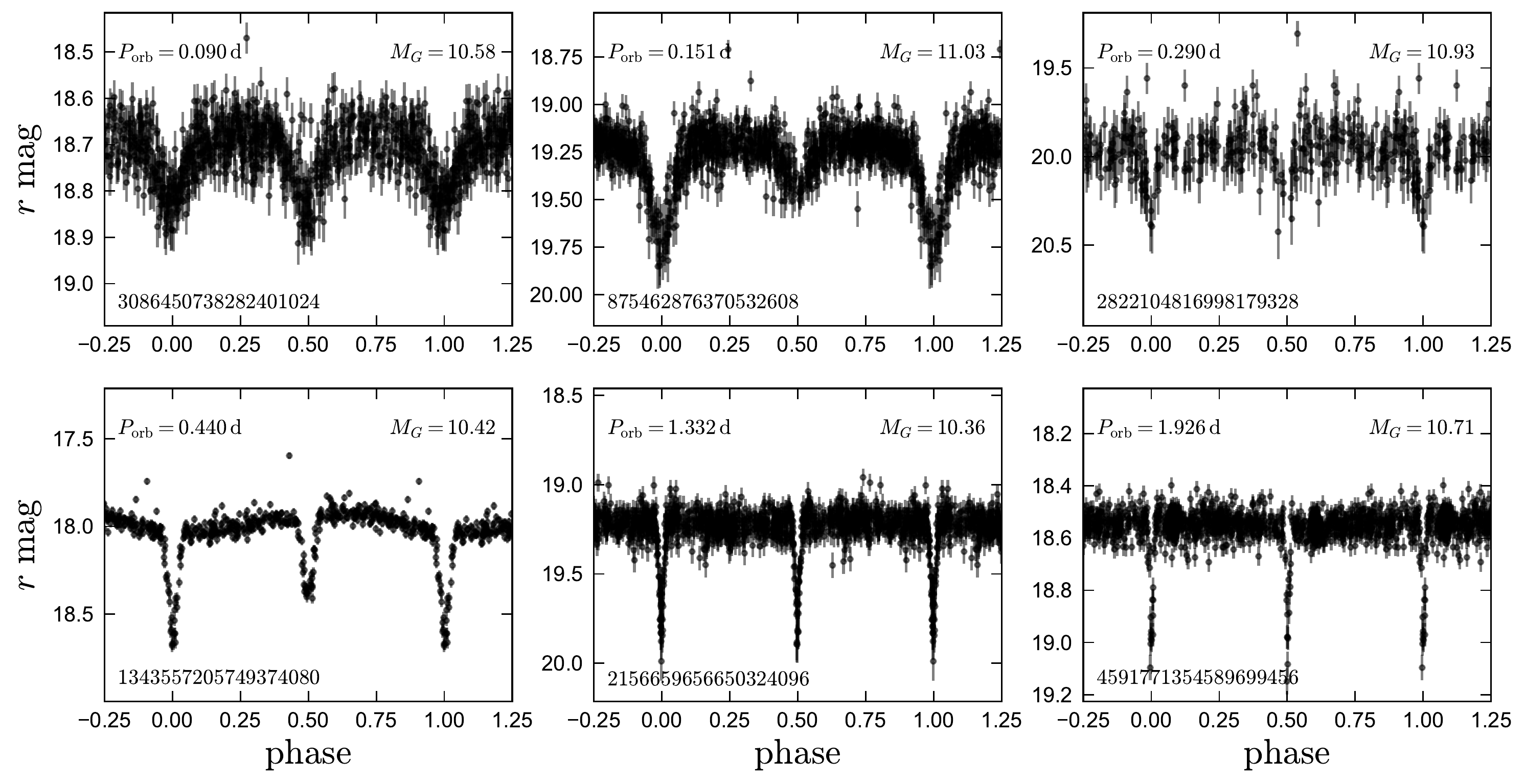}
    \caption{ZTF $r-$band light curves of a quasi-random subset of detached eclipsing binaries in the low-mass sample (fully-convective stars with $M\lesssim 0.3\,M_{\odot}$; red points in Figure~\ref{fig:samples}). These objects have periods, magnitudes, and light curve morphologies typical of the sample. Objects are ordered by increasing orbital period and labeled by absolute magnitude and {\it Gaia} DR3 source ID. }
    \label{fig:examples}
\end{figure*}

\subsubsection{Search for eclipsing binaries}
\label{sed:eb_search}
Having obtained ZTF light curves for $\approx 6$ million {\it Gaia} sources, we used a multi-step process to identify EBs. We began by computing a box least squares (BLS;~\citealt{Kovacs2002}) periodogram for each $r-$ band light curve, using a frequency grid running from 45 minutes to 10 days. The 45 minute lower limit was chosen to be half of the shortest plausible orbit inside which a main-sequence star would fit inside its Roche lobe. We used the GPU-accelerated implementation of BLS in \texttt{cuvarbase}\footnote{https://github.com/johnh2o2/cuvarbase} and a linear frequency grid with an oversample factor of 30; this typically yields about $10^6$ trial frequencies per source. We used a logarithmic grid of 20 eclipse phase durations between 0.005 and 0.5, selecting the ``best'' duration as the one that yielded the highest likelihood. In order to mitigate the effects of flares and spuriously bright photometric epochs, we masked points that exceeded the median flux by more than 1.5 times the inter-quartile range.

Because low-mass main-sequence stars are red, the $r-$ band light curves have smaller uncertainties and more photometric points in most cases than the $g-$ band light curves. The $i-$band is generally sparsely sampled. We thus only calculated periodogram metrics for the $r-$band, though we also inspected the phased $g-$ and $i-$band data during the visual inspection phase.  

In addition to the BLS periodogram, we calculated the conditional entropy periodogram \citep[CE;][]{Graham2013} on the same frequency grid. Unlike the BLS periodgram, which assumes a specific light curve shape, CE quantifies how much the level of disorder improves in phased vs. unphased light curves. In particular, CE is often more robust than traditional periodograms against spurious periods associated with the observing cadence. We used 20 phase bins and 10 magnitude bins in calculating CE.

We then selected candidate EBs as sources satisfying the following: 

\begin{itemize}
    \item Best-fit BLS eclipse depth of at least 10\% in flux. 
    \item BLS power of at least 50.
    \item At least 6 distinct eclipses covered by the $r-$band data. Here the best-fit BLS model determines whether a data point is ``in eclipse''. The number of ``in eclipse'' datapoints is often larger than the number of eclipses covered by the data, because multiple epochs can fall into a single eclipse. 
    \item Period not close to 1 sidereal day: $\left|P-{\rm 1\,day}\right|>0.005$. This eliminates a large fraction of spurious detections, including sources displaying aperiodic long-term variability. It causes the search to miss real EBs with periods between 0.995 and 1.005 days, but this is of little consequence for our analysis.
    \item For sources with best-fit periods less than 1 day, we additionally require that the CE associated with the best-fit BLS period is at least 10\% smaller than the median of the CE periodogram. We do not impose this requirement for longer periods, because there the eclipse phase duration becomes short enough that the eclipse is sometimes not well-resolved with the 20 phase bins used to calculate CE. 
\end{itemize}

These criteria were chosen with the goal of minimizing both the number of spurious detections and the number of genuine EBs missed. We chose the detection thresholds based on injection and recovery tests with simulated eclipse light curves injected into the data (see Section~\ref{sec:completeness}). 

We also experimented with using {\it Gaia} variability statistics to identify variable stars, as described by \citet{Guidry2020}. We found that cuts on variability statistics successfully identify most EBs with $P_{\rm orb} < 1$ day. However, at longer periods, many sources that are unambiguous EBs have unremarkable {\it Gaia} variability statistics and would be missed with cuts on such statistics. This reflects the fact that the eclipse duty cycle falls with increasing period, such that the finite sampling of {\it Gaia} light curves might never catch longer-period binaries in eclipse. 

\subsubsection{Visual inspection}
\label{sec:visual}
The final stage of our EB search is visual inspection of candidates identified in the automated search described above. In short, we looked at the phased ZTF light curves of each source in all available bands to decide whether the light curve shape could best be attributed to a detached main-sequence EB or to another source of variability, either astrophysical or instrumental. 

In Figure~\ref{fig:examples} and~\ref{fig:non_ebs}, we show examples of periodic variables that are (Figure~\ref{fig:examples}) and are not (Figure~\ref{fig:non_ebs}) likely detached main-sequence EBs. All the examples we show are from the lowest-mass bin. Objects in the higher-mass bins are generally brighter and thus have higher-quality light curves. Common false-positives that were removed during our visual inspection include detached white dwarf + main sequence binaries, cataclysmic variables, non-eclipsing ellipsoidal variables, contact binaries, rotating stars with spots, and young stellar objects. In addition to these astrophysically periodic variables, we also removed sources in which variability was likely due to spurious instrumental effects, such as stars falling on a bad column in the ZTF detector. Many of the short-period variables we reject as rotators are likely still  binaries, in many cases with the rotation period tracking the orbital period \citep[e.g.][]{Simonian2019}. However, we limit our sample to detached EBs in the interests of maintaining a selection function that is straightforward to model. 

We also experimented with a range of automated tools for distinguishing EBs from other periodic variables, including the random forest classifier Upsilon \citep[e.g.][]{Kim2016} and a variety of bespoke cuts on light curve summary statistics. Such approaches can achieve reasonably good purity and completeness in a small fraction of the time required by human classifiers. However, all approaches we tested performed worse than human classifiers. Because our main goal -- constraining the intrinsic period distribution at short periods -- can be achieved with a modest sample size (and because we find visual inspection of light curves a pleasantly meditative activity), we opted for the more labor-intensive approach of inspecting candidates visually.

The false-positive rate of our initial automated selection varies significantly between absolute magnitude  bins. In the highest-mass bin, about half of all candidates that pass our automated selection criteria also pass visual inspection. In the lowest-mass bin, the success rate of the automated selection is lower than 20\%. This owes primarily to the poorer signal-to-noise ratio in the fainter light curves, which results in many spurious periodic detections due to correlated noise. 

\subsubsection{Removing contact binaries}
\label{sec:contact}
\begin{figure*}
    \centering
    \includegraphics[width=\textwidth]{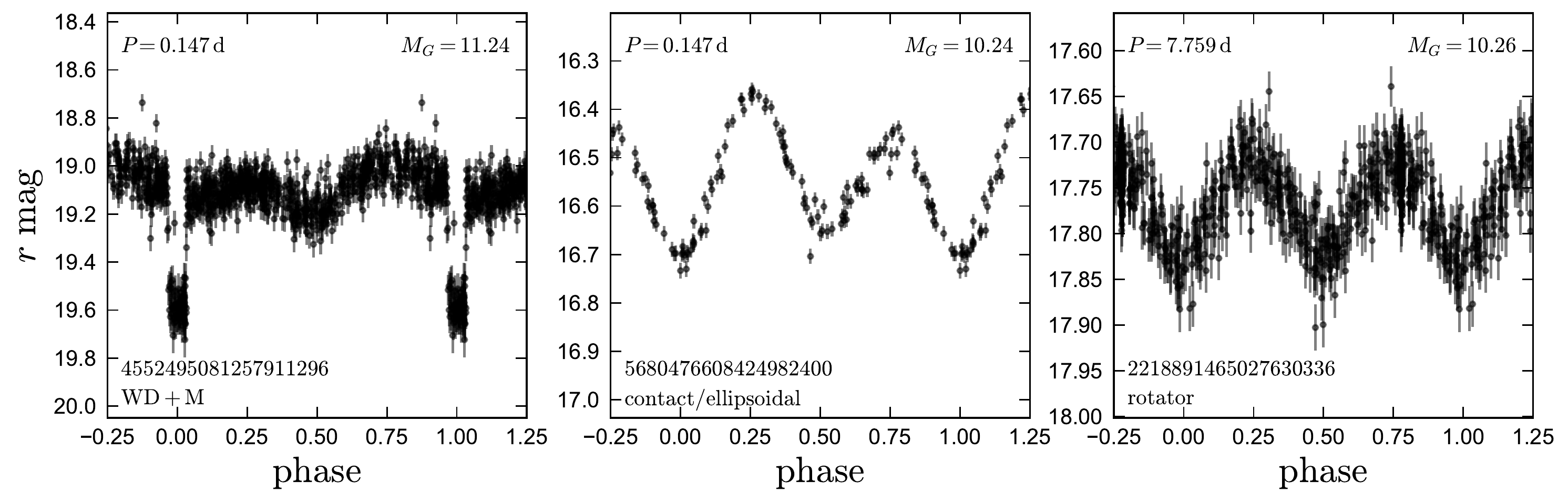}
    \caption{ZTF $r-$band light curves of representative periodic variables in the lowest-mass sample that are {\it not} detached main sequence eclipsing binaries and were removed during visual inspection. Left panel shows a white dwarf + main sequence binary, which can be distinguished from main sequence binaries by the sharp ingress/egress and lack of secondary eclipse. Middle panel shows a suspected ellipsoidal variable or contact binary, characterized by its smoothly varying light curve shape. Right panel shows a suspected rotating spotted star, which has sinusoidal variability and too long of a period for the variation to be due to tidal distortion of a low-mass main-sequence star.}
    \label{fig:non_ebs}
\end{figure*}

At the shortest periods in each absolute magnitude subsample (e.g. $P_{\rm orb}\lesssim 0.1\,\rm day$ for $M_{G,0} > 10$;  $P_{\rm orb}\lesssim 0.3\,\rm day$ for $5 < M_{G,0} < 7$), we find candidate EBs with smoothly varying light curves (i.e., no clear ingress or egress and no flat region between eclipses). In most cases, this is due to tidal deformation of component stars that either nearly or completely fill their Roche lobes. In at least some such binaries, mass transfer leads to a long-lived, stable common envelope phase during which binaries are observed as ``W UMa'' contact systems \citep[e.g.][]{Lucy1968}. Since mass transfer complicates the angular momentum evolution, we exclude such binaries from the sample. We find that they are common in the highest-mass bin, where there is a clear pileup in the period distribution at $P_{\rm orb} \lesssim 0.3\,\rm days$, but are rare (and possibly completely absent) in the three lower-mass bins. \citet{Jiang2012} have suggested that contact binaries are unstable, or at least relatively short lived, at primary masses below $\approx 0.6\,M_{\odot}$. Our period distributions are consistent with such a scenario, which predicts a pileup of binaries at the period where main-sequence stars overflow their Roche lobes. Although there are some systems with light curves resembling contact binaries in the lower mass samples (e.g., the rejected source in the center panel of Figure~\ref{fig:non_ebs}), follow-up by \citet{Drake2014} found that at least a large fraction of such systems are M dwarfs with white dwarf companions.

Although we attempt to exclude contact binaries, some cases are ambiguous: detached binaries with components that are nearly Roche lobe filling can have light curves quite similar to contact binaries. Fortunately, the orbital period at which a main-sequence star will overflow its Roche lobe is primarily a function of mass. To prevent unrecognized contact binaries from biasing our results, we only consider binaries in subsequent sections that have periods well above the longest Roche lobe overflow period for their mass.

\subsubsection{Period aliases}
\label{sec:period_alias}
As is common in analysis of EBs, it is sometimes unclear whether the peak of the BLS periodogram corresponds to $P_{\rm orb}$ or $P_{\rm orb}/2$. To decide, we phase each light curve to both periods during our visual inspection. In most cases, a secondary minimum is evident with different depth from the primary minimum, removing the ambiguity. 

For cases where no secondary eclipse is evident, we performed simulations of eclipsing binaries using the \texttt{ellc} code (see Section~\ref{sec:completeness}). We found that in all but the highest-mass bin, a stellar companion large enough to cause an eclipse with $>10\%$ depth would also produce a detectable secondary eclipse; this reflects the fact that temperature varies slowly with mass along the lower main sequence. For the highest mass bin, there is some degeneracy between an edge-on orbit with $q\approx 0.3$ and a grazing orbit with $q=1$ and twice the orbital period. Given the precise {\it Gaia} distances to objects in our sample, we can distinguish between these two scenarios based on  CMD position: a $q \approx 0.3$ companion will increase the luminosity of the unresolved binary by only a few percent compared to the primary in isolation, so such binaries are expected to be found near the main sequence. On the other hand, a $q\approx 1$ companion will double the luminosity of the unresolved source, shifting it $\approx 0.75$ mag above the main sequence. This allows us to determine the orbital period unambiguously in most cases, though we expect that some period aliases still exist in the catalog. 

In the course of visual inspection, we identified one short-period EB in which there is no secondary eclipse, but the orbital period is still well-determined due to ellipsoidal variation. This object, {\it Gaia} DR3 2085085710293765120, has the 2nd shortest period in our sample, with $P_{\rm orb} = 1.90$ hours. We suspect that the companion is a brown dwarf and will analyze this object in more detail in future work.

\subsubsection{Eccentricities}
\label{sec:ecc}
To assess whether EBs have circular or eccentric orbits, we checked the relative spacing of the primary and secondary eclipses. In EBs with circular orbits, these will always be separated by half an orbit, whereas in eccentric EBs, the spacing will in general be unequal. Searching the full catalog, we identified only a few short-period systems with clearly eccentric orbits. The shortest-period eccentric system is {\it Gaia} DR3 518184643170677376, with $P_{\rm orb}= 2.15$ days and absolute magnitude $M_{G,0} = 8.2$. Most importantly for our purposes, we did not find any eccentric EBs in the period range where we expect MB to be efficient ($P_{\rm orb}\lesssim 2$ days). This validates our assumption that binaries at these periods are tidally synchronized, a requirement for MB to remove orbital angular momentum.

\subsubsection{Completeness}
\label{sec:completeness}
\begin{figure*}
    \centering
    \includegraphics[width=\textwidth]{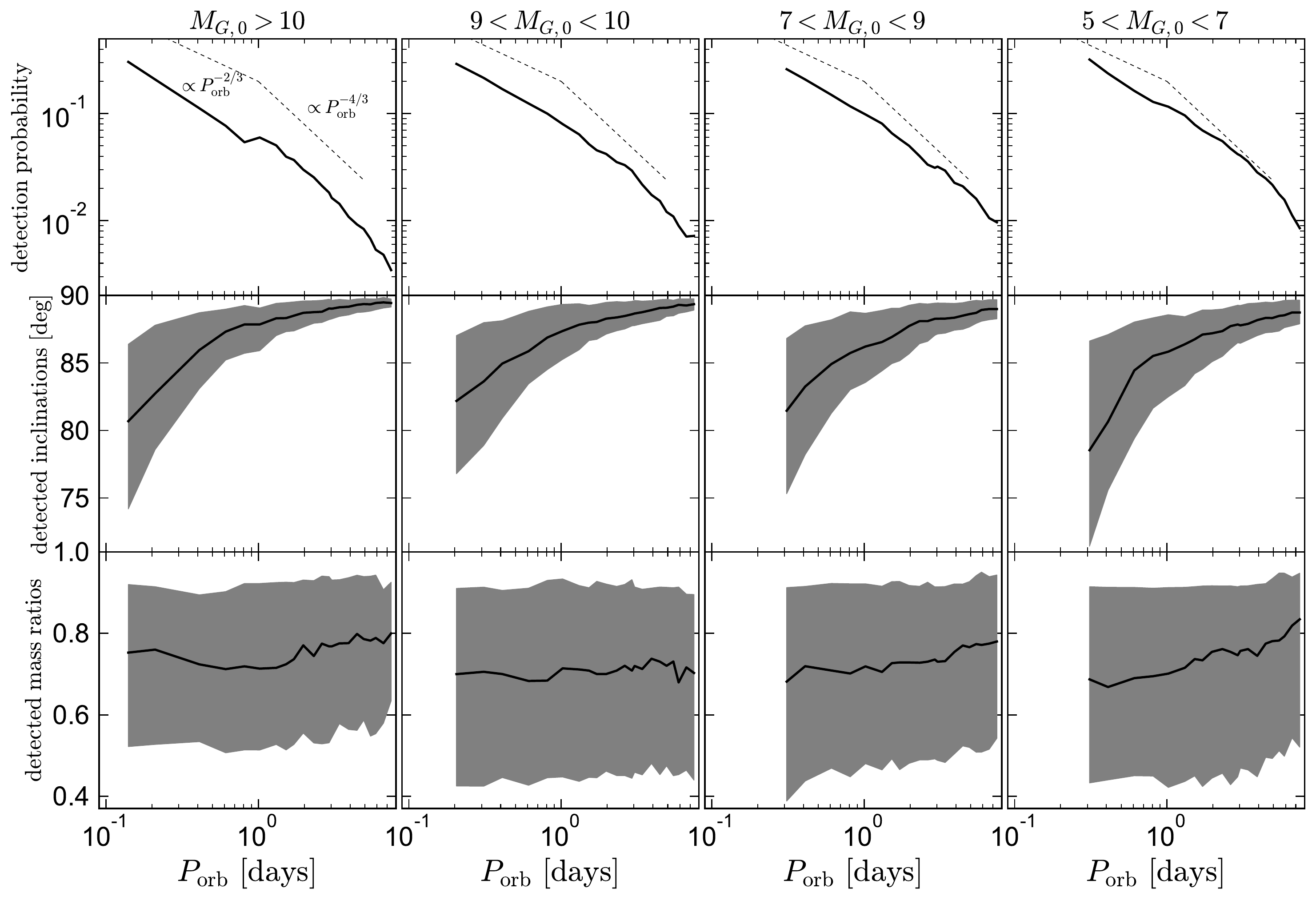}
    \caption{Sensitivity analysis. We inject synthetic light curves of eclipsing binaries into observed ZTF light curves of sources in each absolute magnitude bin, and then repeat the search used to identify real EBs. Top panel shows the fraction of all binaries that are detected, with the denominator including systems that do not eclipse. Bottom panels show the median and middle 68\% the inclinations and mass ratios of the injected binaries that were detected.  }
    \label{fig:inject_recovery}
\end{figure*}

To quantify the selection function of our search, we performed injection-and-recovery simulations using the  ZTF light curves. For each absolute magnitude subsample, we simulated a population of close binaries, injected them into a random subset of the observed light curves (chosen to have apparent magnitudes and cadences that are representative of the parent sample), and then repeated the search procedure we used with the real data, including both the automated search and visual inspection. We assumed a $p(i)\,{\rm d}i=\sin(i)\,{\rm d}i$ distribution of orbital inclinations and a uniform mass ratio distribution $\mathcal{U}(q_{\rm min},1)$, where $q_{\rm min}=0.1\,M_{\odot}/M_{1}$. We assumed both components are on the zero-age main sequence, with radii calculated from solar-metallicity MIST models \citep[][]{Choi_2016}. We calculated model light curves for each binary using the \texttt{ellc} code \citep[][]{Maxted2016}, assuming a uniform distribution of orbital phases and gravity and limb darkening coefficients appropriate for the component stars \citep{Claret2004}. For each bin of absolute magnitude and orbital period shown in Figure~\ref{fig:inject_recovery}, we proceeded until at least 20 simulated binaries had been recovered. 

Figure~\ref{fig:inject_recovery} presents the results of these experiments. The top panels show the overall recovery rate of all binaries, including those which do not eclipse because their inclinations are too low. The middle and bottom panels show the median and middle 68\% of the inclinations and mass ratios of the detected binaries. As expected, the range of inclinations that result in a detectable eclipse is largest at the shortest periods. The range of detectable mass ratios varies only weakly with primary mass and period. 
At short periods, the detection efficiency is set primarily by the eclipse probability, which scales as $P_{\rm orb}^{-2/3}$ at fixed mass and radius. The eclipse duty cycle (i.e., the fraction of the orbit spent in eclipse) also scales as $P_{\rm orb}^{-2/3}$, and the combination of these factors leads to a roughly $P_{\rm orb}^{-4/3}$ scaling at longer periods. 

\subsubsection{Wide tertiaries}
\label{sec:tertiaries}
It is well-established \citep[e.g.][]{Tokovinin2006} that most close binaries have distant tertiary companions. Indeed, outer tertiaries likely play a decisive role in the formation of almost all close binaries \citep[e.g.][]{Mazeh1979, Fabrycky2007}. These tertiaries are expected to have a separation distribution similar to that of normal binaries (a lognormal distribution peaking at $\log(P_{\rm orb}/{\rm days})\approx 5$ with logarithmic dispersion of 2.3), which is truncated at the closest separations due to dynamical stability limits \citep[e.g.][]{Tokovinin2014}.

Given a typical distance of 300 pc and the {\it Gaia} angular resolution of $\approx 1$ arcsec, most tertiaries to EBs in our sample will be unresolved at separations less than $\approx 300$\,AU. This means that, absent other selection effects, we expect roughly 30\% of tertiaries to be resolved, and 70\% to be unresolved. In the three lower mass bins, the unresolved fraction may be even higher, since the separation distribution is skewed toward closer separations in low-mass binaries compared to solar-type binaries \citep[e.g.][]{Moe2017}.

\begin{figure*}
    \centering
    \includegraphics[width=\textwidth]{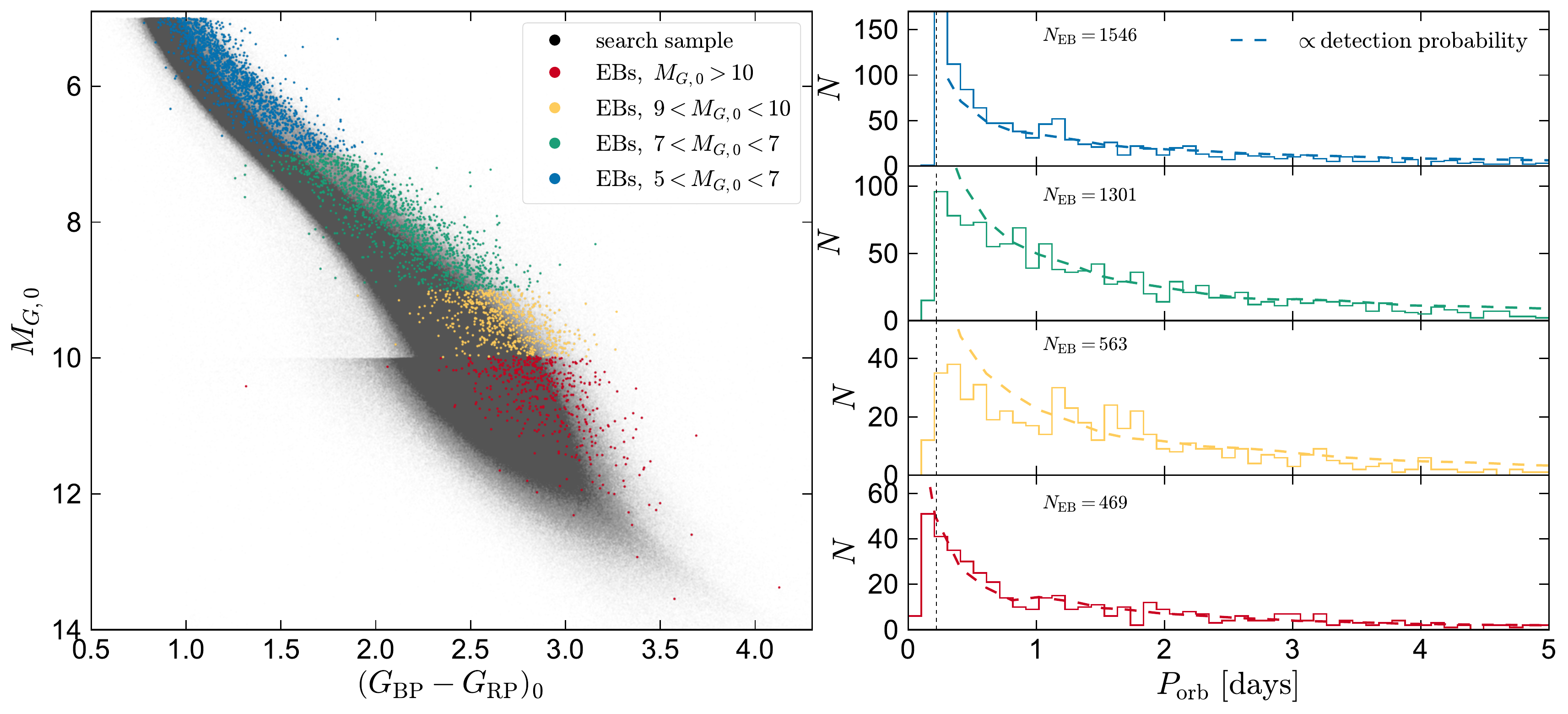}
    \caption{Left panel compares the color-magnitude diagram of detected EBs to the full sample of sources whose light curves were searched for eclipses. As expected, most EBs are overluminous for their color. Right panel shows period distributions of the detected EBs. In the top panel, the shortest-period bin extends beyond the plot limits: it contains almost 600 EBs, which are probably mostly misclassified contact binaries. Dashed lines show the detection probability, which is measured via injection/recovery simulations. The period distributions in all samples are skewed toward short periods, but this is primarily due to selection effects. Dashed vertical line shows a period of 0.22 days, where a period cutoff has been reported. Such a cutoff is present in the highest-mass bin, but not at lower masses.}
    \label{fig:detections}
\end{figure*}

There is, however, an additional selection effect: we remove sources with bright, close companions from the parent sample in order to minimize light curve issues related to blending (Section~\ref{sec:parent}). This means that sources with a bright companion separated by $1 \lesssim \rho/{\rm arcsec}\lesssim 5$ are missing from the sample, with the exact limit depending on luminosity ratio. As a result, we expect almost half of all systems with a resolved companion (corresponding to $\approx 15\%$ of all systems) to be removed due to the \texttt{norm\_dG} (blending) cut. 

To compare the wide binary fraction of sources in the EB sample to that of the Galactic field, we cross-matched the EB sample with the wide binary catalog constructed by \citet[][]{El-Badry2021_gaia}. Considering the subsample with $M_{G,0}>10$, we found that of the 162 EBs with $P_{\rm orb} < 0.5$ days, 24 have a wide companion with projected physical separation $s > 2000\,\rm AU$. We focus on companions with  $s > 2000\,\rm AU$ because our blending cut will preferentially remove resolved companions with closer separations.  This implies a detectable wide binary fraction of $\approx$\,15\% for the short-period EBs. This is consistent with a scenario in which most of the short-period EBs have an outer tertiary.

On the other hand, among the 207,436 sources with $M_{G} > 10$ in the {\it Gaia} Catalog of Nearby Stars \citep[GCNS;][]{GaiaCollaboration2021b}, 2694 are in a wide binary with $s > 2000\,\rm AU$ -- a wide binary fraction of only 1.3\%, compared to 15\% for the EBs. Completeness to wide companions is likely {\it higher} in the GCNS sample due to its $\lesssim 100\,\rm pc$ distance limit. It is thus clear that tertiaries play an important role in the formation of the short-period EBs in our sample. Most of the tertiaries that are spatially resolved are sufficiently wide that the timescale for Kozai-Lidov oscillations in the current orbital configurations likely  exceed the Hubble time, but this may not have been the case in the past (see also \citealt{Hwang2022}).
Since only companions with separations of $300\lesssim s/{\rm AU} \lesssim 1500$ are removed by the \texttt{norm\_dG} cut, we expect the population properties of EBs in the sample to be reasonably representative of the population. 

\section{Results}
\label{sec:results}

\subsection{Summary of the eclipsing binary sample}
\label{sec:ztf_description}

The results of our search are summarized in Figure~\ref{fig:detections}. The left panel compares our EB candidates (after visual inspection) to the parent sample, which is shown in black. The apparent discontinuity at $M_{G,0}=10$ is a consequence of the looser distance and apparent magnitude limits we use in the faintest sample. As expected, most of the EB candidates are overluminous compared to the main sequence. Those which are not overluminous are primarily systems with weak secondary eclipses, indicative of an unequal mass ratio system in which the secondary contributes little additional light. 

The right panels show the period distributions of visually vetted EBs. In all samples, these rise steeply toward shorter periods. This is primarily an observational selection effect: the dashed lines in each panel show the detection efficiency for that mass bin (Section~\ref{sec:completeness}), with arbitrary rescaling to match the normalization of the data. Hints of a discontinuity at $P_{\rm orb} = 1$ day can be seen in both the observed period distributions and detection efficiencies; this is a result of the conditional entropy cut we impose only for systems with inferred periods below 1 day. 

Table~\ref{tab:summary} lists all the binaries in all absolute magnitude subsamples. The full catalog contains 3,879 EBs with $P_{\rm orb} < 10$ days. The number in each absolute magnitude subsample is shown in Figure~\ref{fig:detections}. The reported $P_{\rm orb}$ values represent the peak of the BLS periodogram, or twice this value. We have not attempted to model light curves in detail or account for apparent period variations due to light travel time effects in triples, so more accurate periods can likely be inferred with careful modeling. 
Several of the binaries in the lowest-mass subsample have shorter orbital periods than any other main-sequence binaries discovered to date \citep[e.g.][]{Maceroni2004, Nefs2012, Soszynski2015, Koen2022, Kurtenkov2022}. The shortest-period system has an orbital period of 109 minutes. Follow-up analysis of some of the shortest-period systems will be presented in future work.  

The three lower-mass subsamples contain many binaries well below 0.22 day ``cutoff'' widely reported for main sequence binaries \citep[e.g.][]{Rucinski1992, Paczynski2006}. Indeed, in the lowest-mass bin, the {\it peak} of the observed distribution is well below this limit. The cutoff is, however, observed in the period distribution of the highest-mass subsample. We suspect that the cutoff found in previous works thus primarily reflects the fact that these surveys had poor sensitivity to faint, low-mass stars. Some previous works \citep[e.g.][]{Stepien2006} interpreted the lack of observed binaries with periods below 0.22 days as a result of inefficient MB in M dwarfs.\footnote{These authors used a MB prescription in which the MB torque is exponentially suppressed at high Rossby number, causing inspiral to stall at short periods.} Our results suggest that this is unlikely to be the full story, since short-period binaries are indeed observed at all periods down to the contact limit at all masses.

In the highest-mass sample, the shortest-period bin in Figure~\ref{fig:detections} with $P_{\rm orb} < 0.3$ days extends well beyond the plot limits, containing almost 600 EBs. This is very likely due to a pileup of unrecognized long-lived contact binaries, which are absent in the lower-mass bins. This population does not affect our inference in later sections, where for this absolute magnitude bin we only consider EBs with $P_{\rm orb} > 0.35$ days. 

\begin{figure*}
    \centering
    \includegraphics[width=\textwidth]{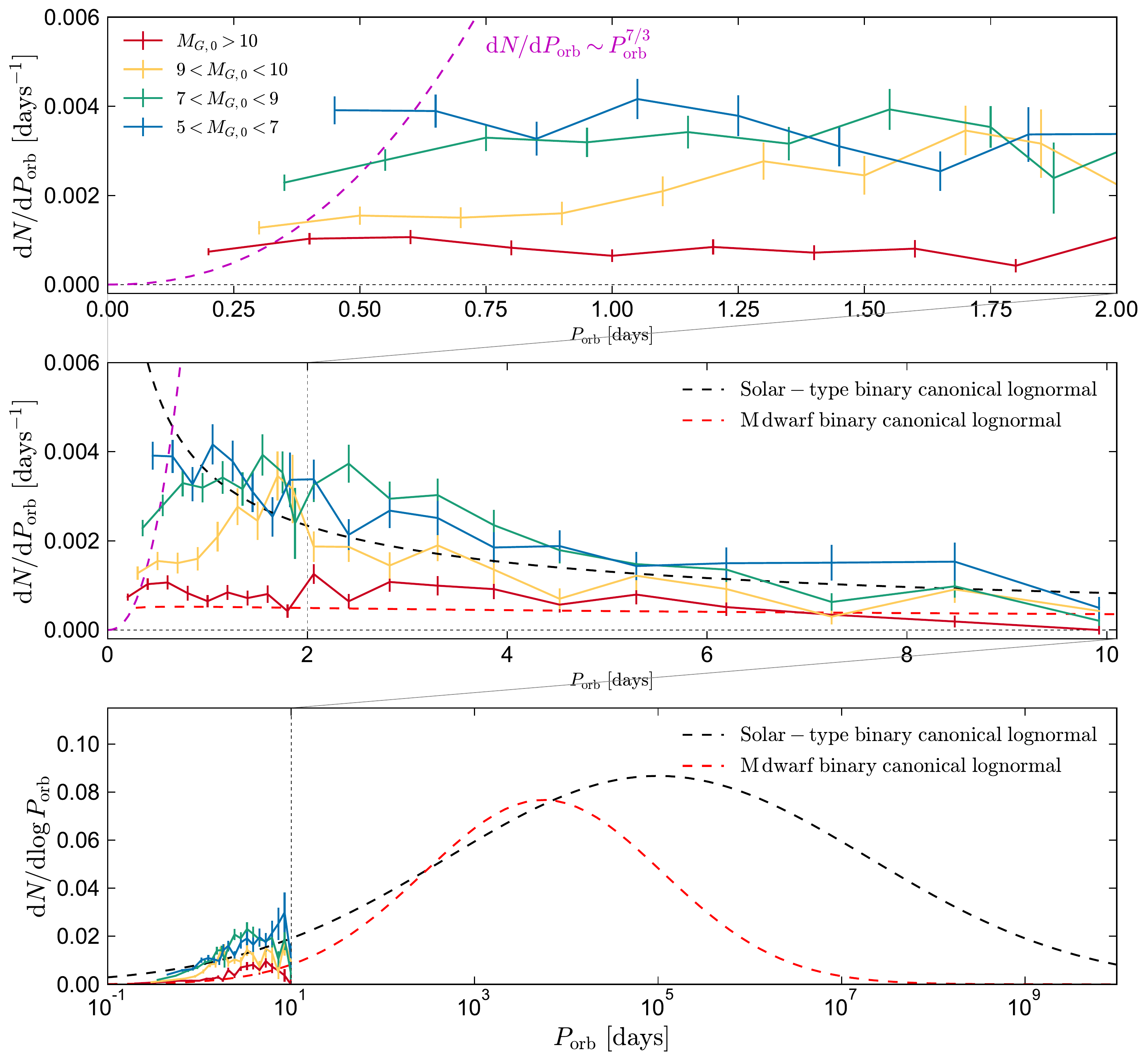}
    \caption{Incompleteness-corrected orbital period distribution. Lines with errorbars show different absolute magnitude  samples. The upper panels are zoomed-in versions of the lower panels. Dashed magenta line shows the distribution expected if the period distribution reaches equilibrium at short periods due to a Skumanich-like MB law. The data differ dramatically from this prediction: they are all flat. Bottom panel shows a wider period range, with logarithmic rather than linear period bins. Dashed lines show the canonical lognormal period distributions for solar-type stars and M dwarfs.} 
    \label{fig:corrected_period_distributions}
\end{figure*}

\subsection{Intrinsic period distribution}
\label{sec:intrinsic}
From the observed period distributions and the detection efficiencies from simulations (both shown in Figure~\ref{fig:detections}), we can calculate the intrinsic period distributions of short-period main sequence binaries. In particular, we define ${\rm d}N/{\rm d}P_{\rm orb}$ to be the mean number of binary companions per linear interval of $P_{\rm orb}$ per unresolved source in the sample. Given $N_{{\rm binaries,\,obs}}$ observed binaries in a given sample and period bin selected from a parent sample of $N_{{\rm tot,\,obs}}$ unique light curves in the sample, we calculate 

\begin{align}
    \label{eq:dNdp}
    \frac{{\rm d}N}{{\rm d}P_{{\rm orb}}}=\frac{N_{{\rm binaries,\,obs}}}{N_{{\rm tot,\,obs}}\Delta P_{{\rm orb,\,bin}}},
\end{align}
where $\Delta P_{{\rm orb,\,bin}}$ is the bin width. The results are shown in Figure~\ref{fig:corrected_period_distributions}, with the uncertainties calculated from Poisson statistics. Different panels show different period ranges, with the ``zoom'' increasing from top to bottom. 

The top panel shows $P_{\rm orb} < 2$ days, roughly the regime in which we expect the effects of MB to be important. As discussed in Section~\ref{sec:contact}, we only show results for orbital periods at which main-sequence binaries must be detached in order to avoid complications from mass transfer and a potentially long-lived contact phase. This is why the minimum period for which we calculate  ${\rm d}N/{\rm d}P_{\rm orb}$ increases from lower to higher mass samples.

The dashed magenta line shows  ${\rm d}N/{\rm d}P_{\rm orb}\propto P_{\rm orb}^{7/3}$, the predicted equilibrium period distribution expected if all binaries evolve according the \citetalias{Rappaport1983} MB law (Equation~\ref{eq:jdot}). In stark disagreement with this prediction, the observed period distributions in all absolute magnitude bins are rather flat. Fitting a power law model ${\rm d}N/{\rm d}P_{\rm orb}\propto P_{\rm orb}^{\alpha}$, we find $\left|\alpha \right | < 0.4$ in for all absolute magnitude bins. This remains true if we limit our analysis to $P_{\rm orb} < 1$ day. The steepest trend is found in the $9 < M_{\rm G,0} < 10$ bin, where we find  ${\rm d}N/{\rm d}P_{\rm orb}\propto P_{\rm orb}^{0.35}$ at $P_{\rm orb} < 2$ days, but this trend flattens again at longer periods. 

In the middle and lower panels, we zoom out further to show the separation distribution at longer periods. Here the dashed black line shows the canonical lognormal period distribution observed for solar-type stars in the solar neighborhood \citep[][]{Duquennoy1991, Raghavan2010, Duchene2013, Moe2017}, which has a peak at $10^5$ days, logarithmic dispersion of 2.3 dex, and a total binary fraction of $\approx$ 0.5. The red dashed line shows the analogous distribution for M dwarfs, which we model with a peak at $10^{3.75}$ days, logarithmic dispersion of 1.3 dex, and binary fraction 0.25 \citep[][]{Fischer1992, Duchene2013}. We note that while ${\rm d}N/{\rm d}\log P_{{\rm orb}}$ (bottom panel) decreases at short periods, ${\rm d}N/{\rm d}P_{\rm orb}$ (top and middle panels) increases. In the period range probed by our data, the observed period distributions in the two higher-mass bins (early M and K dwarfs) are basically consistent with the solar-type lognormal distribution. In the lowest-mass bin, the binary fraction at all periods probed by the data is lower than for solar-type stars. This is basically consistent with trends found in previous work at longer periods \citep[e.g.][]{Duchene2013, Moe2017}. 

\subsection{Implications for magnetic braking}
\label{sec:implications}

We now compare the observed, incompleteness-corrected period distributions to Monte Carlo simulations, which we carry out using the same approach used in Figure~\ref{fig:simulations}. We assume a uniform age distribution between 0 and 10 Gyr, and experiment with a range of birth period distributions and several of the MB laws summarized in Section~\ref{sec:other_MB}.

\subsubsection{Birth period distribution}

\begin{figure}
    \centering
    \includegraphics[width=\columnwidth]{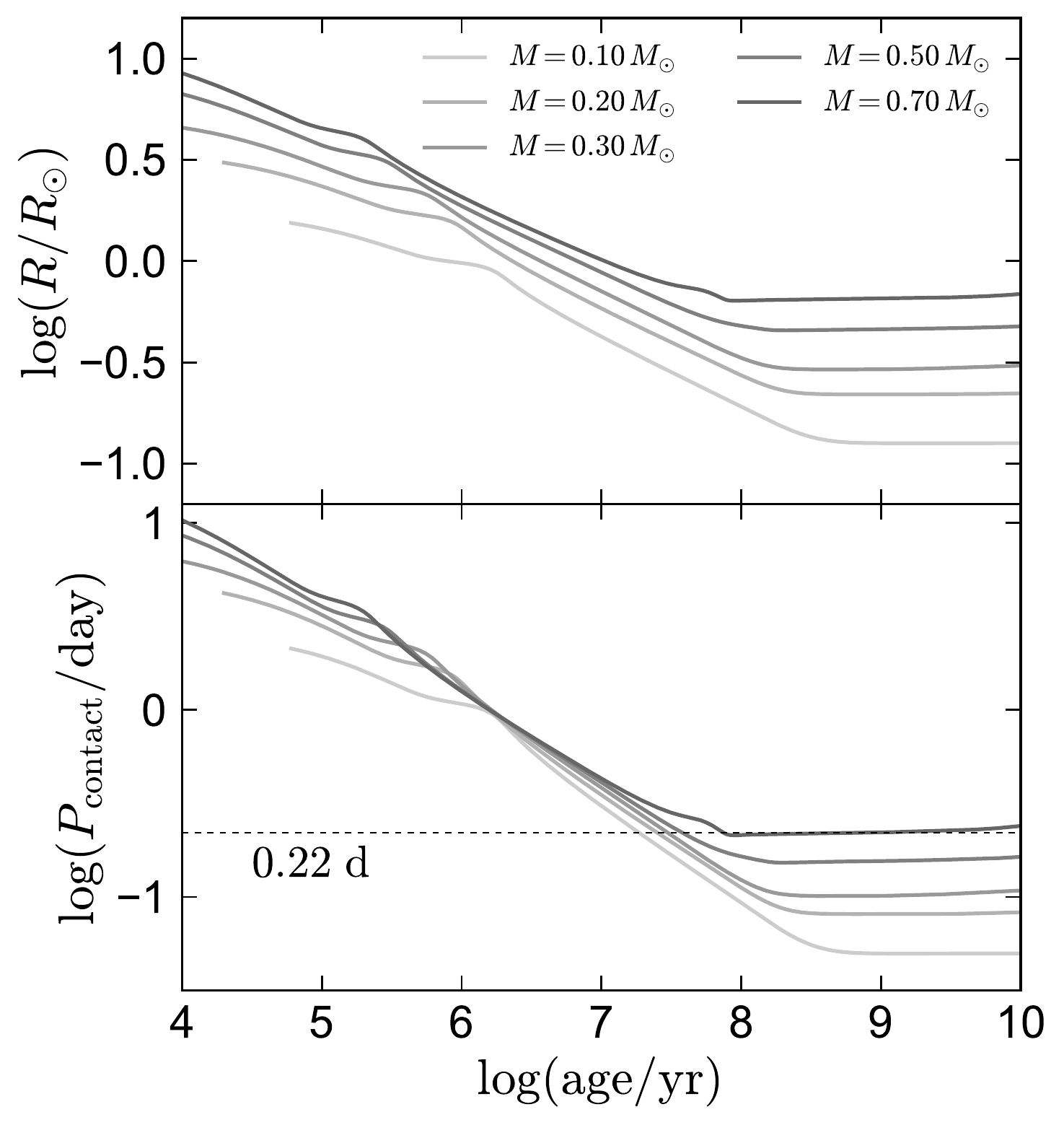}
    \caption{Top: radius evolution of single-star MIST models with several masses. Bottom: orbital period at which each model would overflow its Roche lobe in an equal-mass binary. For main-sequence binaries with masses $M\lesssim 0.9\,M_{\odot}$, binaries with $P_{\rm orb}\gtrsim 0.3$ days will be detached. The shortest achievable period for a detached main-sequence binary is $\approx 85$ minutes.
    Stars are larger during their pre-main sequence evolution, and thus cannot reach the shortest observed periods until they reach the main sequence. For example, binaries of any mass would be in contact at $P_{\rm orb}\lesssim 1$ day at an age of $10^6$ years.}
    \label{fig:mist_Pmin}
\end{figure}

The birth period distribution -- i.e., the period distribution {\it before} MB-driven orbital evolution --  is uncertain. At the periods where MB is expected to be important, observational constraints are scarce \citep[e.g.][]{Kounkel2019}. Pre-main sequence stars are larger and less dense than main-sequence stars, and thus would overflow their Roche lobes at the orbital periods of greatest interest. This is illustrated in Figure~\ref{fig:mist_Pmin}, where we show as a function of age the orbital period at which main-sequence stars with a range of masses would overflow their Roche lobes. We calculate Roche lobe radii using the fit of \citet[][]{Eggleton_1983}. At an age of $3\,\rm Myr$ -- comparable to the lifetime of a typical young stellar association in the Galactic disk -- there can be no binaries with $P_{\rm orb}\lesssim 0.75\,\rm day$. This has led some studies to assume a minimum birth period set by protostar radii at a fixed age. For example, \citet{Stepien2006} modeled the evolution of short-period binaries under the assumption that the minimum birth period is 2 days. 

Even after the dissolution of their birth environments, binary orbits can evolve due to three-body interactions. In particular, excitation of eccentricity through Kozai cycles, with subsequent dissipation of orbital energy through tides, can shrink orbits after stars have reached the main sequence \citep[``Kozai cylces with tidal friction'' (KCTF); e.g.][]{Mazeh1979, Fabrycky2007}. Since most of the EBs in our sample likely have distant tertiaries (Section~\ref{sec:tertiaries}), interactions with them are likely important in setting the birth period distribution. \citet{Fabrycky2007} performed simulations to assess the effects of KCTF on the period distribution of main-sequence binaries, assuming an initial period distribution similar to the canonical lognormal distribution for solar-type stars. Fitting a power law to their predicted final period distribution, we find that at short periods it is well-described by ${\rm d}N/{\rm d}\log P_{\rm orb} \propto P_{\rm orb}^{1/2}$; i.e., ${\rm d}N/{\rm d}P_{\rm orb} \propto P_{\rm orb}^{-1/2}.$

Given the uncertainties discussed above, we show results for three different birth distributions: 
\begin{itemize}
    \item Uniform distribution between the contact limit for a main-sequence star (bottom panel of Figure~\ref{fig:mist_Pmin}), and 5 days. The 5 day upper limit is chosen to be large enough that increasing it has no effect on the distribution at $P_{\rm orb} < 2$ days. 
    \item Uniform distribution between 0.75 and 5 days. The lower limit corresponds to the minimum possible period for a pre-main sequence star with age of a few Myr (Figure~\ref{fig:mist_Pmin}).
    \item Power-law distribution, ${\rm d}N/{\rm d}P_{\rm orb}\propto P^{-1/2}$ between 3 times the contact limit and 5 days, motivated by the results of simulations by \citet[][]{Fabrycky2007}. Here the lower limit of 3 times the contact period is chosen to represent the period where tides become very strong, suppressing Kozai cycles. The typical period at which Kozai cycles are suppressed is likely longer than this. However, we do not find any dearth of short-period EBs among kinematically young stars (Appendix~\ref{sec:ages_appendex}), suggesting that either MB or KCTF produces a population of short-period EBs on a timescale that is short compared to the Hubble time.
\end{itemize}

\subsubsection{Simulations}
\begin{figure*}
    \centering
    \includegraphics[width=\textwidth]{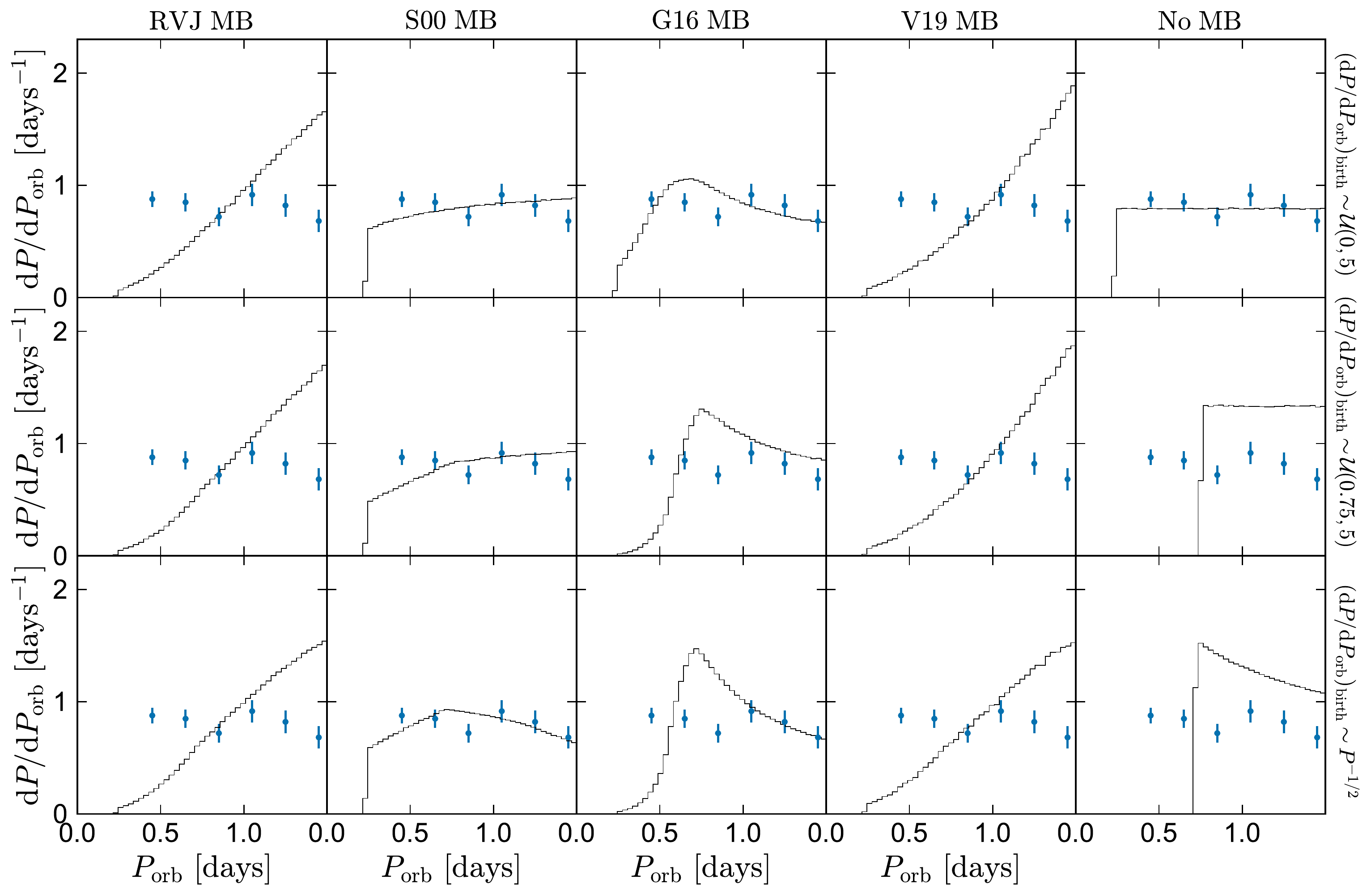}
    \caption{Comparison of simulated and observed period distributions for the $5 < M_{G,0} < 7$ bin, corresponding to primary masses $0.65 \lesssim M_1/M_{\odot} \lesssim 0.9$. Different rows show different birth period distributions (which are shown in the rightmost column); different columns show different MB laws. The flat observed period distribution is in strong tension with the predictions of the \citetalias{Rappaport1983} and \citetalias[][]{Van2019} MB models for any plausible birth period distribution. The saturated MB model from \citetalias[][]{Sills2000} reproduces the data reasonably well for all the simulated birth period distributions. The \citetalias{Garraffo2016} model predicts MB to become inefficient at short periods and thus matches the data only for a flat birth period distribution extending to short periods. }
    \label{fig:simulation_0.75}
\end{figure*}

\begin{figure*}
    \centering
    \includegraphics[width=\textwidth]{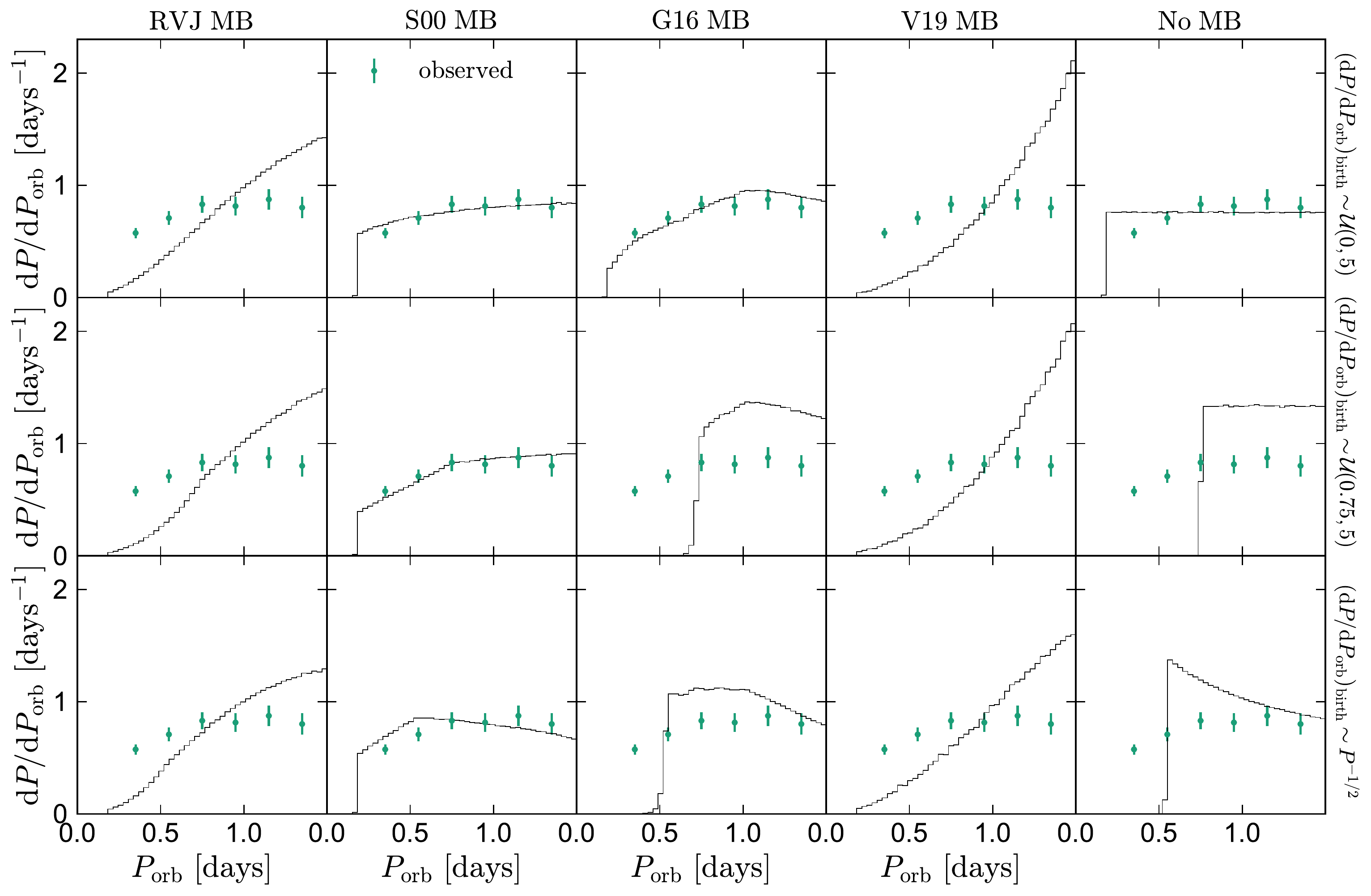}
    \caption{Same as Figure~\ref{fig:simulation_0.75}, but for the  $7 < M_{G,0} < 9$ bin,  corresponding to primary masses $0.4 \lesssim M_1/M_{\odot} \lesssim 0.65$. The observed period distribution is close to flat, and the \citetalias{Rappaport1983} and \citetalias[][]{Van2019} MB prescriptions predict a steeper fall-off at short periods than observed for all birth period distributions.  The \citetalias[][]{Sills2000} prescription produces a period distribution similar to that observed for all birth distributions. The \citetalias[][]{Garraffo2016} model predicts weak MB at these masses, and reproduces the data only if the birth period distribution extends to short periods. }
    \label{fig:simulation_0.55}
\end{figure*}

\begin{figure*}
    \centering
    \includegraphics[width=\textwidth]{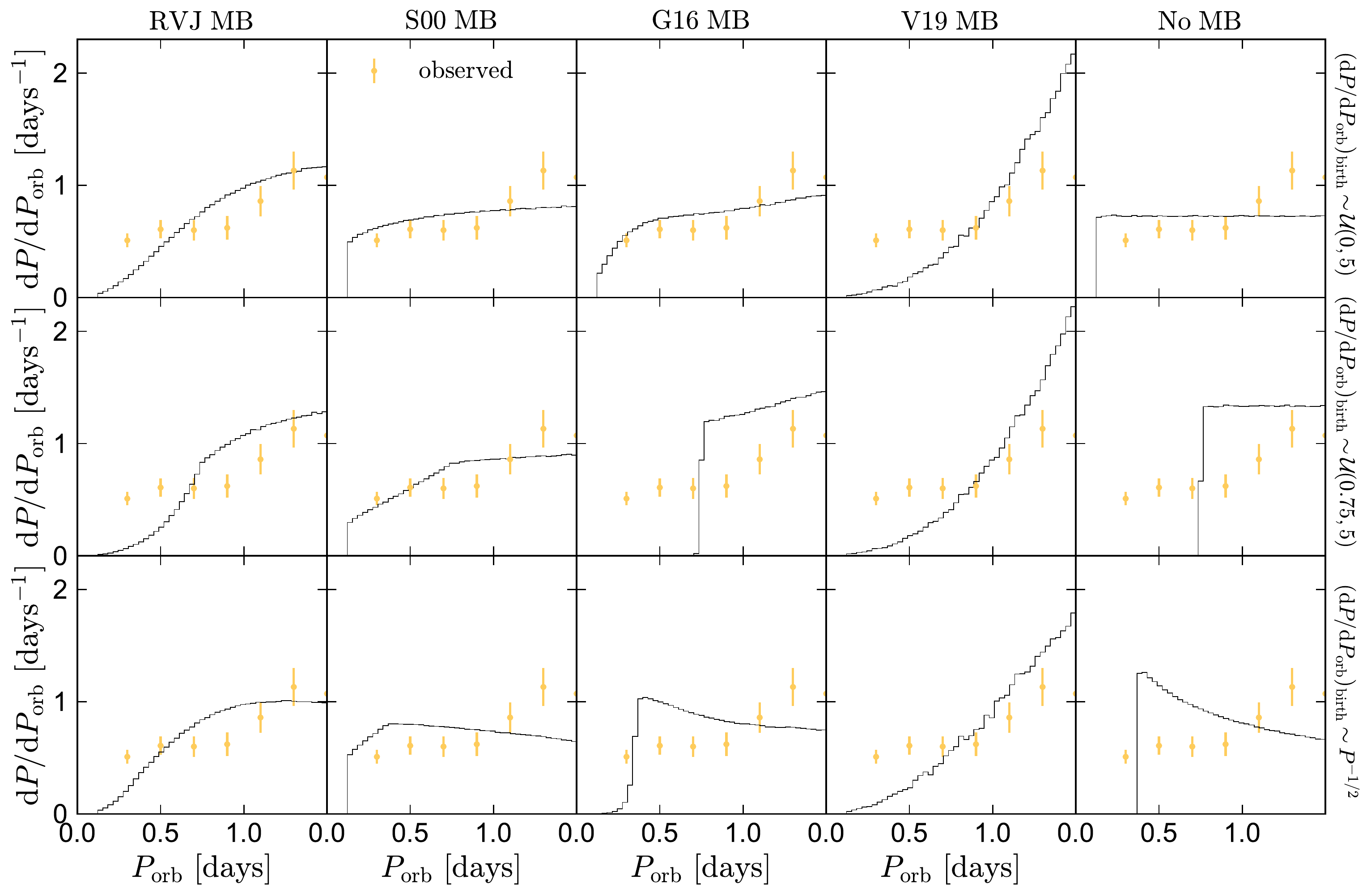}
    \caption{Same as Figure~\ref{fig:simulation_0.75}, but for the  $9 < M_{G,0} < 10$ bin,  corresponding to primary masses $0.3 \lesssim M_1/M_{\odot} \lesssim 0.4$. These binaries straddle the fully/partially convective boundary. As in other mass bins, the observed period distribution is close to flat. The \citetalias{Rappaport1983} and \citetalias[][]{Van2019} MB prescriptions predict a steeper fall-off at short periods than observed. The \citetalias{Sills2000} and \citetalias{Garraffo2016} prescriptions can approximately match the data for birth period distributions that are flat well-populated at short periods.  }
    \label{fig:simulation_0.35}
\end{figure*}

\begin{figure*}
    \centering
    \includegraphics[width=\textwidth]{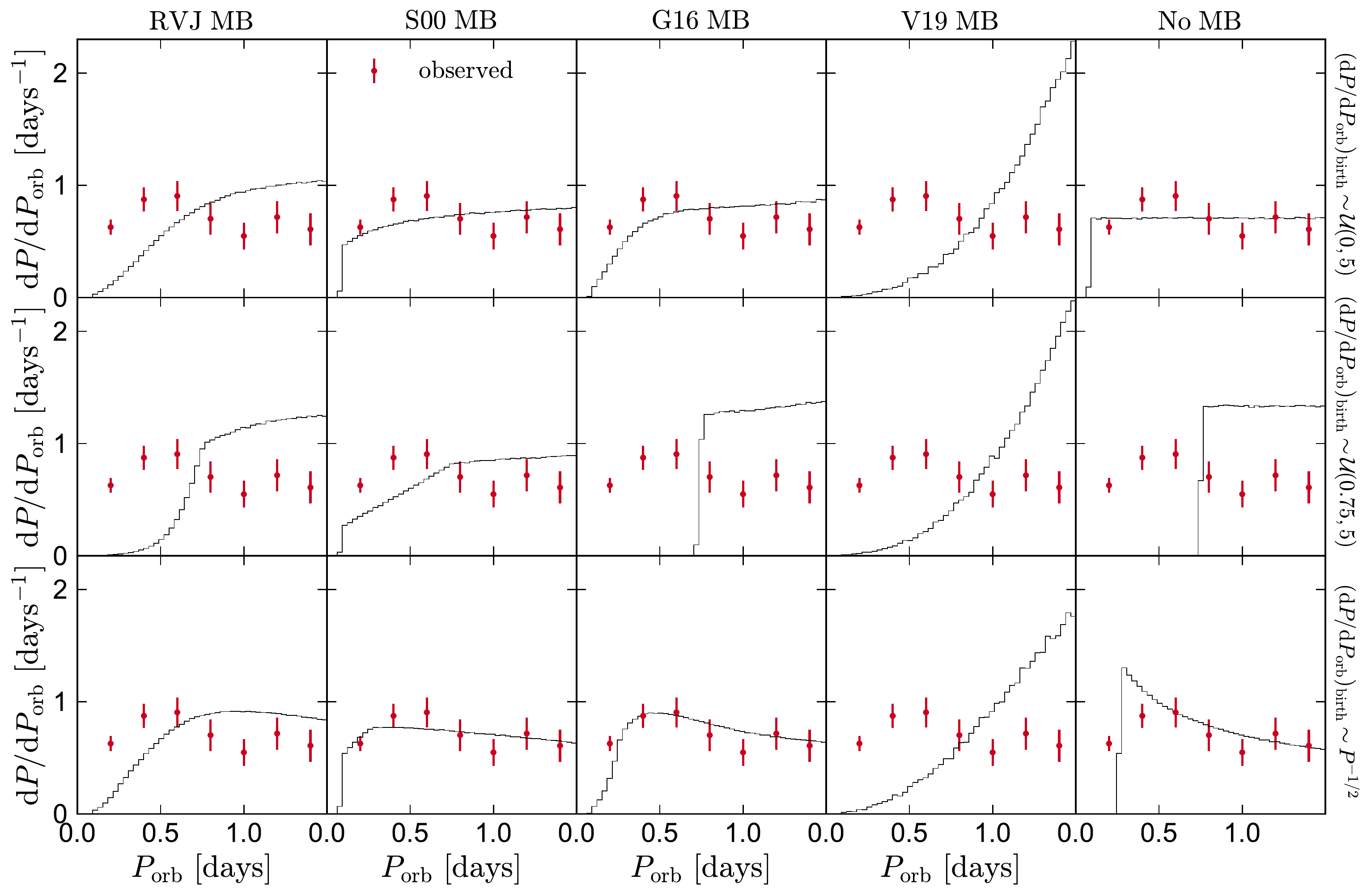}
    \caption{Same as Figure~\ref{fig:simulation_0.75}, but for the  $M_{G,0} > 10$ bin, corresponding to primary masses $ M_1 \lesssim 0.3\,M_{\odot}$. Both components in these binaries are expected to be fully convective. As in the higher-mass bins, the observed period distribution is basically flat. We do not ``turn off'' MB below the fully convective limit for any of the simulations. As in the higher mass bins, the \citetalias{Rappaport1983} and \citetalias{Van2019} prescriptions predict a dearth of binaries at the shortest periods, which is not observed. The \citetalias[][]{Sills2000} and \citetalias{Garraffo2016} models predict MB to be weak and these masses, and thus match the data only if the birth period distribution included binaries down to near the contact limit. }
    \label{fig:simulation_0.25}
\end{figure*}

For each choice of birth period distribution and MB law, we simulate a population of $10^6$ binaries with a uniform age distribution between 0 and 10 Gyr, representative of the solar neighborhood \citep[e.g.][]{Cukanovaite2022}. We assume a primary mass distribution matching the observed samples in each bin, and a uniform mass ratio distribution between $0.1\,M_{\odot}/M_1$ and 1, consistent with observations for close binaries \citep[e.g.][]{Duchene2013}. The lower limit of $0.1\,M_{\odot}$ is chosen to exclude companions below the hydrogen burning limit, which are rare or absent in the observed sample. We remove binaries from the simulation when their primaries overflow their Roche lobes, after which they would appear as contact binaries or merge. The results are shown in Figures~\ref{fig:simulation_0.75}-\ref{fig:simulation_0.25}. Each figure shows a different primary mass bin and compares the results of five MB prescriptions (including no MB) for three different birth period distributions. We normalize the final period distributions for all simulations and for the data, such that the slope is informative, but not the normalization. For the observational constraints, we again exclude the shortest-period bin, only showing systems that are unambiguosly detached.

The clearest message from these experiments is that for any plausible birth distribution, the \citetalias{Rappaport1983} and \citetalias[][]{Van2019} MB prescriptions produce a period distribution that, at the short-period end, increases steeply toward longer periods. The predicted slope of the distribution is $\sim P_{\rm orb}^{7/3}$ for the \citetalias{Rappaport1983} law, and similar for the \citetalias{Van2019} law (Figure~\ref{fig:mb_laws}). The data do not display any such trend in any of the mass bins we analyze and are thus inconsistent with these MB laws. 

For the \citetalias{Sills2000} and \citetalias{Garraffo2016} laws, satisfactory agreement the data can be achieved if the birth period distribution was roughly uniform at short periods and extended to close to the contact limit. The birth period distribution with a sharp truncation at 0.75 days does not reproduce the data for the lowest-mass EBs with these MB laws (Figure~\ref{fig:simulation_0.35} and Figure~\ref{fig:simulation_0.25}) because MB there is predicted to be weak in the  \citetalias{Sills2000} and \citetalias{Garraffo2016} models, such that binaries born with $P_{\rm orb}> 0.75$ days would not reach contact within a Hubble time. This suggests that either MB is stronger at low masses than assumed in these models (but still has a saturated dependence on orbital period), or another process, such as KCTF, is responsible for shrinking of binary orbits to near the contact limit.

Our comparison of simulations and data in figures~\ref{fig:simulation_0.75}-\ref{fig:simulation_0.25} is focused on the slope of the period distribution; i.e., the relative number of EBs at shorter vs. longer periods. We have not attempted to model the absolute number of binaries in a given period bin because we do not precisely know the initial binary fraction at the relevant periods ($P_{\rm orb} < \rm few\,days$). The absolute number of short-period EBs can, however, be used as a constraint on some MB models even with very conservative assumptions. In particular, the CARB model introduced by \citetalias{Van2019} predicts that the pre-contact lifetime of an equal-mass binary with $M_1= 0.4\,M_{\odot}$ and initial period of 0.5 days is only 0.002 Gyr -- this is a consequence of the very large MB torques it predicts at low masses (see Figure~\ref{fig:mb_laws}).

The observed period distribution (Figure~\ref{fig:corrected_period_distributions}) implies that about 1 in 2000 stars with $M\approx 0.4\,M_{\odot}$ is a close binary with $P_{\rm orb} < 0.5$ days. Even in the extreme scenario where {\it every} 0.4\,$M_{\odot}$ star is born with a companion with $P_{\rm orb} < 0.5$ days (a scenario clearly ruled out by measurements of the binary populations in young stellar associations), one would only expect $\approx (0.002\,\rm Gyr)/(10\,\rm Gyr)\approx$ 1 in 5000 binaries to be observed with $P_{\rm orb} < 0.5$ days if their angular momentum loss were governed by CARB MB. Put simply, the model predicts that short-period EBs should almost instantly inspiral and become contact binaries or merge. It thus cannot explain the longer-term survival of a large population of short-period, low-mass main-sequence EBs. We stress that this is a statement about the typical MB law experienced by most short-period EBs. The data do not rule out a scenario in which a small fraction of binaries experience much stronger-than-average MB.

\subsubsection{Other MB models}
We briefly comment on the implications of our analysis for two other MB laws: the quasi-saturated model from \citet{Ivanova2003}, and the flexible model from \citet{Kawaler1988}.

The \citet{Ivanova2003} law predicts a MB torque whose scaling with period, $\dot{J}\propto P^{-1.3}_{\rm orb}$, is only slightly steeper than in the \citetalias[][]{Sills2000} law. The predicted MB at $M=1\,M_{\odot}$ is thus similar to that law, and the model can reasonably reproduce the observed period distributions in our highest-mass EB sample. However, at $M\lesssim 0.5\,M_{\odot}$, the model predicts MB that is too weak to significantly shrink the periods of most EBs within a Hubble time, even at periods near the contact limit. The model can thus only reproduce the data if the birth period distribution is already flat.

The model from \citet[][their Equation 10]{Kawaler1988} has a flexible parameterization that allow for a wide variety of different scalings with $P_{\rm orb}$. We find that with $a = 0$ and $n = 0.5$, their parameterization produces a MB law similar to the \citetalias{Sills2000} and \citet{Matt2015} parameterizations in the saturated regime, and thus predicts period distributions similar to those shown for the \citetalias[][]{Sills2000} law. Models with $a=0$ have a magnetic field strength that is independent of rotation period, and thus are saturated; they also predict $\dot{J}\propto P_{\rm orb}^{-1}$.

\section{Discussion}
\label{sec:discussion}

This work is not the first to suggest that a Skumanich-like MB law overestimates the strength of MB in rapidly-rotating, low mass stars. Indeed, many studies of cluster M dwarfs over the past three decades have found that rapid-rotators are more abundant than predicted by the Skumanich relation, which would predict them to spin down in a few Myr \citep[e.g.][]{Stauffer1987, Stauffer1997, Barnes2003, Newton2016, Medina2020}. These data are better explained by ``saturated'' MB laws, in which the magnetic field strength ceases to increase with rotation rate above a critical rotation rate, leading to a weaker MB torque \citep[e.g.][]{Sills2000}. The simplest such models predict a torque that scales as $\dot{J}\propto P_{\rm orb}^{-1}$, not far from the $\dot{J}\propto P_{\rm orb}^{-2/3}$ we infer.
Our results thus imply that the same saturation occurs in tidally locked, close binaries.

\subsection{Implications for CVs and LMXBs}
\label{sec:cvs}
Two decades ago, \citet{Andronov2003} argued that the observed rotation periods of isolated M dwarfs in clusters are inconsistent with classical MB models assumed in the CV literature, which predict $\dot{J}\propto P_{\rm orb}^{-3}$.  They advocated a saturated MB torque that scales as $\dot{J}\propto P_{\rm orb}^{-1}$, basically consistent with our findings. Andronov et al. found that such a MB torque would require significant revision of the standard evolutionary model for CVs\footnote{By ``standard model'', we loosely include models in which (a) the CV period gap at 2-3 hours results from a rapid weakening of MB as donors cross the the fully convective boundary, (b) above the period gap, MB follows an RVJ-like law and donors are out of thermal equilibrium, and (c) below the period gap, gravitational radiation dominates angular momentum losses  \citep[][]{Knigge_2011}. } for two reasons: (a) the evolutionary timescales and inferred ages of CVs would increase by an order of magnitude and, (b) the donor stars in CVs above the period gap would be in thermal equilibrium and thus would not detach even if MB became weaker at the fully convective boundary. This would render untenable the disrupted MB model for the CV period gap. This tension remains unresolved, and most evolutionary models for CVs in the last two decades have continued to use a $\dot{J}\propto P_{\rm orb}^{-3}$ prescription above the period gap.

Our constraints on the MB law are consistent with the model assumed by \citet[][]{Andronov2003}, at least as concerns the scaling of $\dot{J}$ with $P_{\rm orb}$: the MB law they assumed (which was taken from \citetalias{Sills2000}) can reproduce the observed EB period distributions for a suitable choice of birth period distribution, while the \citetalias{Rappaport1983} law and other laws that scale as $\dot{J}\propto P_{\rm orb}^{-3}$ cannot. It thus seems worth revisiting whether other uncertain aspects of the CV standard model can change, such that the observed period distribution and mass transfer rates can be matched with a saturated MB law. 

We also do not detect any significant difference between the period distributions of binaries containing fully and partially convective M dwarfs. Various works have proposed that MB ``turns off'' below the fully convective boundary \citep[e.g.][]{Spruit1983, Rappaport1983, Howell2001, Knigge_2011}; this is the reason for the ``period gap'' in the CV standard model. The shortest periods we observe in the fully convective EB sample (down to $P_{\rm orb} < 0.1$ days) cannot be the birth periods: pre-main sequence M dwarfs are large and would not fit in orbits with periods less than $\approx 1$ day (e.g. Figure~\ref{fig:mist_Pmin}). The abundance of observed binaries at these short periods, including systems in which the presence of a stellar tertiary can be excluded, would appear to suggest that MB continues to operate efficiently below the fully-convective limit.

Previous works \citep[e.g.][]{Schreiber2010, Zorotovic2016} have found evidence for a reduction in MB at the fully convective boundary in the period and mass distributions of detached white dwarf + M dwarf binaries. Although the binaries studied in those works are detached, it is not easy to distinguish detached CVs (those which have temporarily paused mass transfer) from pre-CVs (those which have not yet come into contact). Inference about the MB law from detached white dwarf + M dwarf binaries thus depends on modeling of the CV population and on assumptions about the post-common envelope period distribution.

The rotation period distributions of single stars (both in clusters and in the field) display no discontinuity at the fully convective boundary, in tension with expectations in the disrupted MB paradigm. On the other hand, there are multiple lines of evidence that CV donors above the period gap are out of thermal equilibrium \citep[e.g.][]{Beuermann_1998, Knigge_2006}, which \citet{Andronov2003} showed is not expected if the only source of angular momentum loss is saturated MB following the prescription from \citetalias[][]{Sills2000}.

The long CV lifetimes predicted for a saturated MB law could help explain the lack of period-bouncers in observed CV samples \citep[e.g.][]{Pala2020, Belloni2020}, but would also increase the fraction of CVs observed to have evolved donors \citep[e.g.][]{El-Badry2021_cvs}.

Clearly, further work is needed to reconcile the apparently discrepant MB laws inferred from CVs and related objects, and those inferred from single stars and main-sequence binaries. One possibility is that MB operates differently in binaries with ongoing mass transfer than in detached binaries. Some phenomenological models have indeed been introduced to add an additional source of angular momentum loss associated with mass transfer (e.g. CAML; \citealt{King1995}), and these models have made significant progress in explaining several aspects of the observed CV population \citep[e.g.][]{Schreiber2016}.
However, among the small number of CVs with well-measured mass transfer rates at periods where MB is expected to dominate, mass transfer rates are generally {\it lower} than predicted by the standard model \citep[e.g.][]{Pala2022}, and an additional source of angular momentum loss would exacerbate this. It is possible that the observed mass transfer rates are consistent with the saturated MB laws we examined, with the addition of consequential angular momentum loss. For example, much of the total angular momentum loss could occur during or shortly after novae \citep[e.g.][]{Nelemans2016, Ginzburg2021}, or be driven by circumbinary disks \citep[e.g.][]{Spruit2001}.
In this case, the total angular momentum loss in CVs could even be similar to the predictions of \citetalias[][]{Rappaport1983}, but this would be a coincidence, and there would still be little reason to expect that law's predictions to hold over a range of periods.

It should in principle be possible to infer the strength of MB in CVs from measurements of their mass transfer rates. The most reliable estimates of CV mass transfer rates are likely those inferred from temperatures of the accreting white dwarfs, which are assumed to be set by compressional heating and trace the long-term average accretion rate \citep[e.g.][]{Townsley2003, Pala2017}. The inferred accretion rates from such studies are, however, difficult to interpret. In the period range of 3 to 5 hours -- where MB is expected to be the dominant driver of CV evolution -- there is a enormous dispersion in the inferred mass transfer rates. A few systems -- mostly nova-like variables just above the period gap -- have mass transfer rates even larger than predicted even in the standard model with the \citetalias{Rappaport1983} MB law. The remaining systems, mostly dwarf novae with periods above 4 hours, have mass transfer rates more than an order of magnitude below the predictions of the standard model. A full explanation of this phenomenology is still lacking \citep[][]{Belloni2020}. Some of the systems with unexpectedly high inferred mass transfer rates could be young systems that recently underwent thermal-timescale mass transfer or a recent novae, but the data may simply indicate that the strength of MB varies significantly from CV to CV at fixed period. 

The situation is similar for LMXBs. At orbital periods $4\lesssim P_{\rm orb}/\rm hr \lesssim 24$, there are several persistent LMXBs with near-Eddington accretion rates \citep[e.g.][]{Van2019b}. It is the existence of these systems that motivated the ``boosted'' CARB MB prescription \citep[][]{Van2019}, which is even stronger than the \citetalias{Rappaport1983} model and which we found inconsistent with the observed period distribution of main-sequence binaries. However, alongside these high-accretion rate systems, there is a larger population of LMXBs with mass transfer rates that are 2-3 orders of magnitude lower at fixed $P_{\rm orb}$. The mass transfer rates of these systems are too low to explain with an unsaturated MB law, except in rather fine-tuned scenarios with an evolved donor just below the bifurcation period \citep[][]{Podsiadlowski2002, Podsiadlowski_2003}. Unlike the CVs with high mass transfer rates, most of the known persistent LMXBs have sufficiently long periods that the donors must be somewhat evolved, and probably above the bifurcation period. In this case, their high mass transfer rates could be driven mainly by the donor's nuclear evolution and expansion, rather than by MB. As with CVs, it is possible that high mass transfer rates in LMXBs are due to angular momentum loss mechanisms that are a consequence of the mass transfer process \citep[e.g.][]{Ma2009, Chen2016}. 

Several millisecond pulsars have been observed with helium white dwarf companion in close orbits, and these systems presumably formed from LMXBs with evolved donors. 
It has been suggested \citep[e.g.][]{Chen2021, Soethe2021} that the CARB MB prescription can alleviate the ``fine-tuning problem'' for the formation of these system \citep{Istrate2014b}: CARB produces sufficiently strong MB that neutron star + main-sequence binaries with a wide range of post-common envelope periods can evolve to short periods. Our results imply that if the MB law is universal, CARB MB is unlikely to be the right solution to that problem, since it dramatically overpredicts the inspiral rate of close main-sequence binaries. Consequential angular momentum loss is also unlikely to help in this case, because stronger MB appears to be required {\it before} the onset of mass transfer. Models in which weak MB below the bifucation period (due, e.g., to loss of the donor's convective envelope) leads to detachment may be more promising.

In summary, there are multiple lines of evidence that MB in close detached binaries does not follow a Skumanich-like law with $\dot{J}\propto P_{\rm orb}^{-3}$, and a saturated MB law with $\dot{J}\propto P_{\rm orb}^{-1}$ is more capable of matching the data. In binaries with ongoing mass transfer, such as CVs and LMXBs, the situation may be more complicated.  Observationally-inferred mass transfer rates in these systems are in most cases consistent with theory at short periods (where MB is expected to be subdominant), but not at long periods \citep[e.g.][]{Belloni2020, Pala2022}. We conclude that the role of MB in driving angular momentum loss in LMXBs and CVs above the period gap is ripe for reassessment. Models in which much of the angular momentum loss in these systems is a consequence of mass transfer provide a potential solution to some, but not all, of these puzzles.

\section{Summary}
\label{sec:summary}
Using light curves from ZTF, we have constrained the period distribution of detached low-mass main-sequence eclipsing binaries (EBs) with primary masses $0.1 \lesssim M/M_{\odot} \lesssim 0.9$ and periods $0.1\lesssim P_{\rm orb} \lesssim 10$ days. Although many previous works have assembled large samples of eclipsing binaries using photometric surveys, this is to our knowledge the first study to model the survey selection function and thus to infer the intrinsic period distribution in this range of masses and periods. We then compared the observed period distribution to simple population models with different assumptions about how magnetic braking (MB) removes angular momentum from binary orbits. Our main results are as follows:

\begin{enumerate}
    \item {\it The period distribution extends down to the contact limit at all masses.} Many previous studies have reported a minimum period of $\approx 0.22$ days for main-sequence binaries. This is roughly the period at which a $0.7\,M_{\odot}$ main-sequence star will overflow its Roche lobe and form a contact binary (Figure~\ref{fig:mist_Pmin}). However, lower-mass main sequence binaries can reach significantly shorter periods, down to $\approx 0.06$ days. When we split the data into bins of primary mass, EBs are abundant down to the contact limit in each mass bin (Table~\ref{tab:summary}; Figure~\ref{fig:detections}).     Several of the binaries in the lowest-mass subsample have shorter orbital periods than any other main-sequence binaries discovered to date. This suggests that the 0.22 day cutoff found in previous studies was primarily a result of the dearth of low-mass  main-sequence binaries in magnitude-limited samples.

    \item {\it The incompleteness-corrected period distribution of detached binaries is flat at $P_{\rm orb} \lesssim 10\,\rm days$.}  We characterize the search's sensitivity using injection and recovery simulations (Figure~\ref{fig:inject_recovery}). Unsurprisingly, the search is most sensitive to short-period EBs, which have higher eclipse probabilities and duty cycles.  
    In all mass bins we consider, the incompleteness-corrected period distribution at short periods is nearly flat; i.e., ${\rm d}N/{\rm d}P\propto P^0$ (Figure~\ref{fig:corrected_period_distributions}; Fitting a power law model ${\rm d}N/{\rm d}P_{\rm orb}\propto P_{\rm orb}^{\alpha}$, we find $\left|\alpha \right | < 0.4$ in all subsamples at $P_{\rm orb} < 2$ days.) This remains true for detached binaries at all periods down to the contact limit. In contrast, contact binaries pile up at the contact limit. This is consistent with distributions hinted at by previous studies of EBs \citep[][]{Maceroni1999}, but is here unambiguous over a range of masses and robust to selection effects. 
    
    Although the absolute number of close binaries per unresolved source declines with stellar mass (Figure~\ref{fig:corrected_period_distributions}), the shape of the period distribution at close periods varies little between mass bins. There is no obvious difference between the period distributions of binaries containing fully-- and partially--convective stars. 
    
    \item {\it Observed period distributions are inconsistent with the MB models most widely used in binary evolution calculations}: MB removes angular momentum from binary orbits, leading to inspiral. How rapidly this inspiral occurs depends on how the MB torque varies with a star's rotation period. The most widely used MB laws in the binary evolution literature, which are based on the \citet{Skumanich1972} spin-down relation, predict a torque that scales as $\dot{J}\propto P_{\rm orb}^{-3}$ (e.g. \citealt{Verbunt1981}; \citetalias{Rappaport1983}). This leads to rapidly accelerating inpiral (Figure~\ref{fig:porb_tracks}) and a dearth of binaries at the shortest periods. At periods where MB is predicted to be efficient, this prediction is insensitive to the initial period distribution (Figure~\ref{fig:simulations}).
    
    A steeply falling period distribution at short periods is not found in the data in any mass bin (Figures~\ref{fig:simulation_0.75}-\ref{fig:simulation_0.25}). We conclude that the data cannot be understood as a result of a Skumanich-like MB law, and the scaling of $\dot{J}$ with orbital period must be shallower than $P_{\rm orb}^{-3}$.
    A flat observed period distribution is most naturally produced by a MB torque that scales as $\dot{J}\propto P_{\rm orb}^{-2/3}$, which would lead to linear period decay with time. For such a MB law, the observed period distribution is also sensitive to the birth period distribution. 
    Given uncertainties in initial short-period binary fraction, the magnitude of the MB torque is more difficult to measure with our approach than how it scales with period. The scalings we adopt (e.g. Equation~\ref{eq:jdot_sat}) are calibrated to the rotation rates of single stars in clusters. 
    
    The observed period distributions are more consistent with models in which the MB torque has a shallower period dependence. Saturated MB laws (e.g., \citetalias{Sills2000}, \citealt[][]{Matt2015}) can reproduce the observed period distributions for a range of birth period distributions at the higher-mass end (i.e., K dwarfs). At the lowest masses, these relations predict that MB is too weak to change binary orbits significantly except at the shortest periods ($P_{\rm orb} < 0.5$ days), so they can explain the data only if the birth period distribution is populated down to near the contact limit. This cannot occur in an isolated binary because pre-main sequence stars cannot fit in such tight orbits, but may occur with the help of three-body interactions. Similar considerations apply for the \citetalias[][]{Garraffo2016} model. Finally, the \citetalias[][]{Van2019} suffers similar challenges in matching the data to the \citetalias[][]{Rappaport1983} model, with the additional complication that it predicts an unrealistically short lifetime for low-mass  detached EBs.  While it cannot apply to a majority of EBs, such a model might still explain rare systems with unusually efficient MB.

\end{enumerate}

Our analysis is focused on low-mass main sequence stars with $M\lesssim 0.9\,M_{\odot}$. While such stars are representative of the donors in typical CVs and LMXBs, there are also several classes of interacting close binaries that are thought to have formed from donors that initially had higher masses. The naive expectation from models is that MB weakens in main-sequence stars with $M\gtrsim 1.2\,M_{\odot}$, which have thin or absent convective envelopes. There is some evidence for such weakening in the rotation period distribution of single stars \citep[e.g.][]{Kraft1967} and in the mass transfer rates observed in evolved CVs \citep[][]{El-Badry2021_cvs}, but it would be useful to search for evidence of it in the orbital period distribution of higher-mass main-sequence binaries. Unlike the analysis in this work, such an analysis will need to account for significant evolution of the component stars during their main-sequence lifetimes.

\section*{Acknowledgements}
We thank Saul Rappaport, Boris Gaensicke, Tom Marsh, Brian Metzger, Ken Shen, Dave Charbonneau, Mercedes López-Morales, Eliot Quataert, Hans-Walter Rix, Selma de Mink, and E. Sterl Phinney, for helpful discussions.

This research was supported in part by the National Science Foundation under Grant No. NSF PHY-1748958.

Based on observations obtained with the Samuel Oschin 48-inch Telescope at the Palomar Observatory as part of the Zwicky Transient Facility project. ZTF is supported by the National Science Foundation under Grant No. AST-1440341 and a collaboration including Caltech, IPAC, the Weizmann Institute for Science, the Oskar Klein Center at Stockholm University, the University of Maryland, the University of Washington, Deutsches Elektronen-Synchrotron and Humboldt University, Los Alamos National Laboratories, the TANGO Consortium of Taiwan, the University of Wisconsin at Milwaukee, and Lawrence Berkeley National Laboratories. Operations are conducted by COO, IPAC, and UW.

\section*{Data Availability}
Data used in this study are available upon request from the corresponding author. 



\bibliographystyle{mnras}

\begin{thebibliography}{}
\makeatletter
\relax
\def\mn@urlcharsother{\let\do\@makeother \do\$\do\&\do\#\do\^\do\_\do\%\do\~}
\def\mn@doi{\begingroup\mn@urlcharsother \@ifnextchar [ {\mn@doi@}
  {\mn@doi@[]}}
\def\mn@doi@[#1]#2{\def\@tempa{#1}\ifx\@tempa\@empty \href
  {http://dx.doi.org/#2} {doi:#2}\else \href {http://dx.doi.org/#2} {#1}\fi
  \endgroup}
\def\mn@eprint#1#2{\mn@eprint@#1:#2::\@nil}
\def\mn@eprint@arXiv#1{\href {http://arxiv.org/abs/#1} {{\tt arXiv:#1}}}
\def\mn@eprint@dblp#1{\href {http://dblp.uni-trier.de/rec/bibtex/#1.xml}
  {dblp:#1}}
\def\mn@eprint@#1:#2:#3:#4\@nil{\def\@tempa {#1}\def\@tempb {#2}\def\@tempc
  {#3}\ifx \@tempc \@empty \let \@tempc \@tempb \let \@tempb \@tempa \fi \ifx
  \@tempb \@empty \def\@tempb {arXiv}\fi \@ifundefined
  {mn@eprint@\@tempb}{\@tempb:\@tempc}{\expandafter \expandafter \csname
  mn@eprint@\@tempb\endcsname \expandafter{\@tempc}}}

\bibitem[\protect\citeauthoryear{{Andronov}, {Pinsonneault}  \&
  {Sills}}{{Andronov} et~al.}{2003}]{Andronov2003}
{Andronov} N.,  {Pinsonneault} M.,   {Sills} A.,  2003, \mn@doi [\apj]
  {10.1086/343030}, \href
  {https://ui.adsabs.harvard.edu/abs/2003ApJ...582..358A} {582, 358}

\bibitem[\protect\citeauthoryear{{Barnes}}{{Barnes}}{2003}]{Barnes2003}
{Barnes} S.~A.,  2003, \mn@doi [\apj] {10.1086/367639}, \href
  {https://ui.adsabs.harvard.edu/abs/2003ApJ...586..464B} {586, 464}

\bibitem[\protect\citeauthoryear{{Bellm} et~al.,}{{Bellm}
  et~al.}{2019}]{Bellm2019}
{Bellm} E.~C.,  et~al., 2019, \mn@doi [\pasp] {10.1088/1538-3873/aaecbe}, \href
  {https://ui.adsabs.harvard.edu/abs/2019PASP..131a8002B} {131, 018002}

\bibitem[\protect\citeauthoryear{{Belloni}, {Schreiber}, {Zorotovic},
  {I{\l}kiewicz}, {Hurley}, {Giersz}  \& {Lagos}}{{Belloni}
  et~al.}{2018}]{Belloni2018}
{Belloni} D.,  {Schreiber} M.~R.,  {Zorotovic} M.,  {I{\l}kiewicz} K.,
  {Hurley} J.~R.,  {Giersz} M.,   {Lagos} F.,  2018, \mn@doi [\mnras]
  {10.1093/mnras/sty1421}, \href
  {https://ui.adsabs.harvard.edu/abs/2018MNRAS.478.5626B} {478, 5626}

\bibitem[\protect\citeauthoryear{{Belloni}, {Schreiber}, {Pala},
  {G{\"a}nsicke}, {Zorotovic}  \& {Rodrigues}}{{Belloni}
  et~al.}{2020}]{Belloni2020}
{Belloni} D.,  {Schreiber} M.~R.,  {Pala} A.~F.,  {G{\"a}nsicke} B.~T.,
  {Zorotovic} M.,   {Rodrigues} C.~V.,  2020, \mn@doi [\mnras]
  {10.1093/mnras/stz3413}, \href
  {https://ui.adsabs.harvard.edu/abs/2020MNRAS.491.5717B} {491, 5717}

\bibitem[\protect\citeauthoryear{{Beuermann}, {Baraffe}, {Kolb}  \&
  {Weichhold}}{{Beuermann} et~al.}{1998}]{Beuermann_1998}
{Beuermann} K.,  {Baraffe} I.,  {Kolb} U.,   {Weichhold} M.,  1998, \aap, \href
  {https://ui.adsabs.harvard.edu/abs/1998A&A...339..518B} {339, 518}

\bibitem[\protect\citeauthoryear{{Brown}}{{Brown}}{2014}]{Brown2014}
{Brown} T.~M.,  2014, \mn@doi [\apj] {10.1088/0004-637X/789/2/101}, \href
  {https://ui.adsabs.harvard.edu/abs/2014ApJ...789..101B} {789, 101}

\bibitem[\protect\citeauthoryear{{Chaboyer}, {Demarque}  \&
  {Pinsonneault}}{{Chaboyer} et~al.}{1995}]{Chaboyer1995}
{Chaboyer} B.,  {Demarque} P.,   {Pinsonneault} M.~H.,  1995, \mn@doi [\apj]
  {10.1086/175408}, \href
  {https://ui.adsabs.harvard.edu/abs/1995ApJ...441..865C} {441, 865}

\bibitem[\protect\citeauthoryear{{Chen} \& {Podsiadlowski}}{{Chen} \&
  {Podsiadlowski}}{2016}]{Chen2016}
{Chen} W.-C.,  {Podsiadlowski} P.,  2016, \mn@doi [\apj]
  {10.3847/0004-637X/830/2/131}, \href
  {https://ui.adsabs.harvard.edu/abs/2016ApJ...830..131C} {830, 131}

\bibitem[\protect\citeauthoryear{{Chen}, {Wang}, {Deng}, {de Grijs}, {Yang}  \&
  {Tian}}{{Chen} et~al.}{2020}]{Chen2020}
{Chen} X.,  {Wang} S.,  {Deng} L.,  {de Grijs} R.,  {Yang} M.,   {Tian} H.,
  2020, \mn@doi [\apjs] {10.3847/1538-4365/ab9cae}, \href
  {https://ui.adsabs.harvard.edu/abs/2020ApJS..249...18C} {249, 18}

\bibitem[\protect\citeauthoryear{{Chen}, {Tauris}, {Han}  \& {Chen}}{{Chen}
  et~al.}{2021}]{Chen2021}
{Chen} H.-L.,  {Tauris} T.~M.,  {Han} Z.,   {Chen} X.,  2021, \mn@doi [\mnras]
  {10.1093/mnras/stab670}, \href
  {https://ui.adsabs.harvard.edu/abs/2021MNRAS.503.3540C} {503, 3540}

\bibitem[\protect\citeauthoryear{{Choi}, {Dotter}, {Conroy}, {Cantiello},
  {Paxton}  \& {Johnson}}{{Choi} et~al.}{2016}]{Choi_2016}
{Choi} J.,  {Dotter} A.,  {Conroy} C.,  {Cantiello} M.,  {Paxton} B.,
  {Johnson} B.~D.,  2016, \mn@doi [\apj] {10.3847/0004-637X/823/2/102}, \href
  {https://ui.adsabs.harvard.edu/abs/2016ApJ...823..102C} {823, 102}

\bibitem[\protect\citeauthoryear{{Claret}}{{Claret}}{2004}]{Claret2004}
{Claret} A.,  2004, \mn@doi [\aap] {10.1051/0004-6361:20041673}, \href
  {https://ui.adsabs.harvard.edu/abs/2004A&A...428.1001C} {428, 1001}

\bibitem[\protect\citeauthoryear{{Cukanovaite}, {Tremblay}, {Toonen},
  {Temmink}, {Manser}, {O'Brien}  \& {McCleery}}{{Cukanovaite}
  et~al.}{2022}]{Cukanovaite2022}
{Cukanovaite} E.,  {Tremblay} P.~E.,  {Toonen} S.,  {Temmink} K.~D.,  {Manser}
  C.~J.,  {O'Brien} M.~W.,   {McCleery} J.,  2022, arXiv e-prints, \href
  {https://ui.adsabs.harvard.edu/abs/2022arXiv220913919C} {p. arXiv:2209.13919}

\bibitem[\protect\citeauthoryear{{Delfosse}, {Forveille}, {Perrier}  \&
  {Mayor}}{{Delfosse} et~al.}{1998}]{Delfosse1998}
{Delfosse} X.,  {Forveille} T.,  {Perrier} C.,   {Mayor} M.,  1998, \aap, \href
  {https://ui.adsabs.harvard.edu/abs/1998A&A...331..581D} {331, 581}

\bibitem[\protect\citeauthoryear{{Derekas}, {Kiss}  \& {Bedding}}{{Derekas}
  et~al.}{2007}]{Derekas2007}
{Derekas} A.,  {Kiss} L.~L.,   {Bedding} T.~R.,  2007, \mn@doi [\apj]
  {10.1086/517994}, \href
  {https://ui.adsabs.harvard.edu/abs/2007ApJ...663..249D} {663, 249}

\bibitem[\protect\citeauthoryear{{Donati} \& {Landstreet}}{{Donati} \&
  {Landstreet}}{2009}]{Donati2009}
{Donati} J.~F.,  {Landstreet} J.~D.,  2009, \mn@doi [\araa]
  {10.1146/annurev-astro-082708-101833}, \href
  {https://ui.adsabs.harvard.edu/abs/2009ARA&A..47..333D} {47, 333}

\bibitem[\protect\citeauthoryear{{Drake} et~al.,}{{Drake}
  et~al.}{2014}]{Drake2014}
{Drake} A.~J.,  et~al., 2014, \mn@doi [\apjs] {10.1088/0067-0049/213/1/9},
  \href {https://ui.adsabs.harvard.edu/abs/2014ApJS..213....9D} {213, 9}

\bibitem[\protect\citeauthoryear{{Duch{\^e}ne} \& {Kraus}}{{Duch{\^e}ne} \&
  {Kraus}}{2013}]{Duchene2013}
{Duch{\^e}ne} G.,  {Kraus} A.,  2013, \mn@doi [\araa]
  {10.1146/annurev-astro-081710-102602}, \href
  {https://ui.adsabs.harvard.edu/abs/2013ARA&A..51..269D} {51, 269}

\bibitem[\protect\citeauthoryear{{Duquennoy} \& {Mayor}}{{Duquennoy} \&
  {Mayor}}{1991}]{Duquennoy1991}
{Duquennoy} A.,  {Mayor} M.,  1991, \aap, \href
  {https://ui.adsabs.harvard.edu/abs/1991A&A...248..485D} {248, 485}

\bibitem[\protect\citeauthoryear{{Eggleton}}{{Eggleton}}{1983}]{Eggleton_1983}
{Eggleton} P.~P.,  1983, \mn@doi [\apj] {10.1086/160960}, \href
  {https://ui.adsabs.harvard.edu/abs/1983ApJ...268..368E} {268, 368}

\bibitem[\protect\citeauthoryear{{El-Badry}, {Rix}  \& {Heintz}}{{El-Badry}
  et~al.}{2021a}]{El-Badry2021_gaia}
{El-Badry} K.,  {Rix} H.-W.,   {Heintz} T.~M.,  2021a, \mn@doi [\mnras]
  {10.1093/mnras/stab323}, \href
  {https://ui.adsabs.harvard.edu/abs/2021MNRAS.tmp..394E} {}

\bibitem[\protect\citeauthoryear{{El-Badry}, {Rix}, {Quataert}, {Kupfer}  \&
  {Shen}}{{El-Badry} et~al.}{2021b}]{El-Badry2021_cvs}
{El-Badry} K.,  {Rix} H.-W.,  {Quataert} E.,  {Kupfer} T.,   {Shen} K.~J.,
  2021b, \mn@doi [\mnras] {10.1093/mnras/stab2583}, \href
  {https://ui.adsabs.harvard.edu/abs/2021MNRAS.508.4106E} {508, 4106}

\bibitem[\protect\citeauthoryear{{Fabrycky} \& {Tremaine}}{{Fabrycky} \&
  {Tremaine}}{2007}]{Fabrycky2007}
{Fabrycky} D.,  {Tremaine} S.,  2007, \mn@doi [\apj] {10.1086/521702}, \href
  {https://ui.adsabs.harvard.edu/abs/2007ApJ...669.1298F} {669, 1298}

\bibitem[\protect\citeauthoryear{{Farinella}, {Luzny}, {Mantegazza}  \&
  {Paolicchi}}{{Farinella} et~al.}{1979}]{Farinella1979}
{Farinella} P.,  {Luzny} F.,  {Mantegazza} L.,   {Paolicchi} P.,  1979, \mn@doi
  [\apj] {10.1086/157581}, \href
  {https://ui.adsabs.harvard.edu/abs/1979ApJ...234..973F} {234, 973}

\bibitem[\protect\citeauthoryear{{Fischer} \& {Marcy}}{{Fischer} \&
  {Marcy}}{1992}]{Fischer1992}
{Fischer} D.~A.,  {Marcy} G.~W.,  1992, \mn@doi [\apj] {10.1086/171708}, \href
  {https://ui.adsabs.harvard.edu/abs/1992ApJ...396..178F} {396, 178}

\bibitem[\protect\citeauthoryear{{Gaia Collaboration} et~al.,}{{Gaia
  Collaboration} et~al.}{2021}]{GaiaCollaboration2021b}
{Gaia Collaboration} et~al., 2021, \mn@doi [\aap]
  {10.1051/0004-6361/202039498}, \href
  {https://ui.adsabs.harvard.edu/abs/2021A&A...649A...6G} {649, A6}

\bibitem[\protect\citeauthoryear{{Gaia Collaboration} et~al.,}{{Gaia
  Collaboration} et~al.}{2022}]{GaiaCollaboration2022}
{Gaia Collaboration} et~al., 2022, arXiv e-prints, \href
  {https://ui.adsabs.harvard.edu/abs/2022arXiv220800211G} {p. arXiv:2208.00211}

\bibitem[\protect\citeauthoryear{{Garraffo}, {Drake}  \& {Cohen}}{{Garraffo}
  et~al.}{2015}]{Garraffo2015}
{Garraffo} C.,  {Drake} J.~J.,   {Cohen} O.,  2015, \mn@doi [\apjl]
  {10.1088/2041-8205/807/1/L6}, \href
  {https://ui.adsabs.harvard.edu/abs/2015ApJ...807L...6G} {807, L6}

\bibitem[\protect\citeauthoryear{{Garraffo}, {Drake}  \& {Cohen}}{{Garraffo}
  et~al.}{2016}]{Garraffo2016}
{Garraffo} C.,  {Drake} J.~J.,   {Cohen} O.,  2016, \mn@doi [\aap]
  {10.1051/0004-6361/201628367}, \href
  {https://ui.adsabs.harvard.edu/abs/2016A&A...595A.110G} {595, A110}

\bibitem[\protect\citeauthoryear{{Garraffo} et~al.,}{{Garraffo}
  et~al.}{2018}]{Garraffo2018}
{Garraffo} C.,  et~al., 2018, \mn@doi [\apj] {10.3847/1538-4357/aace5d}, \href
  {https://ui.adsabs.harvard.edu/abs/2018ApJ...862...90G} {862, 90}

\bibitem[\protect\citeauthoryear{{Ginzburg} \& {Quataert}}{{Ginzburg} \&
  {Quataert}}{2021}]{Ginzburg2021}
{Ginzburg} S.,  {Quataert} E.,  2021, \mn@doi [\mnras]
  {10.1093/mnras/stab2170}, \href
  {https://ui.adsabs.harvard.edu/abs/2021MNRAS.507..475G} {507, 475}

\bibitem[\protect\citeauthoryear{{Giuricin}, {Mardirossian}  \&
  {Mezzetti}}{{Giuricin} et~al.}{1983}]{Giuricin1983}
{Giuricin} G.,  {Mardirossian} F.,   {Mezzetti} M.,  1983, \aap, \href
  {https://ui.adsabs.harvard.edu/abs/1983A&A...119..218G} {119, 218}

\bibitem[\protect\citeauthoryear{{Giuricin}, {Mardirossian}  \&
  {Mezzetti}}{{Giuricin} et~al.}{1984}]{Giuricin1984}
{Giuricin} G.,  {Mardirossian} F.,   {Mezzetti} M.,  1984, \mn@doi [\apjs]
  {10.1086/190938}, \href
  {https://ui.adsabs.harvard.edu/abs/1984ApJS...54..421G} {54, 421}

\bibitem[\protect\citeauthoryear{{Graham}, {Drake}, {Djorgovski}, {Mahabal}  \&
  {Donalek}}{{Graham} et~al.}{2013}]{Graham2013}
{Graham} M.~J.,  {Drake} A.~J.,  {Djorgovski} S.~G.,  {Mahabal} A.~A.,
  {Donalek} C.,  2013, \mn@doi [\mnras] {10.1093/mnras/stt1206}, \href
  {https://ui.adsabs.harvard.edu/abs/2013MNRAS.434.2629G} {434, 2629}

\bibitem[\protect\citeauthoryear{{Green}, {Schlafly}, {Zucker}, {Speagle}  \&
  {Finkbeiner}}{{Green} et~al.}{2019}]{Green2019}
{Green} G.~M.,  {Schlafly} E.,  {Zucker} C.,  {Speagle} J.~S.,   {Finkbeiner}
  D.,  2019, \mn@doi [\apj] {10.3847/1538-4357/ab5362}, \href
  {https://ui.adsabs.harvard.edu/abs/2019ApJ...887...93G} {887, 93}

\bibitem[\protect\citeauthoryear{{Guidry} et~al.,}{{Guidry}
  et~al.}{2021}]{Guidry2020}
{Guidry} J.~A.,  et~al., 2021, \mn@doi [\apj] {10.3847/1538-4357/abee68}, \href
  {https://ui.adsabs.harvard.edu/abs/2021ApJ...912..125G} {912, 125}

\bibitem[\protect\citeauthoryear{{Howell}, {Nelson}  \& {Rappaport}}{{Howell}
  et~al.}{2001}]{Howell2001}
{Howell} S.~B.,  {Nelson} L.~A.,   {Rappaport} S.,  2001, \mn@doi [\apj]
  {10.1086/319776}, \href
  {https://ui.adsabs.harvard.edu/abs/2001ApJ...550..897H} {550, 897}

\bibitem[\protect\citeauthoryear{{Hwang}}{{Hwang}}{2022}]{Hwang2022}
{Hwang} H.-C.,  2022, arXiv e-prints, \href
  {https://ui.adsabs.harvard.edu/abs/2022arXiv220802257H} {p. arXiv:2208.02257}

\bibitem[\protect\citeauthoryear{{Hwang} \& {Zakamska}}{{Hwang} \&
  {Zakamska}}{2020}]{Hwang2020}
{Hwang} H.-C.,  {Zakamska} N.~L.,  2020, \mn@doi [\mnras]
  {10.1093/mnras/staa400}, \href
  {https://ui.adsabs.harvard.edu/abs/2020MNRAS.493.2271H} {493, 2271}

\bibitem[\protect\citeauthoryear{{Istrate}, {Tauris}  \& {Langer}}{{Istrate}
  et~al.}{2014}]{Istrate2014b}
{Istrate} A.~G.,  {Tauris} T.~M.,   {Langer} N.,  2014, \mn@doi [\aap]
  {10.1051/0004-6361/201424680}, \href
  {https://ui.adsabs.harvard.edu/abs/2014A&A...571A..45I} {571, A45}

\bibitem[\protect\citeauthoryear{{Ivanova} \& {Taam}}{{Ivanova} \&
  {Taam}}{2003}]{Ivanova2003}
{Ivanova} N.,  {Taam} R.~E.,  2003, \mn@doi [\apj] {10.1086/379192}, \href
  {https://ui.adsabs.harvard.edu/abs/2003ApJ...599..516I} {599, 516}

\bibitem[\protect\citeauthoryear{{Jao}, {Henry}, {Gies}  \& {Hambly}}{{Jao}
  et~al.}{2018}]{Jao2018}
{Jao} W.-C.,  {Henry} T.~J.,  {Gies} D.~R.,   {Hambly} N.~C.,  2018, \mn@doi
  [\apjl] {10.3847/2041-8213/aacdf6}, \href
  {https://ui.adsabs.harvard.edu/abs/2018ApJ...861L..11J} {861, L11}

\bibitem[\protect\citeauthoryear{{Jayasinghe} et~al.,}{{Jayasinghe}
  et~al.}{2018}]{Jayasinghe2018}
{Jayasinghe} T.,  et~al., 2018, \mn@doi [\mnras] {10.1093/mnras/sty838}, \href
  {https://ui.adsabs.harvard.edu/abs/2018MNRAS.477.3145J} {477, 3145}

\bibitem[\protect\citeauthoryear{{Jiang}, {Han}, {Ge}, {Yang}  \& {Li}}{{Jiang}
  et~al.}{2012}]{Jiang2012}
{Jiang} D.,  {Han} Z.,  {Ge} H.,  {Yang} L.,   {Li} L.,  2012, \mn@doi [\mnras]
  {10.1111/j.1365-2966.2011.20323.x}, \href
  {https://ui.adsabs.harvard.edu/abs/2012MNRAS.421.2769J} {421, 2769}

\bibitem[\protect\citeauthoryear{{Johnstone}, {Bartel}  \&
  {G{\"u}del}}{{Johnstone} et~al.}{2021}]{Johnstone2021}
{Johnstone} C.~P.,  {Bartel} M.,   {G{\"u}del} M.,  2021, \mn@doi [\aap]
  {10.1051/0004-6361/202038407}, \href
  {https://ui.adsabs.harvard.edu/abs/2021A&A...649A..96J} {649, A96}

\bibitem[\protect\citeauthoryear{{Kalomeni}, {Nelson}, {Rappaport}, {Molnar},
  {Quintin}  \& {Yakut}}{{Kalomeni} et~al.}{2016}]{Kalomeni2016}
{Kalomeni} B.,  {Nelson} L.,  {Rappaport} S.,  {Molnar} M.,  {Quintin} J.,
  {Yakut} K.,  2016, \mn@doi [\apj] {10.3847/1538-4357/833/1/83}, \href
  {https://ui.adsabs.harvard.edu/abs/2016ApJ...833...83K} {833, 83}

\bibitem[\protect\citeauthoryear{{Kawaler}}{{Kawaler}}{1988}]{Kawaler1988}
{Kawaler} S.~D.,  1988, \mn@doi [\apj] {10.1086/166740}, \href
  {https://ui.adsabs.harvard.edu/abs/1988ApJ...333..236K} {333, 236}

\bibitem[\protect\citeauthoryear{{Kim} \& {Bailer-Jones}}{{Kim} \&
  {Bailer-Jones}}{2016}]{Kim2016}
{Kim} D.-W.,  {Bailer-Jones} C. A.~L.,  2016, \mn@doi [\aap]
  {10.1051/0004-6361/201527188}, \href
  {https://ui.adsabs.harvard.edu/abs/2016A&A...587A..18K} {587, A18}

\bibitem[\protect\citeauthoryear{{King} \& {Kolb}}{{King} \&
  {Kolb}}{1995}]{King1995}
{King} A.~R.,  {Kolb} U.,  1995, \mn@doi [\apj] {10.1086/175176}, \href
  {https://ui.adsabs.harvard.edu/abs/1995ApJ...439..330K} {439, 330}

\bibitem[\protect\citeauthoryear{{Knigge}}{{Knigge}}{2006}]{Knigge_2006}
{Knigge} C.,  2006, \mn@doi [\mnras] {10.1111/j.1365-2966.2006.11096.x}, \href
  {https://ui.adsabs.harvard.edu/abs/2006MNRAS.373..484K} {373, 484}

\bibitem[\protect\citeauthoryear{{Knigge}, {Baraffe}  \& {Patterson}}{{Knigge}
  et~al.}{2011}]{Knigge_2011}
{Knigge} C.,  {Baraffe} I.,   {Patterson} J.,  2011, \mn@doi [\apjs]
  {10.1088/0067-0049/194/2/28}, \href
  {https://ui.adsabs.harvard.edu/abs/2011ApJS..194...28K} {194, 28}

\bibitem[\protect\citeauthoryear{{Koen}}{{Koen}}{2022}]{Koen2022}
{Koen} C.,  2022, \mn@doi [\mnras] {10.1093/mnras/stab3431}, \href
  {https://ui.adsabs.harvard.edu/abs/2022MNRAS.510.1857K} {510, 1857}

\bibitem[\protect\citeauthoryear{{Kolb}}{{Kolb}}{1993}]{Kolb1993}
{Kolb} U.,  1993, \aap, \href
  {https://ui.adsabs.harvard.edu/abs/1993A&A...271..149K} {271, 149}

\bibitem[\protect\citeauthoryear{{Kounkel} et~al.,}{{Kounkel}
  et~al.}{2019}]{Kounkel2019}
{Kounkel} M.,  et~al., 2019, \mn@doi [\aj] {10.3847/1538-3881/ab13b1}, \href
  {https://ui.adsabs.harvard.edu/abs/2019AJ....157..196K} {157, 196}

\bibitem[\protect\citeauthoryear{{Kov{\'a}cs}, {Zucker}  \&
  {Mazeh}}{{Kov{\'a}cs} et~al.}{2002}]{Kovacs2002}
{Kov{\'a}cs} G.,  {Zucker} S.,   {Mazeh} T.,  2002, \mn@doi [\aap]
  {10.1051/0004-6361:20020802}, \href
  {https://ui.adsabs.harvard.edu/abs/2002A&A...391..369K} {391, 369}

\bibitem[\protect\citeauthoryear{{Kraft}}{{Kraft}}{1967}]{Kraft1967}
{Kraft} R.~P.,  1967, \mn@doi [\apj] {10.1086/149359}, \href
  {https://ui.adsabs.harvard.edu/abs/1967ApJ...150..551K} {150, 551}

\bibitem[\protect\citeauthoryear{{Kurtenkov}}{{Kurtenkov}}{2022}]{Kurtenkov2022}
{Kurtenkov} A.,  2022, arXiv e-prints, \href
  {https://ui.adsabs.harvard.edu/abs/2022arXiv220110637K} {p. arXiv:2201.10637}

\bibitem[\protect\citeauthoryear{{Li}, {Han}  \& {Zhang}}{{Li}
  et~al.}{2004}]{Li2004}
{Li} L.,  {Han} Z.,   {Zhang} F.,  2004, \mn@doi [\mnras]
  {10.1111/j.1365-2966.2004.08457.x}, \href
  {https://ui.adsabs.harvard.edu/abs/2004MNRAS.355.1383L} {355, 1383}

\bibitem[\protect\citeauthoryear{{Lucy}}{{Lucy}}{1968}]{Lucy1968}
{Lucy} L.~B.,  1968, \mn@doi [\apj] {10.1086/149510}, \href
  {https://ui.adsabs.harvard.edu/abs/1968ApJ...151.1123L} {151, 1123}

\bibitem[\protect\citeauthoryear{{Lurie} et~al.,}{{Lurie}
  et~al.}{2017}]{Lurie2017}
{Lurie} J.~C.,  et~al., 2017, \mn@doi [\aj] {10.3847/1538-3881/aa974d}, \href
  {https://ui.adsabs.harvard.edu/abs/2017AJ....154..250L} {154, 250}

\bibitem[\protect\citeauthoryear{{Ma} \& {Li}}{{Ma} \& {Li}}{2009}]{Ma2009}
{Ma} B.,  {Li} X.-D.,  2009, \mn@doi [\apj] {10.1088/0004-637X/698/2/1907},
  \href {https://ui.adsabs.harvard.edu/abs/2009ApJ...698.1907M} {698, 1907}

\bibitem[\protect\citeauthoryear{{Maceroni} \& {Montalb{\'a}n}}{{Maceroni} \&
  {Montalb{\'a}n}}{2004}]{Maceroni2004}
{Maceroni} C.,  {Montalb{\'a}n} J.,  2004, \mn@doi [\aap]
  {10.1051/0004-6361:20040549}, \href
  {https://ui.adsabs.harvard.edu/abs/2004A&A...426..577M} {426, 577}

\bibitem[\protect\citeauthoryear{{Maceroni} \& {Rucinski}}{{Maceroni} \&
  {Rucinski}}{1999}]{Maceroni1999}
{Maceroni} C.,  {Rucinski} S.~M.,  1999, \mn@doi [\aj] {10.1086/301060}, \href
  {https://ui.adsabs.harvard.edu/abs/1999AJ....118.1819M} {118, 1819}

\bibitem[\protect\citeauthoryear{{Maceroni} \& {van't Veer}}{{Maceroni} \&
  {van't Veer}}{1991}]{Maceroni1991}
{Maceroni} C.,  {van't Veer} F.,  1991, \aap, \href
  {https://ui.adsabs.harvard.edu/abs/1991A&A...246...91M} {246, 91}

\bibitem[\protect\citeauthoryear{{Matt}, {Brun}, {Baraffe}, {Bouvier}  \&
  {Chabrier}}{{Matt} et~al.}{2015}]{Matt2015}
{Matt} S.~P.,  {Brun} A.~S.,  {Baraffe} I.,  {Bouvier} J.,   {Chabrier} G.,
  2015, \mn@doi [\apjl] {10.1088/2041-8205/799/2/L23}, \href
  {https://ui.adsabs.harvard.edu/abs/2015ApJ...799L..23M} {799, L23}

\bibitem[\protect\citeauthoryear{{Maxted}}{{Maxted}}{2016}]{Maxted2016}
{Maxted} P.~F.~L.,  2016, \mn@doi [\aap] {10.1051/0004-6361/201628579}, \href
  {https://ui.adsabs.harvard.edu/abs/2016A&A...591A.111M} {591, A111}

\bibitem[\protect\citeauthoryear{{Mazeh} \& {Shaham}}{{Mazeh} \&
  {Shaham}}{1979}]{Mazeh1979}
{Mazeh} T.,  {Shaham} J.,  1979, \aap, \href
  {https://ui.adsabs.harvard.edu/abs/1979A&A....77..145M} {77, 145}

\bibitem[\protect\citeauthoryear{{McQuillan}, {Aigrain}  \&
  {Mazeh}}{{McQuillan} et~al.}{2013}]{McQuillan2013}
{McQuillan} A.,  {Aigrain} S.,   {Mazeh} T.,  2013, \mn@doi [\mnras]
  {10.1093/mnras/stt536}, \href
  {https://ui.adsabs.harvard.edu/abs/2013MNRAS.432.1203M} {432, 1203}

\bibitem[\protect\citeauthoryear{{Medina}, {Winters}, {Irwin}  \&
  {Charbonneau}}{{Medina} et~al.}{2020}]{Medina2020}
{Medina} A.~A.,  {Winters} J.~G.,  {Irwin} J.~M.,   {Charbonneau} D.,  2020,
  \mn@doi [\apj] {10.3847/1538-4357/abc686}, \href
  {https://ui.adsabs.harvard.edu/abs/2020ApJ...905..107M} {905, 107}

\bibitem[\protect\citeauthoryear{{Mestel}}{{Mestel}}{1968}]{Mestel1968}
{Mestel} L.,  1968, \mn@doi [\mnras] {10.1093/mnras/138.3.359}, \href
  {https://ui.adsabs.harvard.edu/abs/1968MNRAS.138..359M} {138, 359}

\bibitem[\protect\citeauthoryear{{Mestel} \& {Spruit}}{{Mestel} \&
  {Spruit}}{1987}]{Mestel1987}
{Mestel} L.,  {Spruit} H.~C.,  1987, \mn@doi [\mnras] {10.1093/mnras/226.1.57},
  \href {https://ui.adsabs.harvard.edu/abs/1987MNRAS.226...57M} {226, 57}

\bibitem[\protect\citeauthoryear{{Moe} \& {Di Stefano}}{{Moe} \& {Di
  Stefano}}{2017}]{Moe2017}
{Moe} M.,  {Di Stefano} R.,  2017, \mn@doi [\apjs] {10.3847/1538-4365/aa6fb6},
  \href {https://ui.adsabs.harvard.edu/abs/2017ApJS..230...15M} {230, 15}

\bibitem[\protect\citeauthoryear{{Nefs} et~al.,}{{Nefs}
  et~al.}{2012}]{Nefs2012}
{Nefs} S.~V.,  et~al., 2012, \mn@doi [\mnras]
  {10.1111/j.1365-2966.2012.21338.x}, \href
  {https://ui.adsabs.harvard.edu/abs/2012MNRAS.425..950N} {425, 950}

\bibitem[\protect\citeauthoryear{{Nelemans}, {Siess}, {Repetto}, {Toonen}  \&
  {Phinney}}{{Nelemans} et~al.}{2016}]{Nelemans2016}
{Nelemans} G.,  {Siess} L.,  {Repetto} S.,  {Toonen} S.,   {Phinney} E.~S.,
  2016, \mn@doi [\apj] {10.3847/0004-637X/817/1/69}, \href
  {https://ui.adsabs.harvard.edu/abs/2016ApJ...817...69N} {817, 69}

\bibitem[\protect\citeauthoryear{{Newton}, {Irwin}, {Charbonneau},
  {Berta-Thompson}, {Dittmann}  \& {West}}{{Newton} et~al.}{2016}]{Newton2016}
{Newton} E.~R.,  {Irwin} J.,  {Charbonneau} D.,  {Berta-Thompson} Z.~K.,
  {Dittmann} J.~A.,   {West} A.~A.,  2016, \mn@doi [\apj]
  {10.3847/0004-637X/821/2/93}, \href
  {https://ui.adsabs.harvard.edu/abs/2016ApJ...821...93N} {821, 93}

\bibitem[\protect\citeauthoryear{{Newton}, {Irwin}, {Charbonneau}, {Berlind},
  {Calkins}  \& {Mink}}{{Newton} et~al.}{2017}]{Newton2017}
{Newton} E.~R.,  {Irwin} J.,  {Charbonneau} D.,  {Berlind} P.,  {Calkins}
  M.~L.,   {Mink} J.,  2017, \mn@doi [\apj] {10.3847/1538-4357/834/1/85}, \href
  {https://ui.adsabs.harvard.edu/abs/2017ApJ...834...85N} {834, 85}

\bibitem[\protect\citeauthoryear{{Norton} et~al.,}{{Norton}
  et~al.}{2011}]{Norton2011}
{Norton} A.~J.,  et~al., 2011, \mn@doi [\aap] {10.1051/0004-6361/201116448},
  \href {https://ui.adsabs.harvard.edu/abs/2011A&A...528A..90N} {528, A90}

\bibitem[\protect\citeauthoryear{{Paczy{\'n}ski}, {Szczygie{\l}}, {Pilecki}  \&
  {Pojma{\'n}ski}}{{Paczy{\'n}ski} et~al.}{2006}]{Paczynski2006}
{Paczy{\'n}ski} B.,  {Szczygie{\l}} D.~M.,  {Pilecki} B.,   {Pojma{\'n}ski} G.,
   2006, \mn@doi [\mnras] {10.1111/j.1365-2966.2006.10223.x}, \href
  {https://ui.adsabs.harvard.edu/abs/2006MNRAS.368.1311P} {368, 1311}

\bibitem[\protect\citeauthoryear{{Pala} et~al.,}{{Pala}
  et~al.}{2017}]{Pala2017}
{Pala} A.~F.,  et~al., 2017, \mn@doi [\mnras] {10.1093/mnras/stw3293}, \href
  {https://ui.adsabs.harvard.edu/abs/2017MNRAS.466.2855P} {466, 2855}

\bibitem[\protect\citeauthoryear{{Pala} et~al.,}{{Pala}
  et~al.}{2020}]{Pala2020}
{Pala} A.~F.,  et~al., 2020, \mn@doi [\mnras] {10.1093/mnras/staa764}, \href
  {https://ui.adsabs.harvard.edu/abs/2020MNRAS.494.3799P} {494, 3799}

\bibitem[\protect\citeauthoryear{{Pala} et~al.,}{{Pala}
  et~al.}{2022}]{Pala2022}
{Pala} A.~F.,  et~al., 2022, \mn@doi [\mnras] {10.1093/mnras/stab3449}, \href
  {https://ui.adsabs.harvard.edu/abs/2022MNRAS.510.6110P} {510, 6110}

\bibitem[\protect\citeauthoryear{{Patterson}}{{Patterson}}{1984}]{Patterson_1984}
{Patterson} J.,  1984, \mn@doi [\apjs] {10.1086/190940}, \href
  {https://ui.adsabs.harvard.edu/abs/1984ApJS...54..443P} {54, 443}

\bibitem[\protect\citeauthoryear{{Paxton} et~al.,}{{Paxton}
  et~al.}{2015}]{Paxton_2015}
{Paxton} B.,  et~al., 2015, \mn@doi [\apjs] {10.1088/0067-0049/220/1/15}, \href
  {https://ui.adsabs.harvard.edu/abs/2015ApJS..220...15P} {220, 15}

\bibitem[\protect\citeauthoryear{{Podsiadlowski}, {Rappaport}  \&
  {Pfahl}}{{Podsiadlowski} et~al.}{2002}]{Podsiadlowski2002}
{Podsiadlowski} P.,  {Rappaport} S.,   {Pfahl} E.~D.,  2002, \mn@doi [\apj]
  {10.1086/324686}, \href
  {https://ui.adsabs.harvard.edu/abs/2002ApJ...565.1107P} {565, 1107}

\bibitem[\protect\citeauthoryear{{Podsiadlowski}, {Han}  \&
  {Rappaport}}{{Podsiadlowski} et~al.}{2003}]{Podsiadlowski_2003}
{Podsiadlowski} P.,  {Han} Z.,   {Rappaport} S.,  2003, \mn@doi [\mnras]
  {10.1046/j.1365-8711.2003.06380.x}, \href
  {https://ui.adsabs.harvard.edu/abs/2003MNRAS.340.1214P} {340, 1214}

\bibitem[\protect\citeauthoryear{{Pr{\v{s}}a} et~al.,}{{Pr{\v{s}}a}
  et~al.}{2011}]{Prsa2011}
{Pr{\v{s}}a} A.,  et~al., 2011, \mn@doi [\aj] {10.1088/0004-6256/141/3/83},
  \href {https://ui.adsabs.harvard.edu/abs/2011AJ....141...83P} {141, 83}

\bibitem[\protect\citeauthoryear{{Raghavan} et~al.,}{{Raghavan}
  et~al.}{2010}]{Raghavan2010}
{Raghavan} D.,  et~al., 2010, \mn@doi [\apjs] {10.1088/0067-0049/190/1/1},
  \href {https://ui.adsabs.harvard.edu/abs/2010ApJS..190....1R} {190, 1}

\bibitem[\protect\citeauthoryear{{Rappaport}, {Verbunt}  \& {Joss}}{{Rappaport}
  et~al.}{1983}]{Rappaport1983}
{Rappaport} S.,  {Verbunt} F.,   {Joss} P.~C.,  1983, \mn@doi [\apj]
  {10.1086/161569}, \href
  {https://ui.adsabs.harvard.edu/abs/1983ApJ...275..713R} {275, 713}

\bibitem[\protect\citeauthoryear{{Reimers}}{{Reimers}}{1975}]{Reimers1975}
{Reimers} D.,  1975, Memoires of the Societe Royale des Sciences de Liege,
  \href {https://ui.adsabs.harvard.edu/abs/1975MSRSL...8..369R} {8, 369}

\bibitem[\protect\citeauthoryear{{Reiners}, {Basri}  \& {Browning}}{{Reiners}
  et~al.}{2009}]{Reiners2009}
{Reiners} A.,  {Basri} G.,   {Browning} M.,  2009, \mn@doi [\apj]
  {10.1088/0004-637X/692/1/538}, \href
  {https://ui.adsabs.harvard.edu/abs/2009ApJ...692..538R} {692, 538}

\bibitem[\protect\citeauthoryear{{Reiners} et~al.,}{{Reiners}
  et~al.}{2022}]{Reiners2022}
{Reiners} A.,  et~al., 2022, \mn@doi [\aap] {10.1051/0004-6361/202243251},
  \href {https://ui.adsabs.harvard.edu/abs/2022A&A...662A..41R} {662, A41}

\bibitem[\protect\citeauthoryear{{Rowan} et~al.,}{{Rowan}
  et~al.}{2022}]{Rowan2022}
{Rowan} D.~M.,  et~al., 2022, arXiv e-prints, \href
  {https://ui.adsabs.harvard.edu/abs/2022arXiv220505687R} {p. arXiv:2205.05687}

\bibitem[\protect\citeauthoryear{{Rucinski}}{{Rucinski}}{1992}]{Rucinski1992}
{Rucinski} S.~M.,  1992, \mn@doi [\aj] {10.1086/116118}, \href
  {https://ui.adsabs.harvard.edu/abs/1992AJ....103..960R} {103, 960}

\bibitem[\protect\citeauthoryear{{Rucinski}}{{Rucinski}}{1998}]{Rucinski1998}
{Rucinski} S.~M.,  1998, \mn@doi [\aj] {10.1086/300644}, \href
  {https://ui.adsabs.harvard.edu/abs/1998AJ....116.2998R} {116, 2998}

\bibitem[\protect\citeauthoryear{{Rucinski}}{{Rucinski}}{2007}]{Rucinski2007}
{Rucinski} S.~M.,  2007, \mn@doi [\mnras] {10.1111/j.1365-2966.2007.12377.x},
  \href {https://ui.adsabs.harvard.edu/abs/2007MNRAS.382..393R} {382, 393}

\bibitem[\protect\citeauthoryear{{Rybizki}, {Green}, {Rix}, {Demleitner},
  {Zari}, {Udalski}, {Smart}  \& {Gould}}{{Rybizki} et~al.}{2021}]{Rybizki2021}
{Rybizki} J.,  {Green} G.,  {Rix} H.-W.,  {Demleitner} M.,  {Zari} E.,
  {Udalski} A.,  {Smart} R.~L.,   {Gould} A.,  2021, arXiv e-prints, \href
  {https://ui.adsabs.harvard.edu/abs/2021arXiv210111641R} {p. arXiv:2101.11641}

\bibitem[\protect\citeauthoryear{{Schatzman}}{{Schatzman}}{1962}]{Schatzman1962}
{Schatzman} E.,  1962, Annales d'Astrophysique, \href
  {https://ui.adsabs.harvard.edu/abs/1962AnAp...25...18S} {25, 18}

\bibitem[\protect\citeauthoryear{{Schreiber} et~al.,}{{Schreiber}
  et~al.}{2010}]{Schreiber2010}
{Schreiber} M.~R.,  et~al., 2010, \mn@doi [\aap] {10.1051/0004-6361/201013990},
  \href {https://ui.adsabs.harvard.edu/abs/2010A&A...513L...7S} {513, L7}

\bibitem[\protect\citeauthoryear{{Schreiber}, {Zorotovic}  \&
  {Wijnen}}{{Schreiber} et~al.}{2016}]{Schreiber2016}
{Schreiber} M.~R.,  {Zorotovic} M.,   {Wijnen} T.~P.~G.,  2016, \mn@doi
  [\mnras] {10.1093/mnrasl/slv144}, \href
  {https://ui.adsabs.harvard.edu/abs/2016MNRAS.455L..16S} {455, L16}

\bibitem[\protect\citeauthoryear{{Schreiber}, {Belloni}, {G{\"a}nsicke},
  {Parsons}  \& {Zorotovic}}{{Schreiber} et~al.}{2021}]{Schreiber2021}
{Schreiber} M.~R.,  {Belloni} D.,  {G{\"a}nsicke} B.~T.,  {Parsons} S.~G.,
  {Zorotovic} M.,  2021, \mn@doi [Nature Astronomy]
  {10.1038/s41550-021-01346-8}, \href
  {https://ui.adsabs.harvard.edu/abs/2021NatAs...5..648S} {5, 648}

\bibitem[\protect\citeauthoryear{{Seabroke} \& {Gilmore}}{{Seabroke} \&
  {Gilmore}}{2007}]{Seabroke2007}
{Seabroke} G.~M.,  {Gilmore} G.,  2007, \mn@doi [\mnras]
  {10.1111/j.1365-2966.2007.12210.x}, \href
  {https://ui.adsabs.harvard.edu/abs/2007MNRAS.380.1348S} {380, 1348}

\bibitem[\protect\citeauthoryear{{Sills}, {Pinsonneault}  \&
  {Terndrup}}{{Sills} et~al.}{2000}]{Sills2000}
{Sills} A.,  {Pinsonneault} M.~H.,   {Terndrup} D.~M.,  2000, \mn@doi [\apj]
  {10.1086/308739}, \href
  {https://ui.adsabs.harvard.edu/abs/2000ApJ...534..335S} {534, 335}

\bibitem[\protect\citeauthoryear{{Simonian}, {Pinsonneault}  \&
  {Terndrup}}{{Simonian} et~al.}{2019}]{Simonian2019}
{Simonian} G. V.~A.,  {Pinsonneault} M.~H.,   {Terndrup} D.~M.,  2019, \mn@doi
  [\apj] {10.3847/1538-4357/aaf97c}, \href
  {https://ui.adsabs.harvard.edu/abs/2019ApJ...871..174S} {871, 174}

\bibitem[\protect\citeauthoryear{{Skumanich}}{{Skumanich}}{1972}]{Skumanich1972}
{Skumanich} A.,  1972, \mn@doi [\apj] {10.1086/151310}, \href
  {https://ui.adsabs.harvard.edu/abs/1972ApJ...171..565S} {171, 565}

\bibitem[\protect\citeauthoryear{{Smith}}{{Smith}}{1979}]{Smith1979}
{Smith} M.~A.,  1979, \mn@doi [\pasp] {10.1086/130579}, \href
  {https://ui.adsabs.harvard.edu/abs/1979PASP...91..737S} {91, 737}

\bibitem[\protect\citeauthoryear{{Soethe} \& {Kepler}}{{Soethe} \&
  {Kepler}}{2021}]{Soethe2021}
{Soethe} L.~T.~T.,  {Kepler} S.~O.,  2021, \mn@doi [\mnras]
  {10.1093/mnras/stab1916}, \href
  {https://ui.adsabs.harvard.edu/abs/2021MNRAS.506.3266S} {506, 3266}

\bibitem[\protect\citeauthoryear{{Soszy{\'n}ski} et~al.,}{{Soszy{\'n}ski}
  et~al.}{2015}]{Soszynski2015}
{Soszy{\'n}ski} I.,  et~al., 2015, \actaa, \href
  {https://ui.adsabs.harvard.edu/abs/2015AcA....65...39S} {65, 39}

\bibitem[\protect\citeauthoryear{{Soszy{\'n}ski} et~al.,}{{Soszy{\'n}ski}
  et~al.}{2016}]{Soszynski2016}
{Soszy{\'n}ski} I.,  et~al., 2016, \actaa, \href
  {https://ui.adsabs.harvard.edu/abs/2016AcA....66..405S} {66, 405}

\bibitem[\protect\citeauthoryear{{Spruit} \& {Ritter}}{{Spruit} \&
  {Ritter}}{1983}]{Spruit1983}
{Spruit} H.~C.,  {Ritter} H.,  1983, \aap, \href
  {https://ui.adsabs.harvard.edu/abs/1983A&A...124..267S} {124, 267}

\bibitem[\protect\citeauthoryear{{Spruit} \& {Taam}}{{Spruit} \&
  {Taam}}{2001}]{Spruit2001}
{Spruit} H.~C.,  {Taam} R.~E.,  2001, \mn@doi [\apj] {10.1086/319030}, \href
  {https://ui.adsabs.harvard.edu/abs/2001ApJ...548..900S} {548, 900}

\bibitem[\protect\citeauthoryear{{Stauffer} \& {Hartmann}}{{Stauffer} \&
  {Hartmann}}{1987}]{Stauffer1987}
{Stauffer} J.~R.,  {Hartmann} L.~W.,  1987, \mn@doi [\apj] {10.1086/165371},
  \href {https://ui.adsabs.harvard.edu/abs/1987ApJ...318..337S} {318, 337}

\bibitem[\protect\citeauthoryear{{Stauffer}, {Caillault}, {Gagne}, {Prosser}
  \& {Hartmann}}{{Stauffer} et~al.}{1994}]{Stauffer1994}
{Stauffer} J.~R.,  {Caillault} J.~P.,  {Gagne} M.,  {Prosser} C.~F.,
  {Hartmann} L.~W.,  1994, \mn@doi [\apjs] {10.1086/191951}, \href
  {https://ui.adsabs.harvard.edu/abs/1994ApJS...91..625S} {91, 625}

\bibitem[\protect\citeauthoryear{{Stauffer}, {Hartmann}, {Prosser}, {Randich},
  {Balachandran}, {Patten}, {Simon}  \& {Giampapa}}{{Stauffer}
  et~al.}{1997}]{Stauffer1997}
{Stauffer} J.~R.,  {Hartmann} L.~W.,  {Prosser} C.~F.,  {Randich} S.,
  {Balachandran} S.,  {Patten} B.~M.,  {Simon} T.,   {Giampapa} M.,  1997,
  \mn@doi [\apj] {10.1086/303930}, \href
  {https://ui.adsabs.harvard.edu/abs/1997ApJ...479..776S} {479, 776}

\bibitem[\protect\citeauthoryear{{Stepien}}{{Stepien}}{2006}]{Stepien2006}
{Stepien} K.,  2006, \actaa, \href
  {https://ui.adsabs.harvard.edu/abs/2006AcA....56..347S} {56, 347}

\bibitem[\protect\citeauthoryear{{Taam} \& {Spruit}}{{Taam} \&
  {Spruit}}{1989}]{Taam1989}
{Taam} R.~E.,  {Spruit} H.~C.,  1989, \mn@doi [\apj] {10.1086/167966}, \href
  {https://ui.adsabs.harvard.edu/abs/1989ApJ...345..972T} {345, 972}

\bibitem[\protect\citeauthoryear{{Tauris} \& {Savonije}}{{Tauris} \&
  {Savonije}}{1999}]{Tauris1999}
{Tauris} T.~M.,  {Savonije} G.~J.,  1999, \aap, \href
  {https://ui.adsabs.harvard.edu/abs/1999A&A...350..928T} {350, 928}

\bibitem[\protect\citeauthoryear{{Tokovinin}}{{Tokovinin}}{2014}]{Tokovinin2014}
{Tokovinin} A.,  2014, \mn@doi [\aj] {10.1088/0004-6256/147/4/87}, \href
  {https://ui.adsabs.harvard.edu/abs/2014AJ....147...87T} {147, 87}

\bibitem[\protect\citeauthoryear{{Tokovinin}, {Thomas}, {Sterzik}  \&
  {Udry}}{{Tokovinin} et~al.}{2006}]{Tokovinin2006}
{Tokovinin} A.,  {Thomas} S.,  {Sterzik} M.,   {Udry} S.,  2006, \mn@doi [\aap]
  {10.1051/0004-6361:20054427}, \href
  {https://ui.adsabs.harvard.edu/abs/2006A&A...450..681T} {450, 681}

\bibitem[\protect\citeauthoryear{{Townsley} \& {Bildsten}}{{Townsley} \&
  {Bildsten}}{2003}]{Townsley2003}
{Townsley} D.~M.,  {Bildsten} L.,  2003, \mn@doi [\apjl] {10.1086/379535},
  \href {https://ui.adsabs.harvard.edu/abs/2003ApJ...596L.227T} {596, L227}

\bibitem[\protect\citeauthoryear{{Townsley} \& {G{\"a}nsicke}}{{Townsley} \&
  {G{\"a}nsicke}}{2009}]{Townsley2009}
{Townsley} D.~M.,  {G{\"a}nsicke} B.~T.,  2009, \mn@doi [\apj]
  {10.1088/0004-637X/693/1/1007}, \href
  {https://ui.adsabs.harvard.edu/abs/2009ApJ...693.1007T} {693, 1007}

\bibitem[\protect\citeauthoryear{{Van} \& {Ivanova}}{{Van} \&
  {Ivanova}}{2019}]{Van2019}
{Van} K.~X.,  {Ivanova} N.,  2019, \mn@doi [\apjl] {10.3847/2041-8213/ab571c},
  \href {https://ui.adsabs.harvard.edu/abs/2019ApJ...886L..31V} {886, L31}

\bibitem[\protect\citeauthoryear{{Van}, {Ivanova}  \& {Heinke}}{{Van}
  et~al.}{2019}]{Van2019b}
{Van} K.~X.,  {Ivanova} N.,   {Heinke} C.~O.,  2019, \mn@doi [\mnras]
  {10.1093/mnras/sty3489}, \href
  {https://ui.adsabs.harvard.edu/abs/2019MNRAS.483.5595V} {483, 5595}

\bibitem[\protect\citeauthoryear{{Verbunt} \& {Zwaan}}{{Verbunt} \&
  {Zwaan}}{1981}]{Verbunt1981}
{Verbunt} F.,  {Zwaan} C.,  1981, \aap, \href
  {https://ui.adsabs.harvard.edu/abs/1981A&A...100L...7V} {100, L7}

\bibitem[\protect\citeauthoryear{{Warner}}{{Warner}}{2003}]{Warner_2003}
{Warner} B.,  2003, {Cataclysmic Variable Stars},
  \mn@doi{10.1017/CBO9780511586491.
}

\bibitem[\protect\citeauthoryear{{Weber} \& {Davis}}{{Weber} \&
  {Davis}}{1967}]{Weber1967}
{Weber} E.~J.,  {Davis} Leverett J.,  1967, \mn@doi [\apj] {10.1086/149138},
  \href {https://ui.adsabs.harvard.edu/abs/1967ApJ...148..217W} {148, 217}

\bibitem[\protect\citeauthoryear{{Wright}, {Drake}, {Mamajek}  \&
  {Henry}}{{Wright} et~al.}{2011}]{Wright2011}
{Wright} N.~J.,  {Drake} J.~J.,  {Mamajek} E.~E.,   {Henry} G.~W.,  2011,
  \mn@doi [\apj] {10.1088/0004-637X/743/1/48}, \href
  {https://ui.adsabs.harvard.edu/abs/2011ApJ...743...48W} {743, 48}

\bibitem[\protect\citeauthoryear{{Zahn}}{{Zahn}}{1977}]{Zahn1977}
{Zahn} J.~P.,  1977, \aap, \href
  {https://ui.adsabs.harvard.edu/abs/1977A&A....57..383Z} {57, 383}

\bibitem[\protect\citeauthoryear{{Zorotovic} et~al.,}{{Zorotovic}
  et~al.}{2016}]{Zorotovic2016}
{Zorotovic} M.,  et~al., 2016, \mn@doi [\mnras] {10.1093/mnras/stw246}, \href
  {https://ui.adsabs.harvard.edu/abs/2016MNRAS.457.3867Z} {457, 3867}

\bibitem[\protect\citeauthoryear{{van Saders}, {Ceillier}, {Metcalfe}, {Silva
  Aguirre}, {Pinsonneault}, {Garc{\'\i}a}, {Mathur}  \& {Davies}}{{van Saders}
  et~al.}{2016}]{vanSaders2016}
{van Saders} J.~L.,  {Ceillier} T.,  {Metcalfe} T.~S.,  {Silva Aguirre} V.,
  {Pinsonneault} M.~H.,  {Garc{\'\i}a} R.~A.,  {Mathur} S.,   {Davies} G.~R.,
  2016, \mn@doi [\nat] {10.1038/nature16168}, \href
  {https://ui.adsabs.harvard.edu/abs/2016Natur.529..181V} {529, 181}

\makeatother
\end{thebibliography}

\appendix
 
\section{Analytic period evolution}
\label{sec:analytic_appendix}
Here we solve for the expected orbital period evolution of a binary losing angular momentum via different MB prescriptions. 

We begin with the \citetalias{Rappaport1983} prescription (Equation~\ref{eq:jdot}).
We work in dimensionless variables $m=M/M_{\odot}$, $r=R/R_{\odot}$, and $p_{\rm orb} = P_{\rm orb}/{1\,\rm day}$. For convenience, we also define a characteristic angular momentum loss timescale $T_{0}=\left(1\,{\rm day}\right)^{1/3}M_{\odot}^{5/3}\left(\frac{G^{2}}{2\pi}\right)^{1/3}\tau_{0}^{-1}\approx5.8\,{\rm Gyr}$, with corresponding dimensionless variable $t=T/T_0$, with $T$ the binary's age. From the chain rule, $\dot{P}_{{\rm orb}}=\dot{J}_{{\rm MB}}\frac{dP_{{\rm orb}}}{dJ_{{\rm orb}}}$. Replacing $\dot{J}_{\rm MB}$ and $J_{\rm orb}$ with appropriate expressions from Equations~\ref{eq:jdot} and~\ref{eq:jorb_circ}, rewriting in terms of dimensionless variables, and including the contributions of both stars to $\dot{J}_{\rm MB}$, we find 
\begin{equation}
\label{eq:dp_dt_normal}
\frac{dp_{{\rm orb}}}{dt}=-\frac{3\left(r_{1}^{4}+qr_{2}^{4}\right)\left(q+1\right)^{1/3}}{qm_{1}^{2/3}}p_{{\rm orb}}^{-7/3}.
\end{equation}
Here $r_1=R_1/R_{\odot}$, $r_2 = R_2/R_{\odot}$, and $m_1 = M_1/M_{\odot}$. Given an initial period $p_{\rm orb,0}$, the predicted evolution is then
\begin{align}
    \label{eq:porb_normal}
    p_{{\rm orb}}(t)=\left[p_{{\rm orb,\,}0}^{10/3}-\frac{10\left(r_{1}^{4}+qr_{2}^{4}\right)\left(q+1\right)^{1/3}}{qm_{1}^{2/3}}t\right]^{3/10}.
\end{align}

A similar calculation can be carried out for saturated MB. In this case, we define a timescale $T_{1}=\frac{\left(1\,{\rm d}\right)^{4/3}M_{\odot}^{5/3}G^{2/3}}{\left(2\pi\right)^{4/3}K_{W}\omega_{{\rm crit,1}}^{2}}$, and $t=T/T_1$. Finally, we define $ \xi=P_{{\rm crit},1}/P_{{\rm crit,2}}$. This leads to predicted period evolution:

\begin{equation}
    \label{eq:porb_sat}
    p_{{\rm orb}}(t)=\left[p_{{\rm orb,\,}0}^{4/3}-\frac{4\left(r_{1}^{1/2}+\xi^{2}\sqrt{r_{2}/q}\right)\left(q+1\right)^{1/3}}{qm_{1}^{13/6}}t\right]^{3/4}.
\end{equation}

For the \citet[][]{Garraffo2016} and \citet{Van2019} MB laws, the analytic equations are more difficult to solve, so we calculate the period evolution numerically.

\section{Kinematic ages}
\label{sec:ages_appendex}

To roughly estimate the ages of the EBs in our sample, we calculated their plane-of-the-sky tangential velocities, $v_{\perp}=4.74\,{\rm km\,s^{-1}}\left(\frac{\mu}{{\rm mas\,yr^{-1}}}\right)\left(\frac{\varpi}{{\rm mas}}\right)^{-1}$. Older stars both have experienced more dynamical heating since their formation and were born from kinematically hotter gas \citep[e.g.][]{Seabroke2007}, and thus have larger typical velocities with respect to the local standard of rest. 

\begin{figure*}
    \centering
    \includegraphics[width=\textwidth]{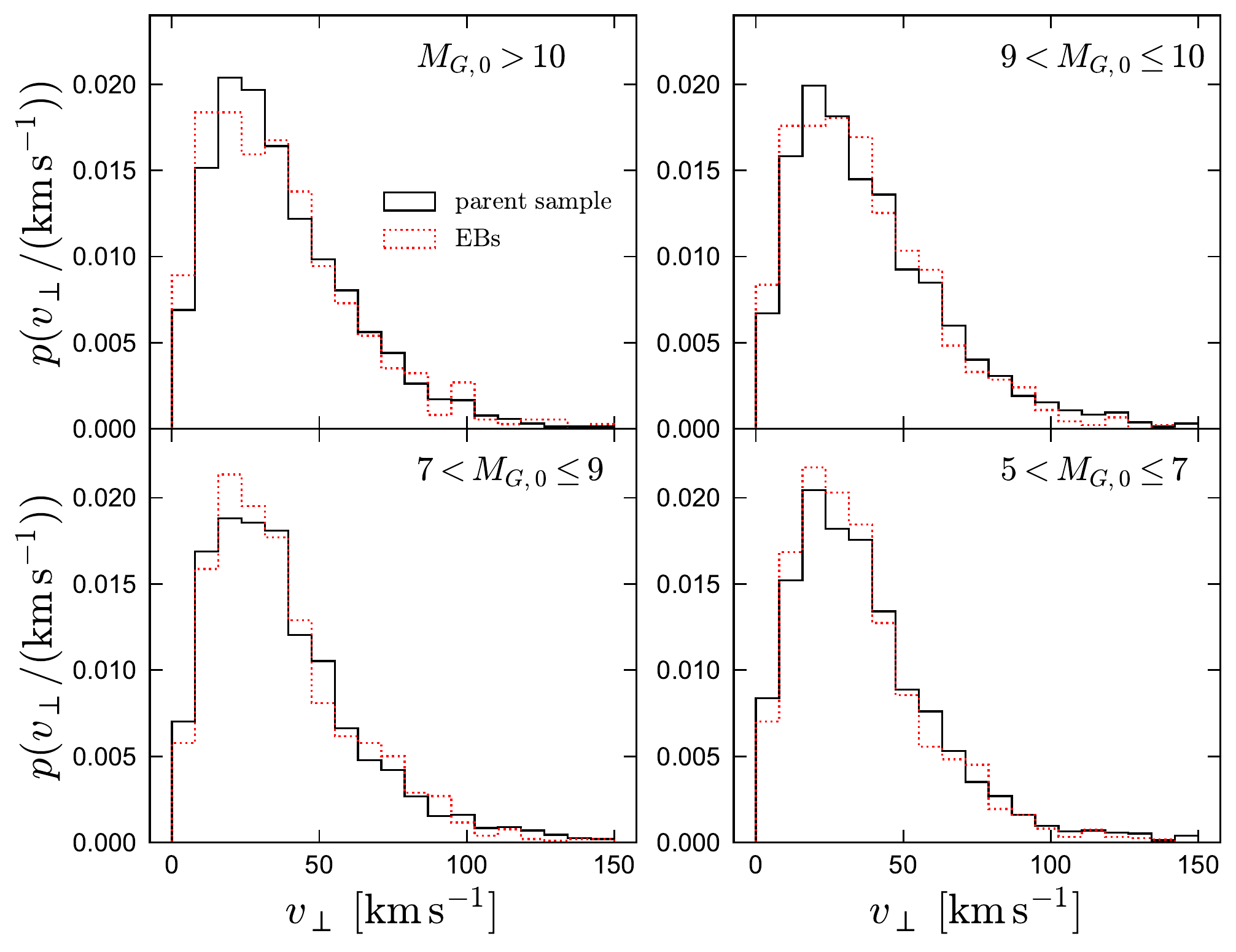}
    \caption{Distributions of plane-of-the-sky tangential velocity, calculated from {\it Gaia} astrometry, for EBs (red dotted) and the parent light curve samples from which they were selected (solid black). Tangential velocity, $v_{\perp}$, is a rough proxy for age. In all the mass bins we consider, the $v_{\perp}$ distributions of EBs are similar to those of the parent samples. This suggests that the age distributions of the EBs is similar to that of the local Galactic disk.}
    \label{fig:vperp_distributions}
\end{figure*}

Figure~\ref{fig:vperp_distributions} compares the $v_{\perp}$ distributions of EBs in each absolute magnitude bin to the $v_{\perp}$ distributions of all stars whose light curves were searched. Overall, the distributions are similar in each absolute magnitude bin. This implies that the EBs in our sample have a similar age distribution to the local field population. 

\begin{figure*}
    \centering
    \includegraphics[width=\textwidth]{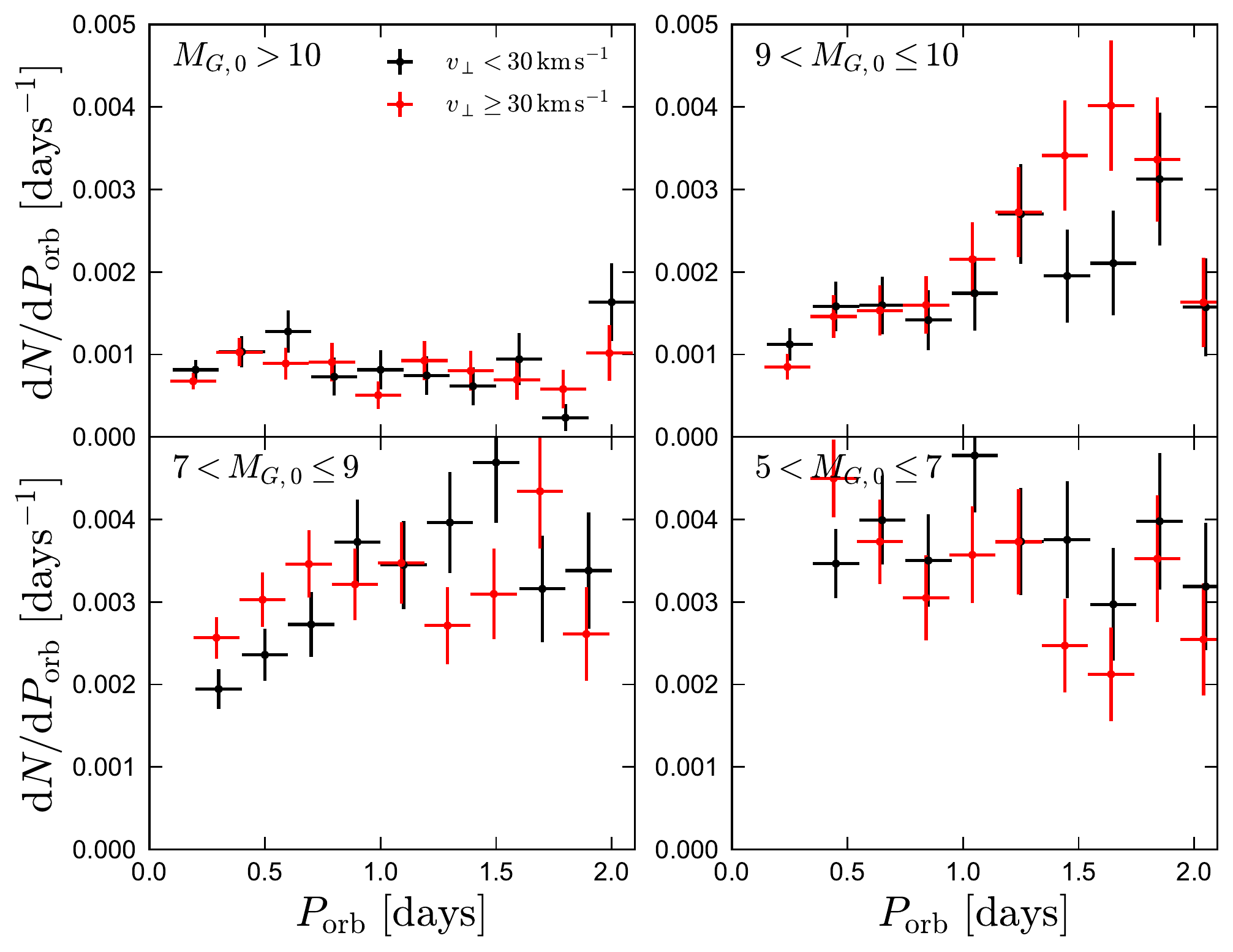}
    \caption{Incompleteness-corrected period distributions (similar to Figure~\ref{fig:corrected_period_distributions}), split by kinematic age. Black and red points correspond to younger and older populations. The period distributions in the young and old populations are broadly similar.   }
    \label{fig:vperp_cut_period_distributions}
\end{figure*}

We also searched for potential difference between the period distributions of kinematically youthful and older binaries. Figure~\ref{fig:vperp_cut_period_distributions} shows the incompleteness-corrected period distributions of binaries with tangential velocities above (red) and below (black) a threshold of $v_{\perp} = 30\,\rm km\,s^{-1}$.  Any differences between the distributions for young and old binaries (in either normalization or slope) are marginal. Most importantly, we can rule out a scenario in which the shortest-period EBs are missing in the young population (not having had sufficient time to reach short periods) or in the older population (having all been depleted by MB). 

Previous work by \citet{Hwang2020} has found an excess of short-period binaries with tangential velocities of 20-40 km/s. They interpreted this as evidence that typical main-sequence close binaries have lifetimes of a few Gyr, and then merge. In our sample, an excess of EBs with intermediate kinematic ages is weakly evident in the two higher-mass bins, and not in the lower-mass bins. We suspect this is primarily because the \citet{Hwang2020} analysis was (a) focused on solar-type binaries, which are more massive than most of our sample, and (b) dominated by contact binaries, whereas we focus on detached binaries and attempt to explicitly remove contact systems. In contrast to the situation at lower masses, short-period solar-type binaries are dominated by contact systems. This suggests that solar-type contact binaries survive in a stable configuration for long timescales -- likely longer than the detached inspiral timescale. 

\bsp	
\label{lastpage}
\end{document}